# Who voted for a No Deal Brexit? A Composition Model of Great Britain's 2019 European Parliamentary Elections

Stephen D Clark,
Consumer Data Research Centre and School of Geography,
University of Leeds,
LEEDS,
LS2 9JT
Email : tra6sdc@leeds.ac.uk

## Abstract

The purpose of this paper is to use the votes cast at the 2019 European elections held in United Kingdom to re-visit the analysis conducted subsequent to its 2016 European Union referendum vote. This exercise provides a staging post on public opinion as the United Kingdom moves to leave the European Union during 2020. A composition data analysis in a seemingly unrelated regression framework is adopted that respects the compositional nature of the vote outcome; each outcome is a share that adds up to 100% and each outcome is related to the alternatives. Contemporary explanatory data for each counting area is sourced from the themes of socio-demographics, employment, life satisfaction and place. The study find that there are still strong and stark divisions in the United Kingdom, defined by age, qualifications, employment and place. The use of a compositional analysis approach produces challenges in regards to the interpretation of these models, but marginal plots are seen to aid the interpretation somewhat.

## 1 Introduction

As a consequence of the inability of the United Kingdom (UK) government to secure Parliamentary approval for its withdrawal deal, the UK took part in the European Union (EU) elections in late May 2019. Such EU elections in the UK are quite untypical of other national elections and these elections were effectively transformed into a referendum on how the UK should leave the EU. The Conservatives were the party in power and officially supported the deal that it had negotiated with the EU, and its manifesto was that the UK should leave the EU with this deal at the earliest opportunity. The position of the Labour party was that in the event of the government failing to get a deal in place there should be a general election to elect a new parliament but, failing this, whatever deal was proposed should be put to a national vote. The Liberal Democrats, the Green Party and the nationalist parties in Scotland and Wales were in support of an immediate second referendum (a People's Vote). A new party, the Brexit party, wanted an exit from the EU, deal or no deal, by 29[th] October 2019, and another new party, Change UK was also formed with a manifesto for a People's Vote. Clarity on how the these various positions on Brexit were viewed by the public is available from May 2019 polling evidence (YouGov, 2019b) which showed 60% of those who intended to vote Conservative supported the notion to Get Brexit Done (with or without a deal), whilst 67% of those who intended to vote Labour, supported remaining in the EU. For comparison, 98% of Brexit party voters wanted to Get Brexit Done whilst 92% of Liberal Democrats wanted to remain in the EU. There is also another sizeable block of the electorate, those who are registered to vote, but decided not to vote. It is difficult to discern what this potentially heterogeneous group of non-voters may think about Brexit.

It may appear to be unfair to case these elections as being this one dimensional, however some points support this perspective. Firstly traditional and social media constantly ran with this Get Brexit Done vs People's Vote narrative, crowding out any other issues. Secondly, the UK electorate generally fails to connect EU parliamentary elections with any other of the broad issues that affect their daily life such as taxation, health, education or defence. Thirdly, to an extent, traditional party loyalties were not in play during these elections, so that assumptions that Conservatives would vote on issues around taxation whilst Labour supporters would vote on public service issues don't apply. Finally, this is the only national election that allows EU citizens resident in the UK to express their view of Brexit and it is likely that many may have used this opportunity to express a view.

Following on from the 2016 EU referendum there were many studies of the outcome, trying to discern what drove individuals or areas to vote how they did (Jackson et al., 2016). Some studies used aggregate data to understand the relationship, using count data from the 2011 Census (Harris and Charlton, 2016; Goodwin and Heath, 2016; Hanretty, 2017; Beecham et al., 2018; Leslie and Arı, 2018; Johnston et al., 2018) or supplementing such counts with results obtained from the Annual Population Survey (Alaimo and Solivetti, 2019; Leslie and Arı, 2018; Matti and Zhou, 2017). Others used a variety of individual survey data, including the British Election Survey (Curtice, 2017; Goodwin and Milazzo, 2017; Hobolt, 2016), Understanding Society (Liberini et al., 2017; Alabrese et al., 2019), Survation (Alabrese and Fetzer, 2018) and the Eurobarometer (Carl et al., 2018). What emerges from all these studies is that Leave reported greatest support amongst those who were older, educated to below degree level and of a white British ethnicity. Those employed in low or semi-skilled occupations and those living in areas that had experienced recent immigration were also more likely to vote Leave. Looking purely at geography, the Welsh voted similarly to the English (Jones, 2017), the Scots voted differently to English (Paddison and Rae, 2017) and (inner) Londoners voted differently to most of the rest of England (Johnston et al., 2018).

These 2019 EU parliamentary elections allows these earlier findings to be re-visited and reproduced, providing an important marker post on what looks like the UK's eventual departure from the EU. What complicates this is that the EU elections were not a binary outcome (Leave/Remain) but can be grouped into three factions (Get Brexit Done; a People's Vote and Did Not Vote) by the aggregation of party support. What is then needed is a modelling framework that models compositions. Such compositions are characterised by each part falling in some range and that the summation of these parts gives a fixed total. To accommodate these features, here a compositional analysis framework is adopted (Aitchison, 1982). These models are applied to the 2019 EU election results using contemporary explanatory data from the domains of demography, economics and place.

## 2  Data and Methods

The 2019 EU votes in each local government district for the Conservatives, the Brexit party, UKIP, the English Democrats and the Independent Network are combined to obtain the Get Brexit Done vote. The votes for Labour, the Liberal Democrats, the Greens, Change UK, the Scottish National Party, Plaid Cymru, the Animal Welfare Party, the Women's Equality Party and the United Kingdom European Union Party are combined for the People's Vote option. This leaves a small proportion of unaccounted votes in these data, but together they only polled around 0.7% of the votes. Figure 1 maps the share of votes for each of these factions and Figure S1 of the supplementary material provides the same information as cartograms.

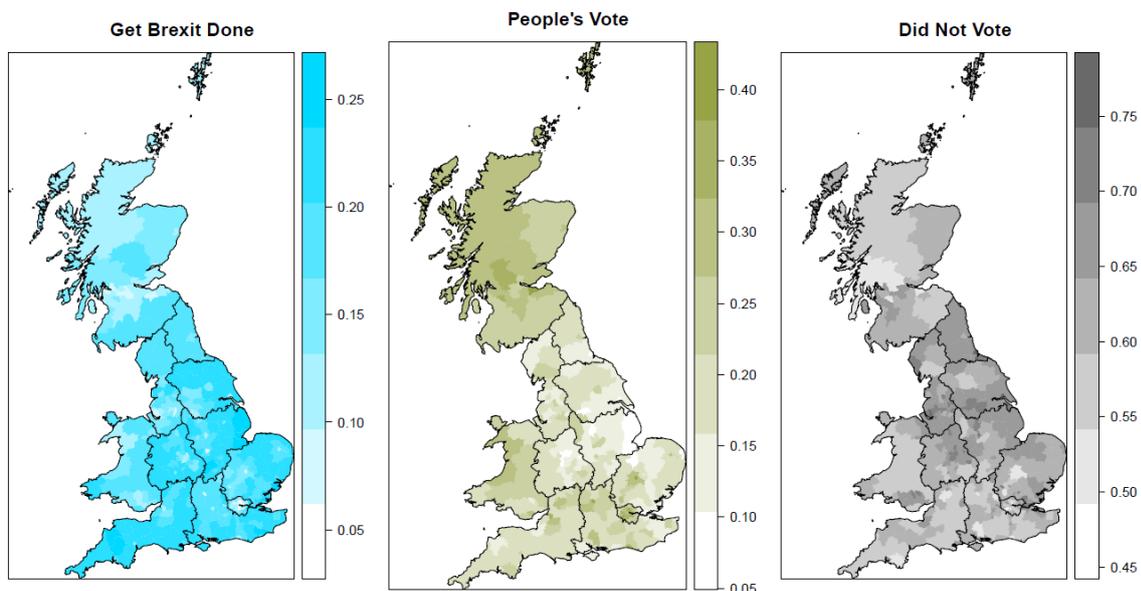

Figure 1 : Geographic distribution of the votes for Get Brexit Done; People's Vote and Did Not Vote.

Demographics are measured using age (Office for National Statistics, 2019b), ethnic composition (Rees et al., 2016), education (NOMIS, 2019), and migration (Office for National Statistics, 2018a). The size of the economically active population, the unemployment rate, the share of employment by industry (NOMIS, 2019) are used to capture the level and nature of employment within the authority. Social conditions are captured using measures of satisfaction with life (Office for National Statistics, 2019c), the degree of austerity (Gray and Barford, 2018), levels of crime (Office for National Statistics, 2018b) and disposable household income (Office for National Statistics, 2019a). The final two variables are place indicators associated with the local authority. The dominant political affiliation for the authority is recorded by a categorical variable and a geographic categorical variable indicates which region or country the authority belongs in. These data are summarised in Table 1 and compiled into an EXCEL workbook as Clark (2020). The relationship between all these potential explanatory variables and the three outcomes are plotted in supplementary figures S2.

Table 1 : Descriptive statistics (% unless otherwise stated)

| Variable | N | Min | Mean | Median | Max |
| --- | --- | --- | --- | --- | --- |
| Get Brexit Done | 369 | 4.2 | 17.3 | 18.2 | 25.7 |
| Hold People's Vote | 369 | 7.3 | 19.6 | 18.2 | 41.0 |
| Did Not Vote | 369 | 46.4 | 63.1 | 63.2 | 77.1 |
| Mean age (years) | 369 | 39.6 | 50.2 | 50.7 | 57.0 |
| Diversity (index) | 369 | 4.9 | 26.4 | 19.3 | 77.5 |
| Degree or better | 369 | 10.6 | 38.3 | 37.5 | 70.7 |
| Some qualification | 369 | 25.9 | 54.3 | 55.1 | 78.8 |
| No qualification | 369 | 0.7 | 7.4 | 6.9 | 20.4 |
| Non-UK born (2008) | 369 | 0.0 | 9.4 | 6.0 | 53.6 |
| Growth in non-UK born (2008-18) | 369 | 0.4 | 1.6 | 1.4 | 6.0 |
| Economically Active | 369 | 64.8 | 79.4 | 79.4 | 91.1 |
| Unemployment | 369 | 0.0 | 3.7 | 3.9 | 10.6 |
| Agriculture | 369 | 0.0 | 1.2 | 0.5 | 9.7 |
| Mining & quarrying | 369 | 0.0 | 1.4 | 1.1 | 12.7 |
| Manufacturing | 369 | 0.7 | 9.8 | 8.8 | 37.5 |
| Construction | 369 | 1.5 | 5.4 | 5.2 | 20.4 |
| Motor trades | 369 | 0.1 | 1.9 | 1.9 | 6.0 |
| Wholesale | 369 | 0.7 | 4.1 | 3.7 | 10.8 |
| Retailer | 369 | 3.4 | 10.1 | 10.0 | 27.2 |
| Transport & storage | 369 | 1.2 | 4.8 | 3.8 | 24.7 |
| Accommodation & food services | 369 | 2.7 | 7.7 | 7.1 | 18.8 |
| Information & communication | 369 | 0.5 | 3.6 | 2.8 | 18.0 |
| Finance & insurance | 369 | 0.4 | 2.2 | 1.4 | 23.4 |
| Property | 369 | 0.3 | 1.5 | 1.4 | 4.5 |
| Professional & scientific | 369 | 2.1 | 7.1 | 6.2 | 23.8 |
| Business adminstration & support | 369 | 1.9 | 8.1 | 7.5 | 35.4 |
| Public administration & defence | 369 | 0.8 | 4.0 | 3.5 | 11.6 |
| Education | 369 | 3.0 | 9.1 | 8.8 | 29.7 |
| Health | 369 | 4.1 | 13.5 | 13.2 | 29.2 |
| Arts, entertainment & recreation | 369 | 1.8 | 4.6 | 4.5 | 12.1 |
| Life satisfaction (score) | 369 | 7.1 | 7.7 | 7.7 | 8.6 |
| Total fiscal cuts | 369 | -46.1 | -17.9 | -17.3 | 1.6 |
| Crime of theft (per 1,000) | 369 | 3.4 | 29.6 | 26.9 | 147.9 |
| Crime of person (per 1,000) | 369 | 6.5 | 23.5 | 21.9 | 62.9 |
| Gross disposable household income (£) | 369 | £12,445 | £19,850 | £18,572 | £64,868 |
| No overall control | 120 | | | | |
| Conservative | 120 | | | | |
| Labour | 97 | | | | |
| Liberal Democrat | 23 | | | | |
| Independent | 8 | | | | |
| Plaid Cymru | 1 | | | | |
| North East | 12 | | | | |
| North West | 39 | | | | |
| Yorkshire & The Humber | 21 | | | | |
| West Midlands | 30 | | | | |
| East Midlands | 40 | | | | |
| East of England | 45 | | | | |
| South East | 67 | | | | |
| Inner London | 12 | | | | |
| Outer London | 20 | | | | |
| South West | 29 | | | | |
| Wales | 22 | | | | |
| Scotland | 32 | | | | |

The methods used to estimate the influence of each of these variables on the vote shares is a composition regression model, using the R package easyCODA (Greenacre, 2017) to construct the Aitchison composition, and the systemfit package (Henningsen and Hamann, 2007) to estimate the parameters. Composition models recognise that the dependant variable is a composition of three parts that sum to 1.0 and that the parts are not independent. Following the guidance in Greenacre (2017), these models are fitted using an Aitchison log-ratio (alr) transformation of the compositions (each part is divided by a common divisor part (here Get Brexit Done) and a logarithm taken of this ratio). This transformation now provides two, rather than three, equations, with Y variables ln(People's Vote/Get Brexit Done) (pPV/pBX) and ln(Did Not Vote/Get Brexit Done) (pDNV/pBX). Applying such a transformation allows traditions models (e.g. ordinary least squares) to be applied to the transformed data so that when transformed back, the fitted or predicted values satisfy the requirements for composition data. Such composition models are commonly applied in the election literature (Katz and King, 1999) and more recent such models have additionally adopted a seemingly related regression framework (Aleman and Kellam, 2008; Basinger et al., 2012; Katz and King, 1999; Magni and Reynolds, 2018; Tomz et al., 2002) which allows for a cross correlation in the errors from all the equations in the system.

Recognising that some of the candidate explanatory variables will explain very little of the variation in the relationships, a forward stepwise approach is used to build up equations. This process starts with two empty equations consisting of just an intercept term, it then trials each unused variable in each of the two equations. The goodness of fit of each trial equation is measured using the AIC value, and the one variable in the one equation that produces the lowest AIC is added to that equation's variables and the, now new, lowest achieved AIC is recorded. The second step then starts with an equation with one variable, found at step one previously, and an empty model for the other equation. As before, all unused variables are trialled in each equation and the one variable in one equation whose addition reduces the AIC the most is added to that equation's variables. This process continues, adding unused variables to one equation until the AIC cannot be reduced further.

## 3   Results

The forward stepwise procedure selected the variables identified in Table 2 with an overall McElroy $R^2$ value form this system of 83.2% and the individual equations having $R^2_{adj}$ values of 85.8% (pPV/pBX) and 82.9% (pDNV/pBX).

Table 2 : Seemingly unrelated regression results (Significance: 0 '***' 0.001 '**' 0.01 '*' 0.05 '.' 0.1 ' ' 1)

| Variable | pPV/pBX | | | pDNV/pBX | | |
|---|---|---|---|---|---|---|
| Constant | 4.2896 | | *** | 8.4301 | | *** |
| Mean age (years) | -0.0545 | -5.3% | *** | -0.0469 | -4.6% | *** |
| Diversity (index) | -0.0013 | -0.1% | | -0.0040 | -0.4% | *** |
| Degree or better | 0.0140 | 1.4% | *** | | | |
| Some qualification | -0.0085 | -0.8% | * | | | |
| *No qualification* | | | | | | |
| Non-UK born (2008) | 0.0150 | 1.5% | *** | 0.0147 | 1.5% | *** |
| Growth in non-UK born (2008-18) | | | | | | |
| Economically Active | | | | -0.0031 | -0.3% | * |
| Unemployment | | | | | | |
| Agriculture | | | | -0.0264 | -2.6% | *** |
| Mining & quarrying | -0.0230 | -2.3% | * | | | |
| Manufacturing | | | | | | |
| Construction | -0.0148 | -1.5% | ** | | | |
| Motor trades | -0.0470 | -4.6% | * | -0.0518 | -5.0% | *** |
| Wholesale | | | | | | |
| Retailer | | | | | | |
| Transport & storage | | | | | | |
| Accommodation & food services | 0.0042 | 0.4% | | -0.0067 | -0.7% | * |
| Information & communication | | | | | | |
| Finance & insurance | 0.0109 | 1.1% | * | | | |
| Property | 0.0450 | 4.6% | * | | | |
| Professional & scientific | | | | | | |
| Business administration & support | | | | | | |
| Public administration & defence | 0.0122 | 1.2% | * | | | |
| Education | 0.0122 | 1.2% | * | -0.0055 | -0.6% | . |
| Health | | | | 0.0023 | 0.2% | |
| Arts, entertainment & recreation | | | | | | |
| Life satisfaction (score) | | | | -0.1036 | -9.8% | |
| Total fiscal cuts | -0.0052 | -0.5% | ** | -0.0033 | -0.3% | ** |
| Crime of theft (per 1,000) | -0.0051 | -0.5% | *** | | | |
| Crime of person (per 1,000) | -0.0050 | -0.5% | * | -0.0045 | -0.5% | *** |
| ln(gross disposable household income (£)) | -0.1614 | -14.9% | | -0.4303 | -35.0% | *** |
| No overall control | -0.1210 | -11.4% | *** | -0.0423 | -4.1% | * |
| Conservative (0/1) | -0.2384 | -21.2% | *** | -0.0849 | -8.1% | *** |
| Labour (0/1) | | | | | | |
| Independent (0/1) | | | | | | |
| Liberal Democrat (0/1) | | | | 0.1120 | 11.9% | ** |
| Plaid Cymru (0/1) | 0.5252 | 69.1% | * | 0.1891 | 20.8% | |
| North East (0/1) | | | | -0.0671 | -6.5% | . |
| North West (0/1) | 0.2160 | 24.1% | *** | 0.0982 | 10.3% | *** |
| Yorkshire & The Humber (0/1) | | | | | | |
| West Midlands (0/1) | | | | | | |
| East Midlands (0/1) | | | | -0.1228 | -11.6% | *** |
| East of England (0/1) | 0.0969 | 10.2% | * | -0.0315 | -3.1% | |
| South East (0/1) | 0.1092 | 11.5% | * | -0.0330 | -3.2% | |
| Inner London (0/1) | 0.2125 | 23.7% | * | 0.1716 | 18.7% | ** |
| Outer London (0/1) | | | | -0.1068 | -10.1% | * |
| South West (0/1) | 0.1343 | 14.4% | ** | -0.1156 | -10.9% | *** |
| Wales (0/1) | 0.3667 | 44.3% | *** | | | |
| Scotland (0/1) | 0.8370 | 131.0% | *** | 0.3214 | 37.9% | *** |

## 3.1 Parameter estimates

To interpret the results in Table 2 the transformed nature of the dependant variable needs to be recognised. To this end, the exponent of the parameter estimate minus one is provided in Table 2, this reverses part of the Aitchison transformation. So for example, the impact of an additional year in the mean age reduces the ratio between the proportions who voted for a People's Vote over those who voted to Get Brexit Done by -5.3%. In all these results the strengthening of the Getting Brexit Done vote over the other option is signified by a negative parameter estimate.

Support for a People's Vote is increases when a large proportion of inhabitants living in the area have degree qualifications or are non-UK born. Also working in the professional sector of the economy and in certain regions or countries of the UK (particularly Inner London, Wales and Scotland) predicts a relatively high support for a People's Vote. Support for Getting Brexit Done is higher for areas with older and diverse populations, whose inhabitants work in skilled and semi-skilled industries, have experienced higher fiscal cuts, crime, with higher incomes and in areas where no party or the Conservatives are in control.

A preference to not actually voting is apparent for areas with a high proportion of non-UK born, who work in the Health sector, are Liberal Democrat controlled, and in the North West, Inner London or Scotland. Higher support for Getting Brexit Done than people not voting is driven by an older and more diverse population, those employed in a range of industries, have high life satisfaction but also an experience of fiscal cuts and crime, with high incomes. Areas where no party or the Conservatives are in control and some regions or countries also show a greater support for turning out to vote and Get Brexit Done.

## 3.2 Marginal plots

The estimates from a compositional model can be difficult to fully interpret, however the use of marginal plots have been found to be illuminating (Basinger et al., 2012). Such plots are constructed by predicting the outcome when the variable of interest varies but the other variables are set to either their mean or modal category. The full set of marginal plots are provided in the supplementary material as figures S3, but here four continuous variables are selected to illustrate their impact on vote share, with the full results for the two categorical variables provided in Table 3.

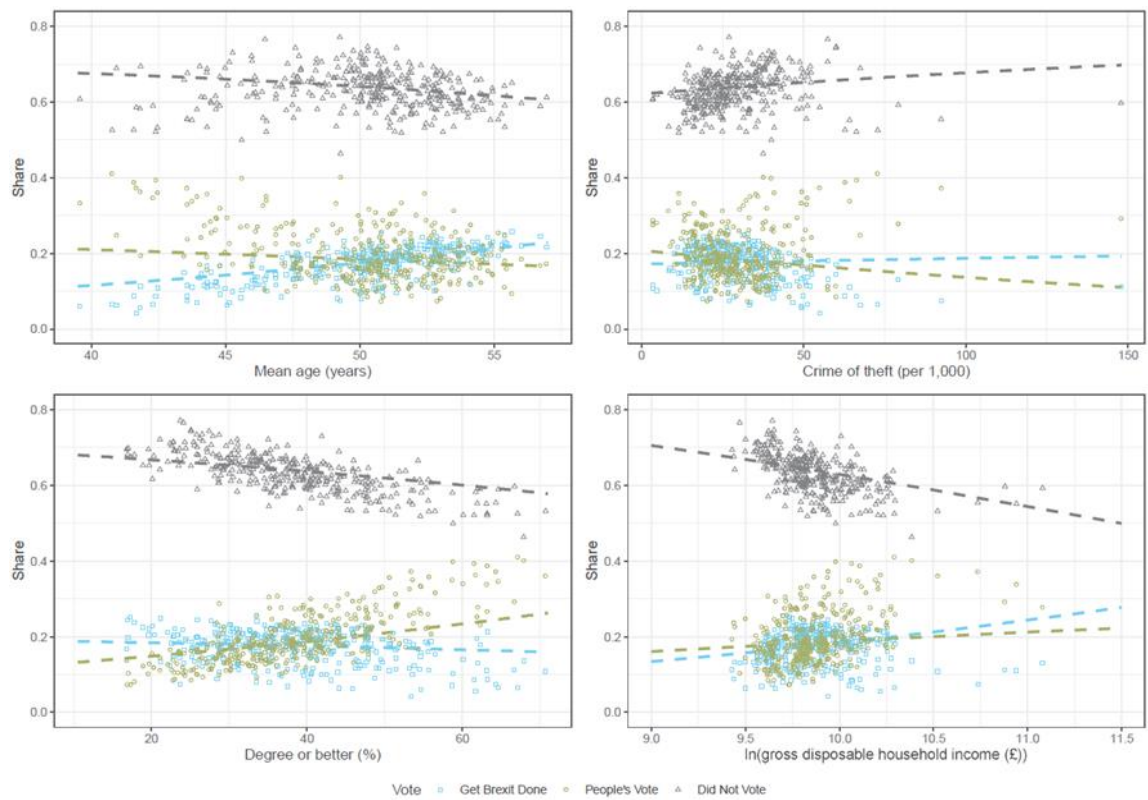

Figure 2 : Marginal plots for (a) Mean Age; (b) Degree Qualification; (c) Theft; and (d) the Disposable Household Income

In Figure 2 the lines represent the trends whilst the points are a random sample of the data points from which the model is estimated. Recall that each line is defined not only by its own points and also all others since the sum of the shares must always equal 1.0. In this figure it is clear that as the mean age increases the share of support for Getting Brexit Done increases and the proportion that support a People's Vote or Did Not Vote decreases. This supports the findings in Table 2 for mean age where the increase in the vote for Getting Brexit Done is at the expense of both a People's Vote and those who Did Not Vote. The plot for degree qualification also illustrates a similar issue, this time the increased support is evident for a People's Vote, primarily at the expense of those who Did Not Vote. The third plot for thefts shows that as the measure of thefts increases, the Did Not Vote share increases and the People's Vote decreases, whilst the Getting Brexit Done share stays stable. The final plot demonstrations that as income increases the share for both Getting Brexit Done and a People's Vote increases and these gains are at the expense of mainly non-voters.

Table 3 : Marginal shares for each categorical outcome

|  | Get Brexit Done | People's Vote | Did Not Vote |
|---|---|---|---|
| **Control** | | | |
| No overall control | 17.6% | 18.4% | 64.0% |
| Conservative | 18.5% | 17.2% | 64.3% |
| Labour | 16.8% | 19.7% | 63.5% |
| Independent | 15.6% | 18.4% | 66.0% |
| Liberal Democrat | 16.8% | 19.7% | 63.5% |
| Plaid Cymru | 13.2% | 26.3% | 60.5% |
| **Region** | | | |
| North East | 18.4% | 17.2% | 64.4% |
| North West | 15.9% | 18.4% | 65.7% |
| Yorkshire & The Humber | 17.6% | 16.5% | 66.0% |
| West Midlands | 17.6% | 16.5% | 66.0% |
| East Midlands | 19.0% | 17.8% | 63.2% |
| East of England | 17.7% | 18.2% | 64.2% |
| South East | 17.6% | 18.4% | 64.0% |
| Inner London | 15.1% | 17.5% | 67.4% |
| Outer London | 18.8% | 17.6% | 63.5% |
| South West | 18.5% | 19.8% | 61.7% |
| Wales | 16.4% | 22.1% | 61.5% |
| Scotland | 12.0% | 25.9% | 62.1% |

The marginal shares for the categorical place variables are given in Table 3. Support for Getting Brexit Done is highest in Conservative controlled authorities and lowest in Independent authorities (Plaid Cymru control just one authority, hence their estimate here and in Table 2 is unreliable). Support for a People's Vote is highest for Labour and Liberal Democrat controlled areas and lowest for Conservative areas. Those who Did Not Vote is highest in Independent controlled areas and lowest (e.g. with higher turnouts) for Labour and Liberal Democrat authorities. Looking at the regions or countries, in the North West, East of England, South East, Inner London, South West, Wales and Scotland, after non-voters, the People's Vote is the most popular option of the two options. This is especially so for Wales and Scotland.

### 3.3   Variable Importance

The order in which variables enter the equations is indicative of their importance, with the variables that reduce the AIC the most at each step considered the next most important. The first 20 variables or categories to be selected are listed in Table 4. The initial variable that gives the greatest reduction in the AIC is having a degree qualification in the equation that explains the proportion for a People's Vote in preference to Getting Brexit Done. The next four steps add the Mean Age and Scotland category to both equations. Further qualifications levels are quickly added to both equations, whilst latterly the industries of Education and Accommodation and Food Services are added. Looking more widely at this top 20, approximately 50% of the variables are various levels of the two categorical variables. The reduction AIC is also not smoothly decreasing, a large drop in AIC occurs when Scotland is added to both equations.

Table 4 : Variable Importance

| Order | Variable (equation) | AIC | Reduction |
|---|---|---|---|
| 0 | Constants (all) | 546.2 | |
| 1 | Degree or better (pPV/pBX) | 186.5 | -359.6 |
| 2 | Mean age (years) (pDNV/pBX) | 76.3 | -110.3 |
| 3 | Mean age (years) (pPV/pBX) | -81.5 | -157.7 |
| 4 | Scotland (0/1) (pPV/pBX) | -115.8 | -34.4 |
| 5 | Scotland (0/1) (pDNV/pBX) | -221.6 | -105.8 |
| 6 | North West (0/1) (pDNV/pBX) | -257.6 | -36.0 |
| 7 | Conservative (0/1) (pPV/pBX) | -289.8 | -32.2 |
| 8 | Conservative (0/1) (pDNV/pBX) | -323.5 | -33.7 |
| 9 | Wales (0/1) (pPV/pBX) | -342.6 | -19.1 |
| 10 | South West (0/1) (pDNV/pBX) | -361.8 | -19.1 |
| 11 | South East (0/1) (pDNV/pBX) | -390.1 | -28.3 |
| 12 | North West (0/1) (pPV/pBX) | -401.5 | -11.4 |
| 13 | Diversity (index) (pPV/pBX) | -415.0 | -13.5 |
| 14 | Education (pPV/pBX) | -429.2 | -14.2 |
| 15 | Economically Active (pDNV/pBX) | -439.3 | -10.1 |
| 16 | Accommodation & food services (pPV/pBX) | -447.5 | -8.2 |
| 17 | Inner London (0/1) (pDNV/pBX) | -457.6 | -10.1 |
| 18 | ln(gross disposable household income (£)) (pDNV/pBX) | -477.9 | -20.3 |
| 19 | Non-UK born (2008) (pDNV/pBX) | -497.7 | -19.7 |
| 20 | Non-UK born (2008) (pPV/pBX) | -507.1 | -9.4 |

## 4 Discussion

The primary impetus for this study has been to understand what factors influenced the vote in the UKs 2019 EU elections for two factions: the desire to Get Brexit Done verse the need for a People's Vote. Outside these two factions there is also the sizeable portion of the electorate who chose not vote to vote. In common with studies of the 2016 referendum vote, this study has identified that the main demographic drivers influencing voters' choice are their education, socio-economic status, age, sense of (dis-)advantage and culture. There are also important influences attributable to place.

A wish to Get Brexit Done is supported by areas that have populations that are older, have lower levels of qualifications and where a large proportion of the population work in skilled or semi-skilled industries. This continues the narrative that areas that are most comfortable with leaving the EU are those whose populations largely feel "left-behind", having not shared in any improvements in wages or living standards (Watson, 2018), who have encountered challenges that limit their opportunities to progress in life (Sensier and Devine, 2017) or who are uncomfortable with the societal shifts (Van de Vyver et al., 2018). A People's Vote is supported in areas where there is a sizeable non-UK born population, the industries of employment are clerical, professional or caring, and in Scotland or Wales.

Looking at a simple tally of vote shares by region, such an analysis suggests that support for Getting Brexit Done is at low levels for Inner London (8.4%), Outer London (13.2%) and Scotland (12.0%). Here however, the marginal share for a People's Vote in London provided in Table 3 here is not actually that dis-similar to the other regions of England. However such a finding needs to be taken with the knowledge that in Table 3 all the other variables are set to the Great British average and London is quite dis-similar to the rest of Great Britain, being younger, more diverse and better educated - the shares quoted here are the estimated shares if the population of London "looked" like the rest of Great Britain (which it doesn't). This interaction of composition and context is also highlighted in the study by Johnston et al. (2018) "… *that once the main individual voter characteristics of places were taken into account there was little geographical variation in support for Brexit.*" (page 175). Here, however, this does not apply to Scotland, which even after controlling for the composition of its population, the support for Getting Brexit Done stays low (Beecham et al. (2018) commented, "… *there is something fundamentally different about Scotland, not accounted for completely by census variables, that lowers preference for Leave …*", page 125).

This study has attempted to gauge public sentiment since the 2016 EU referendum and prior to the December 2019 General election. Other such attempts have been made through the lens of the 2017 General election (Heath and Goodwin, 2017; Jennings and Stoker, 2017) or through surveys (Alabrese and Fetzer, 2018). The impact of the outcome of the 2016 referendum can be discerned in these later elections, with the Conservatives performing well in smaller towns and rural constituencies, all with older and less diverse populations, whose demographic composition is similar to those traits identified here in support of Get Brexit Done. The Labour party has polled well in areas that have high People's Vote demographics (a young, educated and ethnic population) but lost support in the smaller towns with the older, lower educated demographics. These demographic shifts point to further evidence that Brexit has weakened voters' tribal associations (Sanders, 2017).

In conclusion, using what is perhaps the best "revealed preference" information available since the 2016 referendum and one that is strongly themed on Brexit, this study establishes a staging post on the move of the United Kingdom on its journey towards Brexit. Whilst the subsequent 2019 General Election was also influenced by Brexit, other themes emerged in that election too, such as the perception of the party leaders, the national health service or the performance of the economy (YouGov, 2019a). This study identifies that at the time of the 2019 Brexit dominated EU elections, that there are sizeable, distinct, opposed and definable factions within Great Britain on Brexit (Dorling, 2016) and, examined using the notion of protest votes, many voters were unhappy, either that the Brexit they voted for in 2016 had not yet been delivered or conversely that it was an imminent and unstoppable outcome.

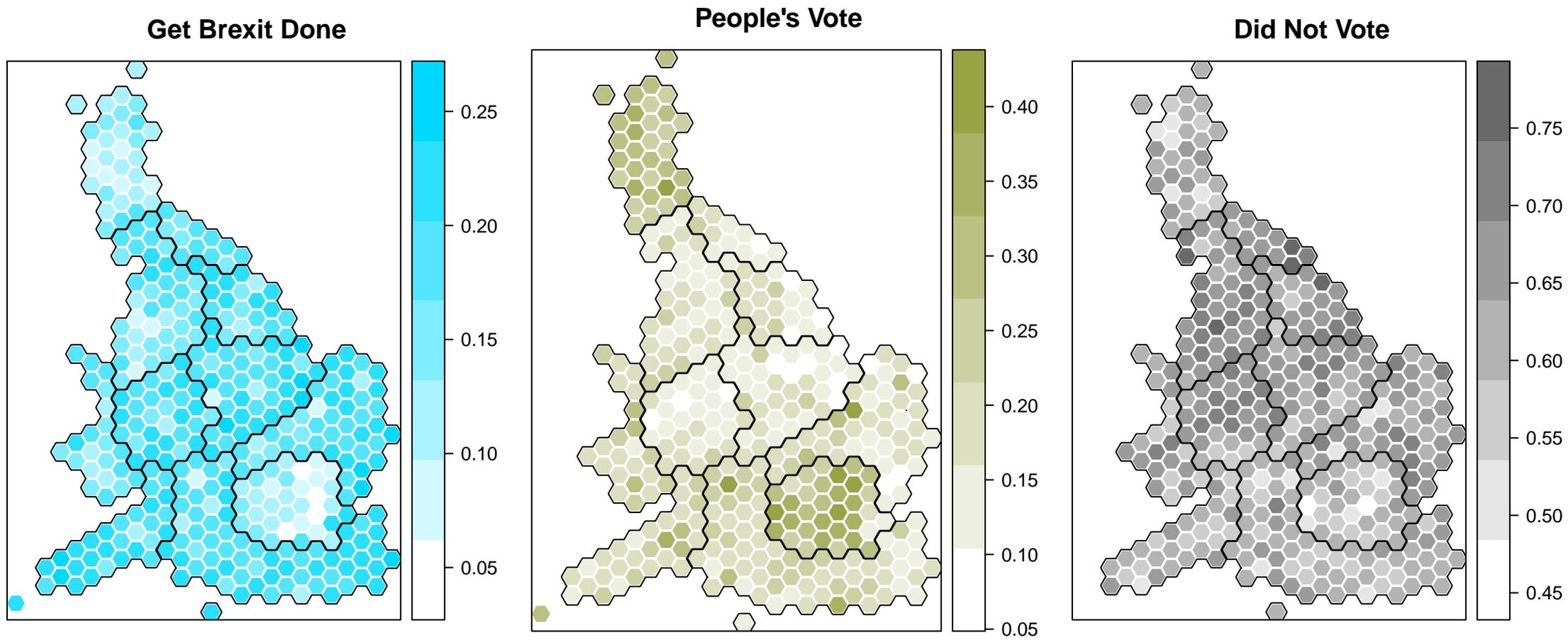

Figure 1S : Distribution of the votes for Get Brexit Done; People's Vote and Did Not Vote

Figure S2 : relationship between all these potential explanatory variables and the three outcomes

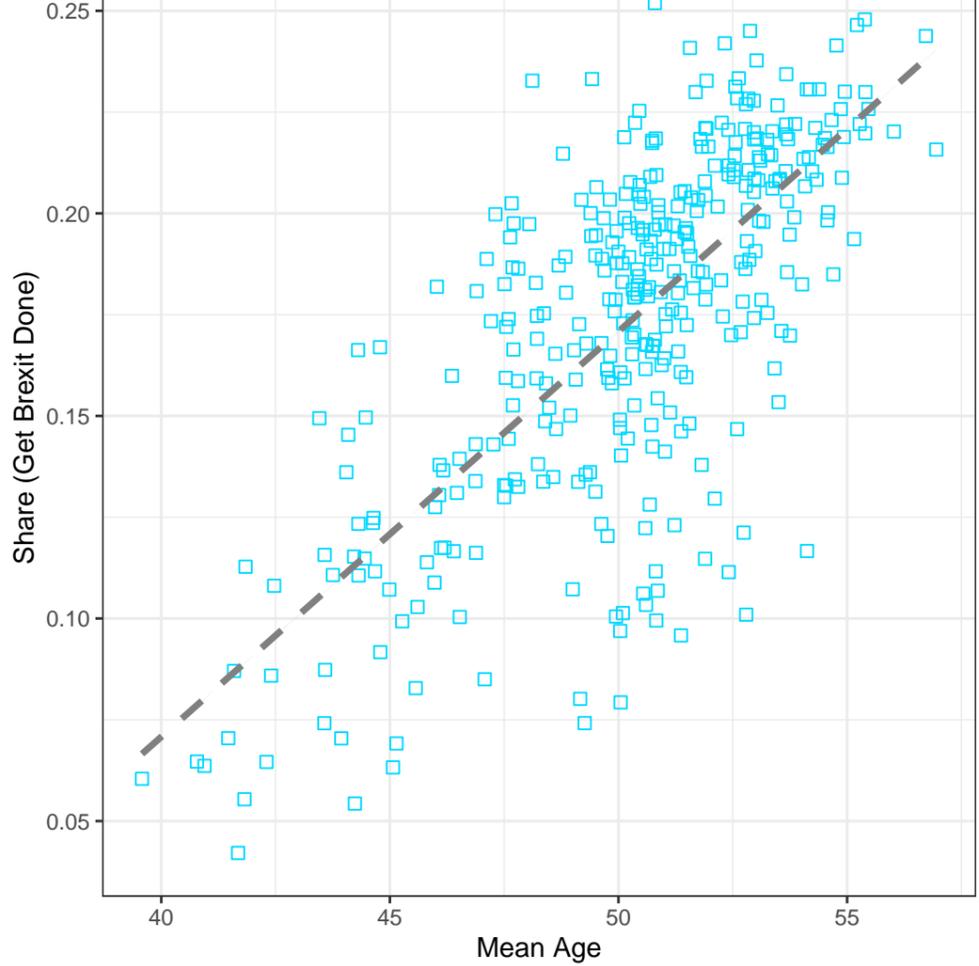
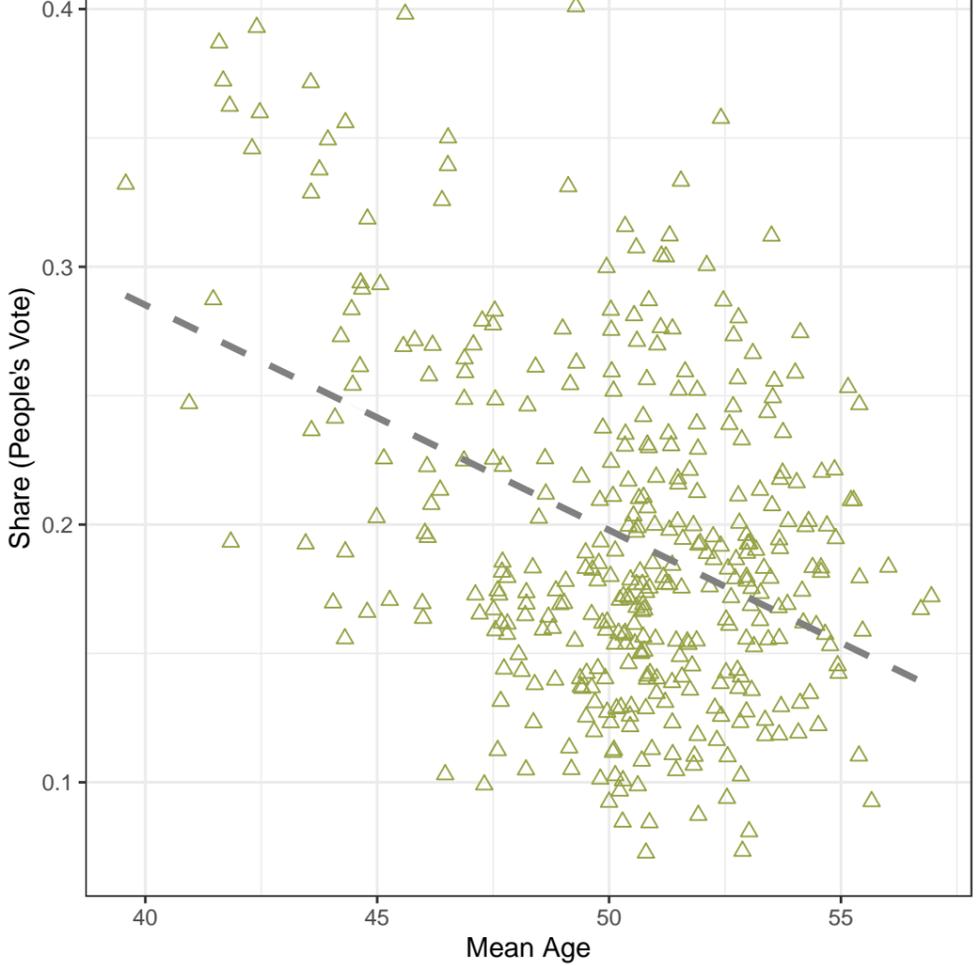
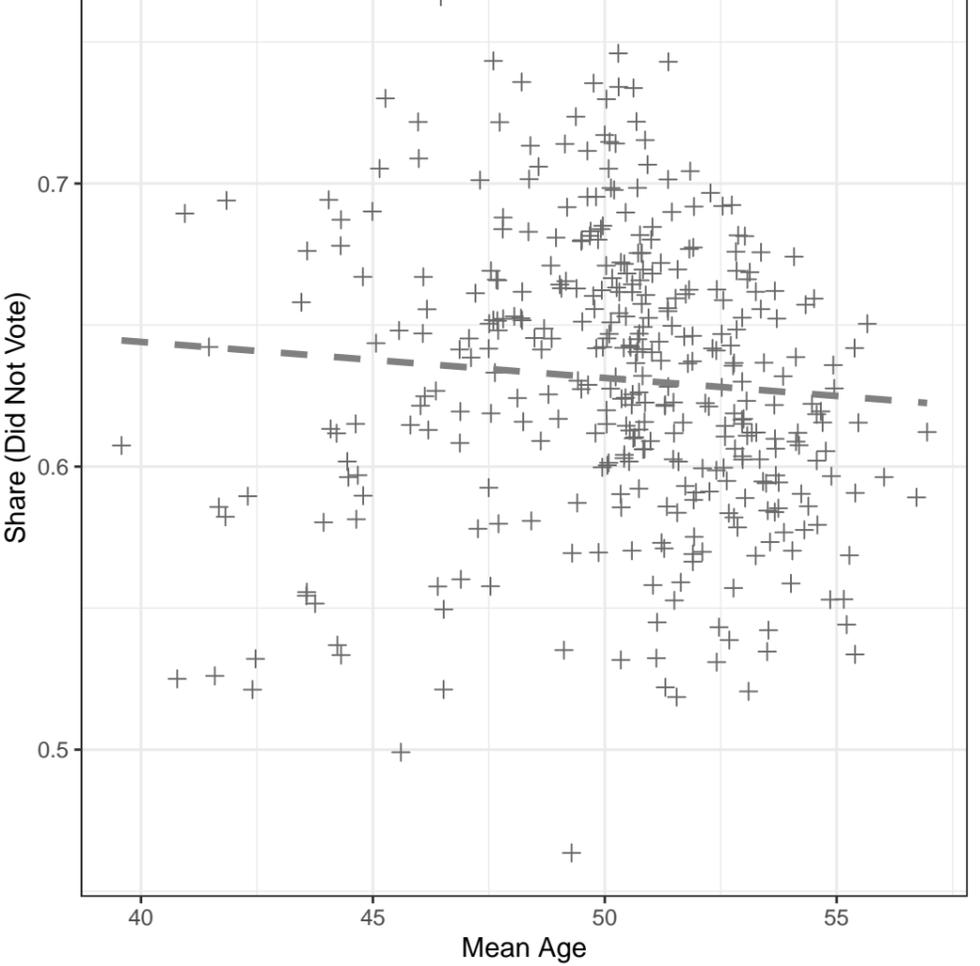
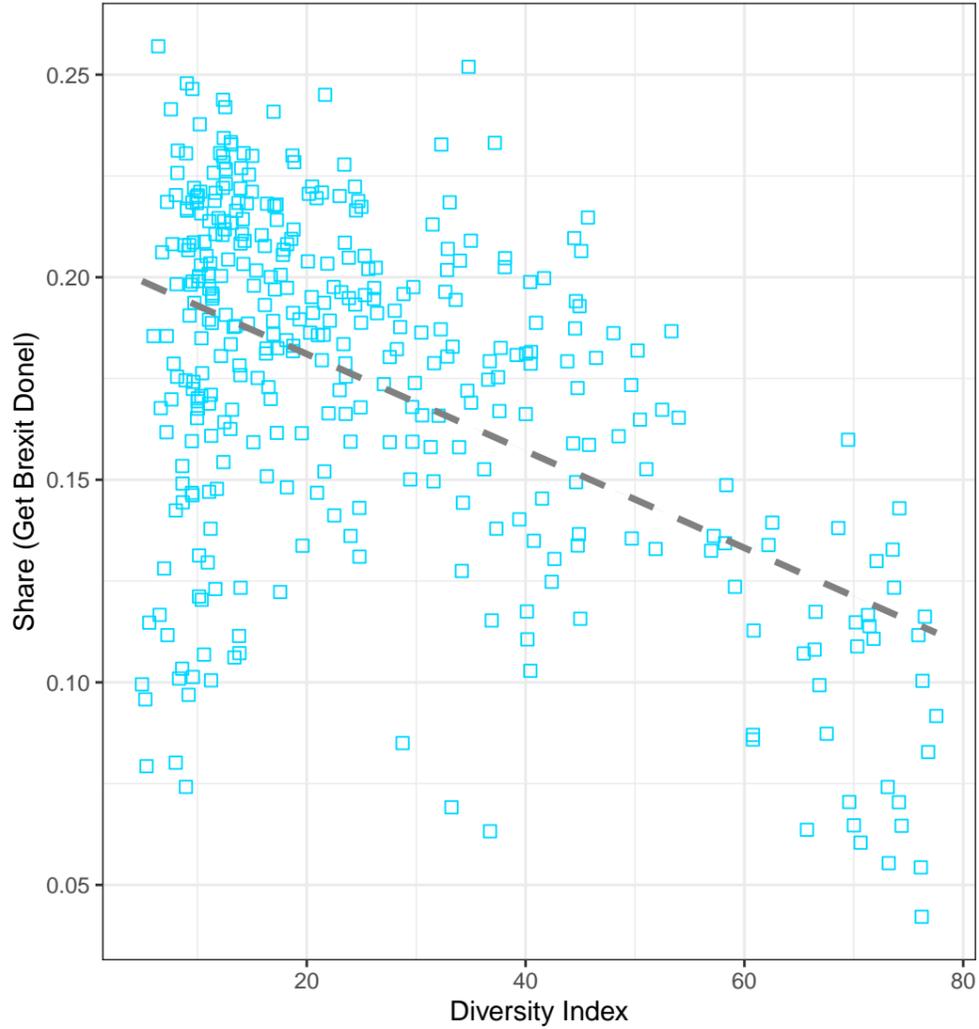
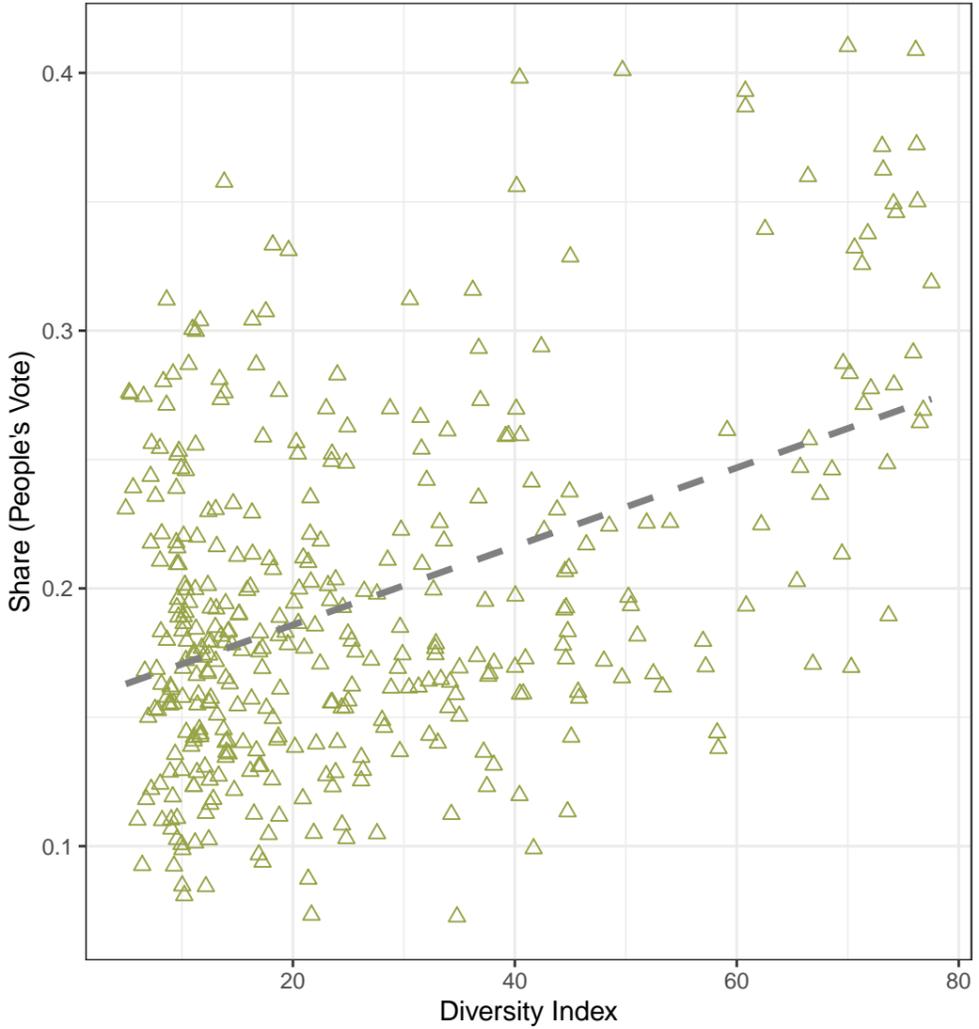
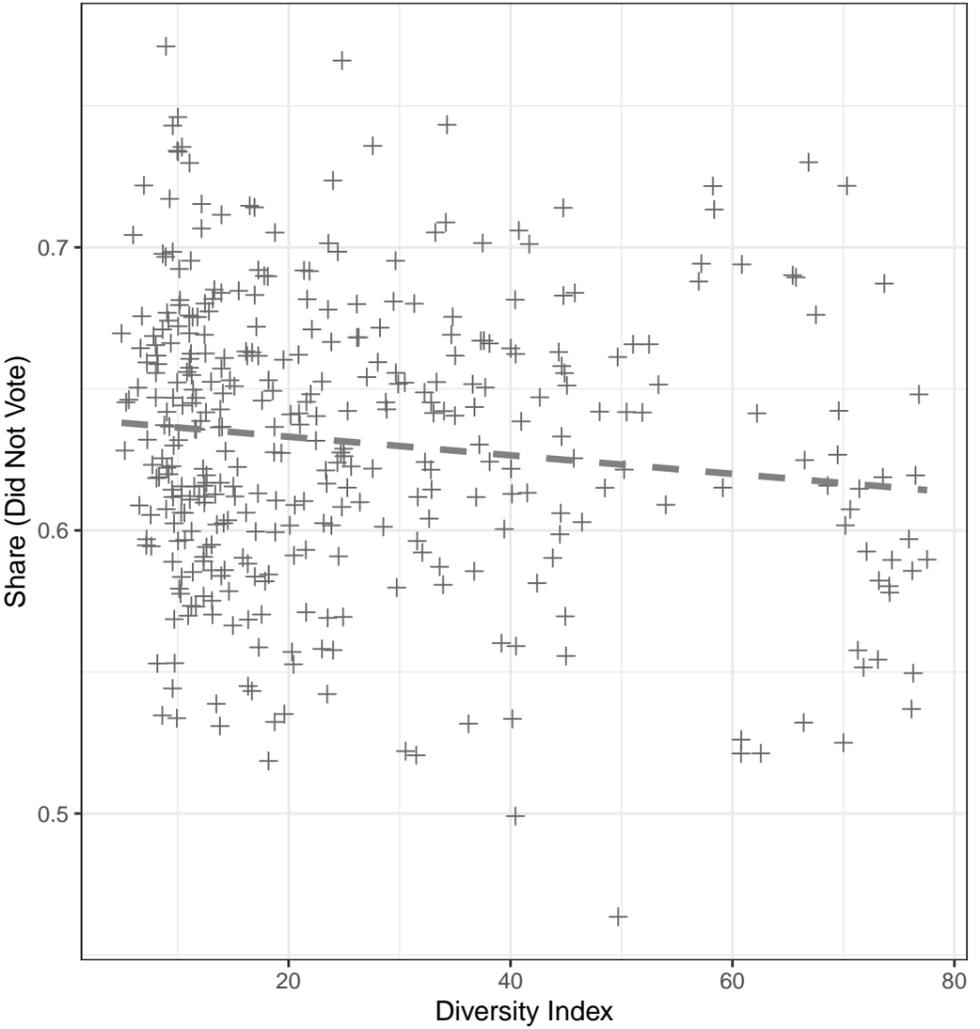

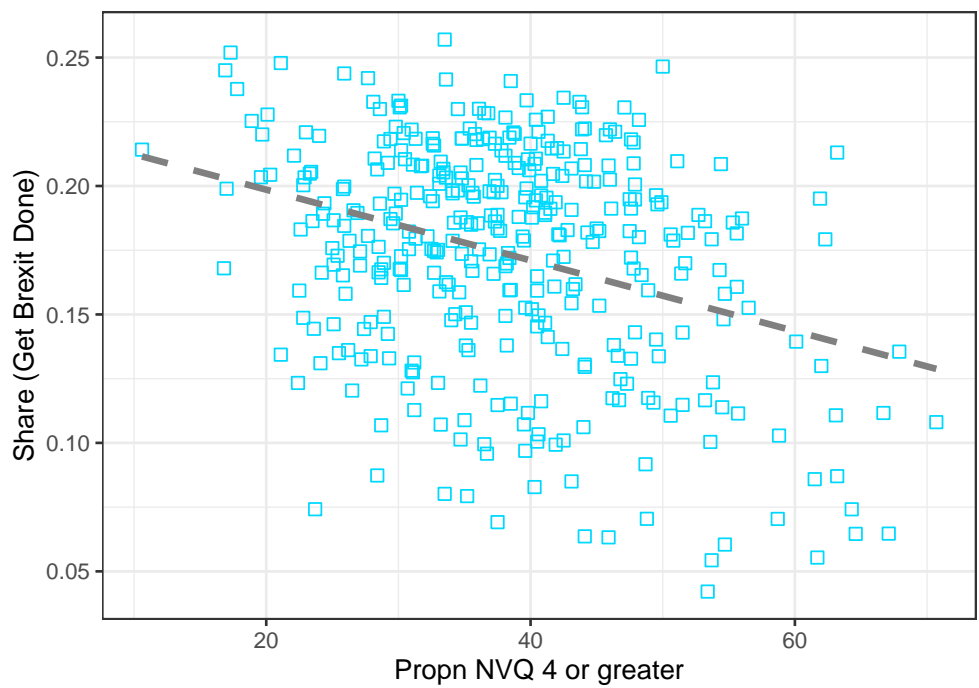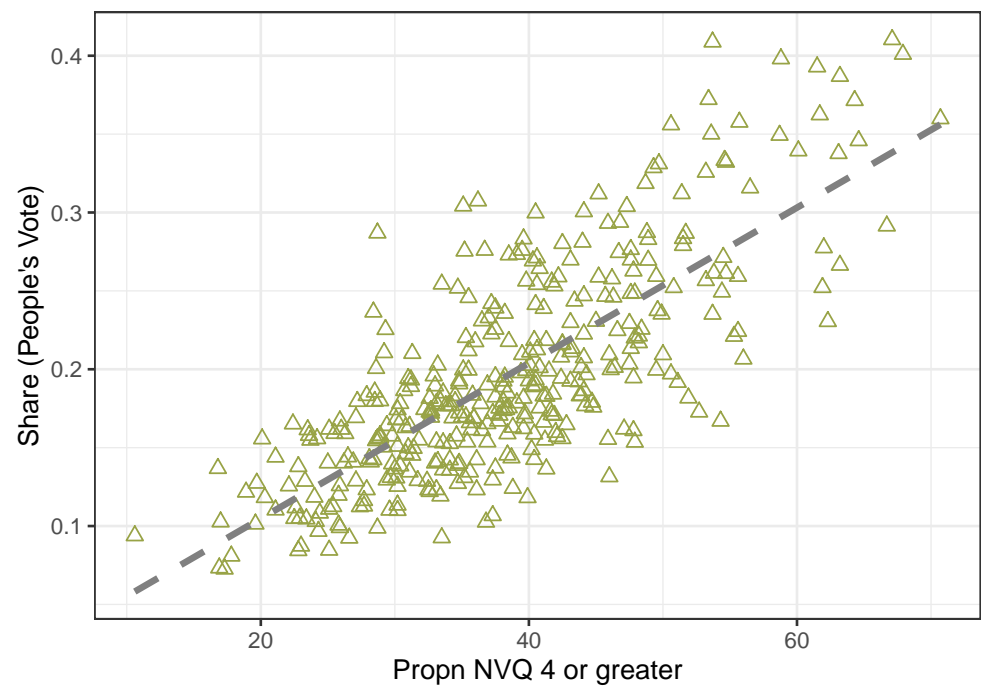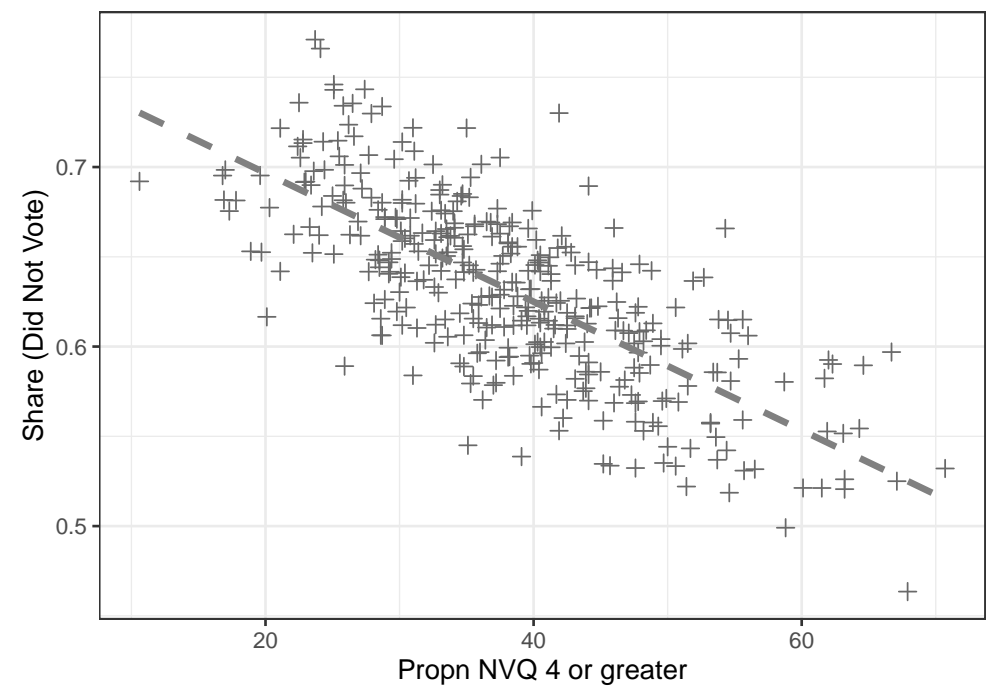
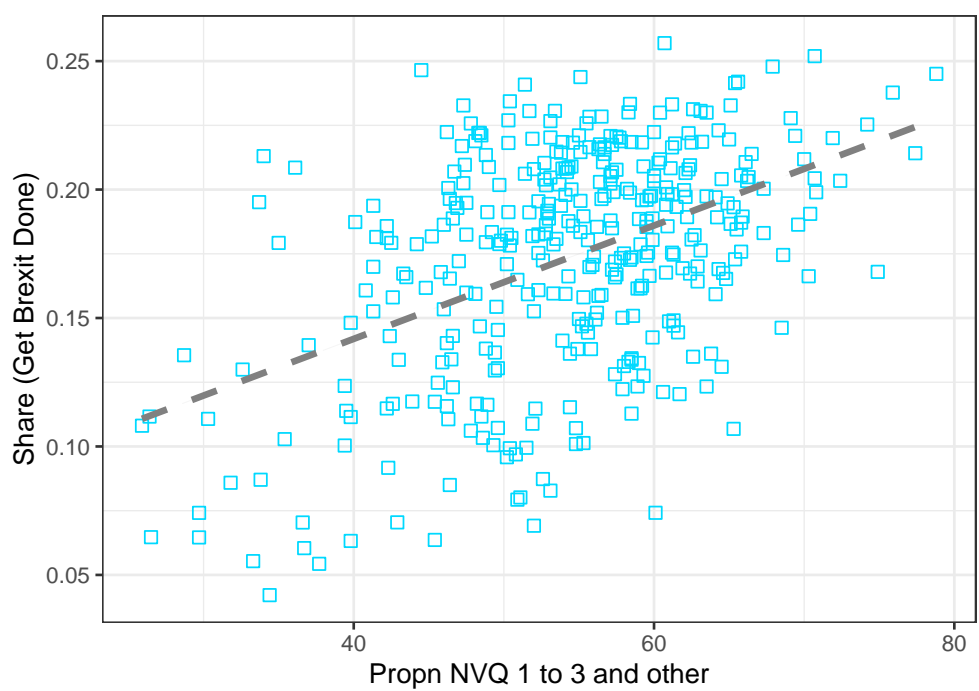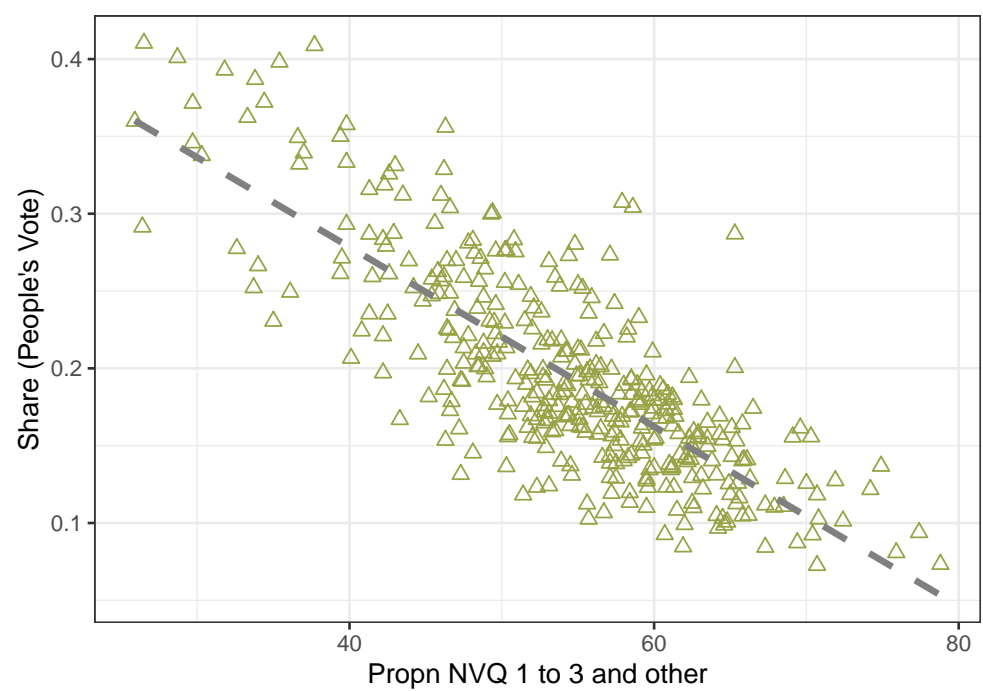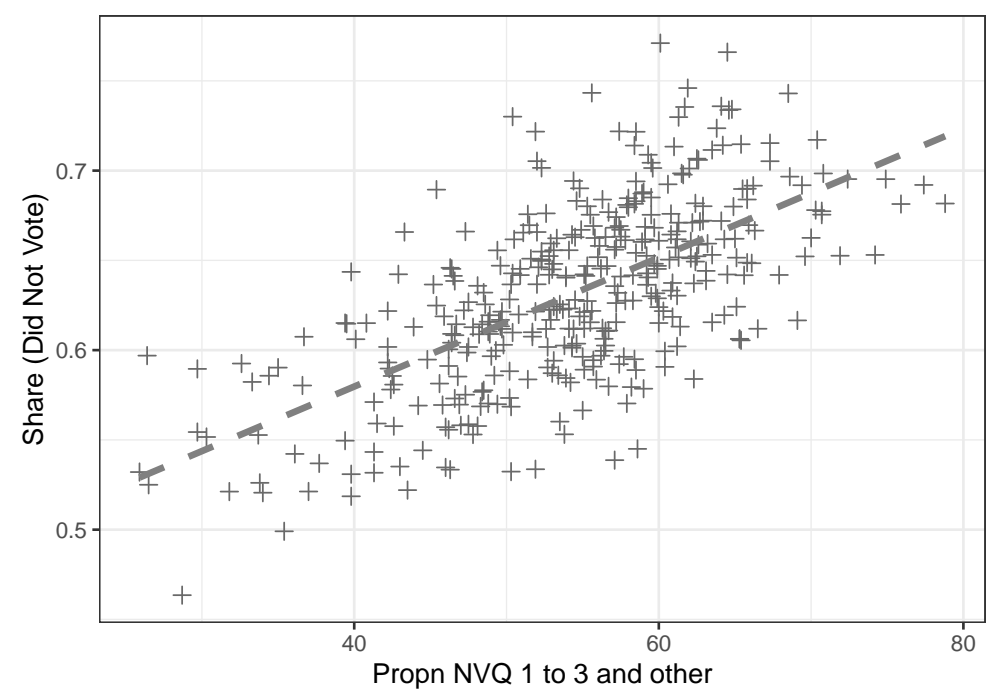
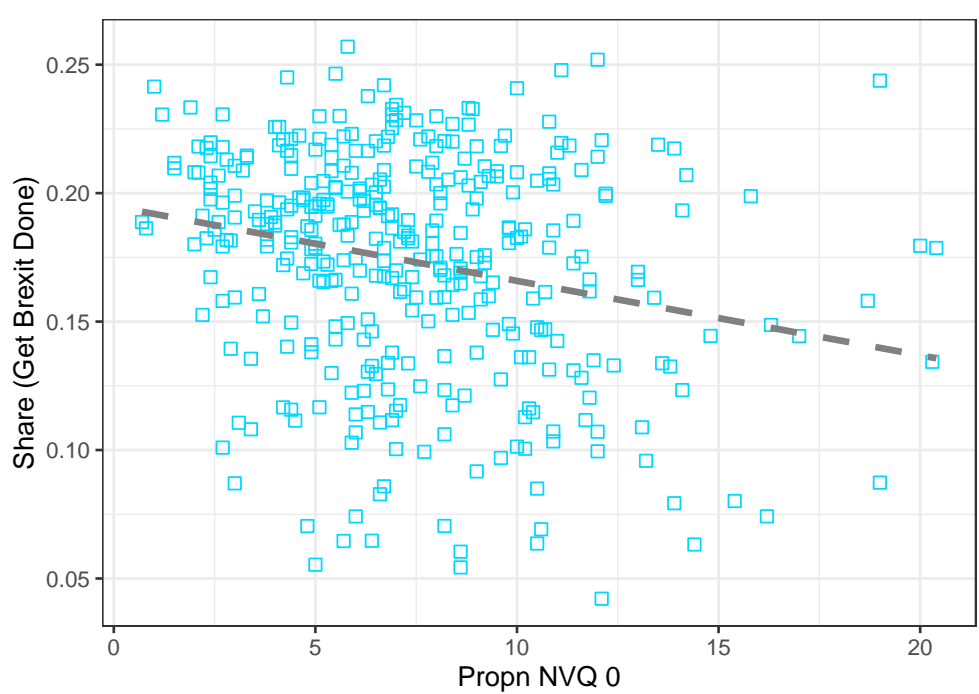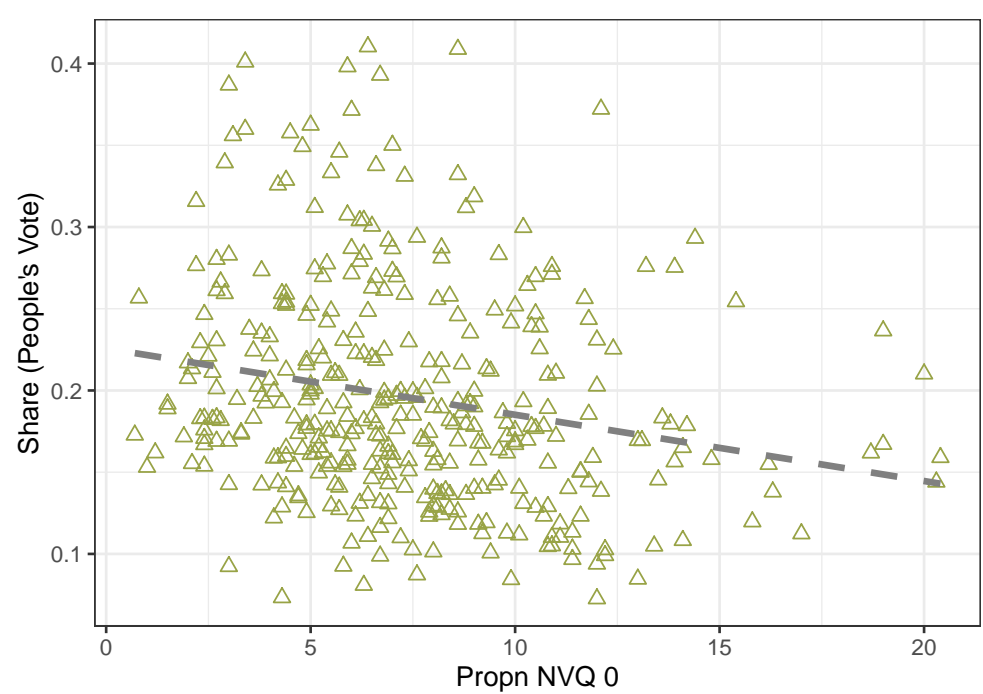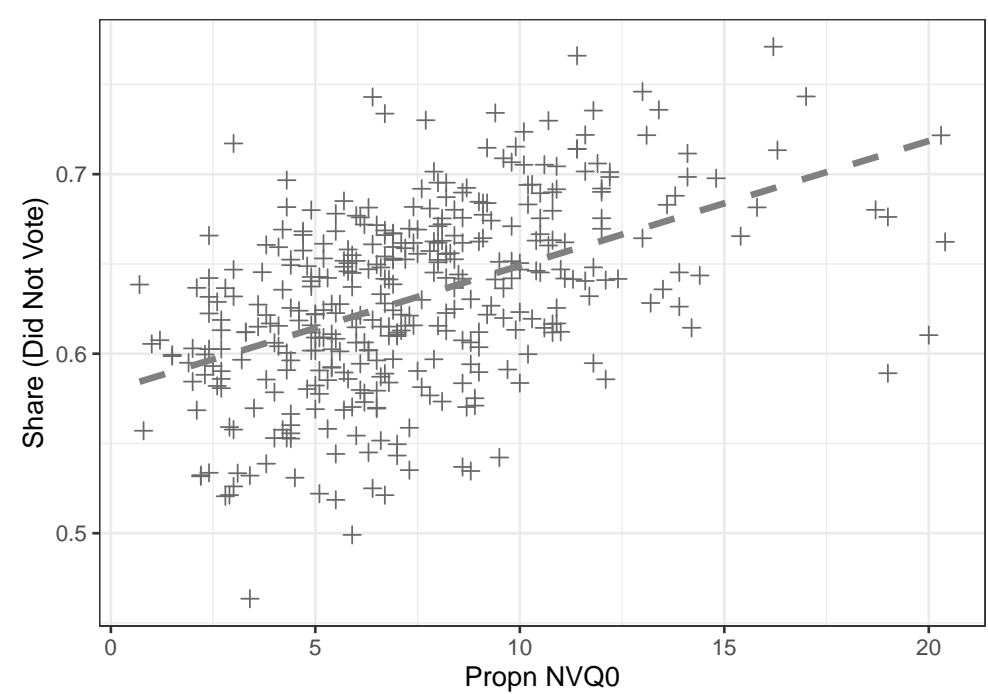

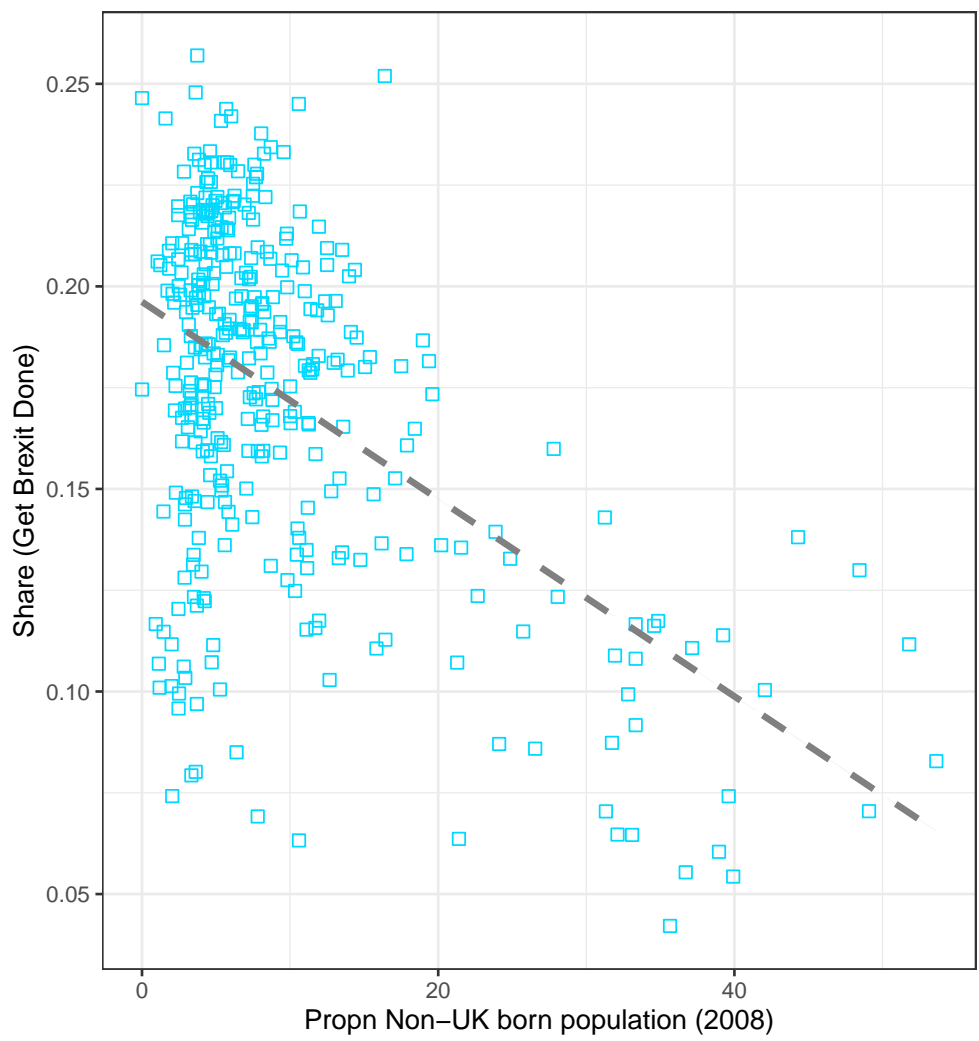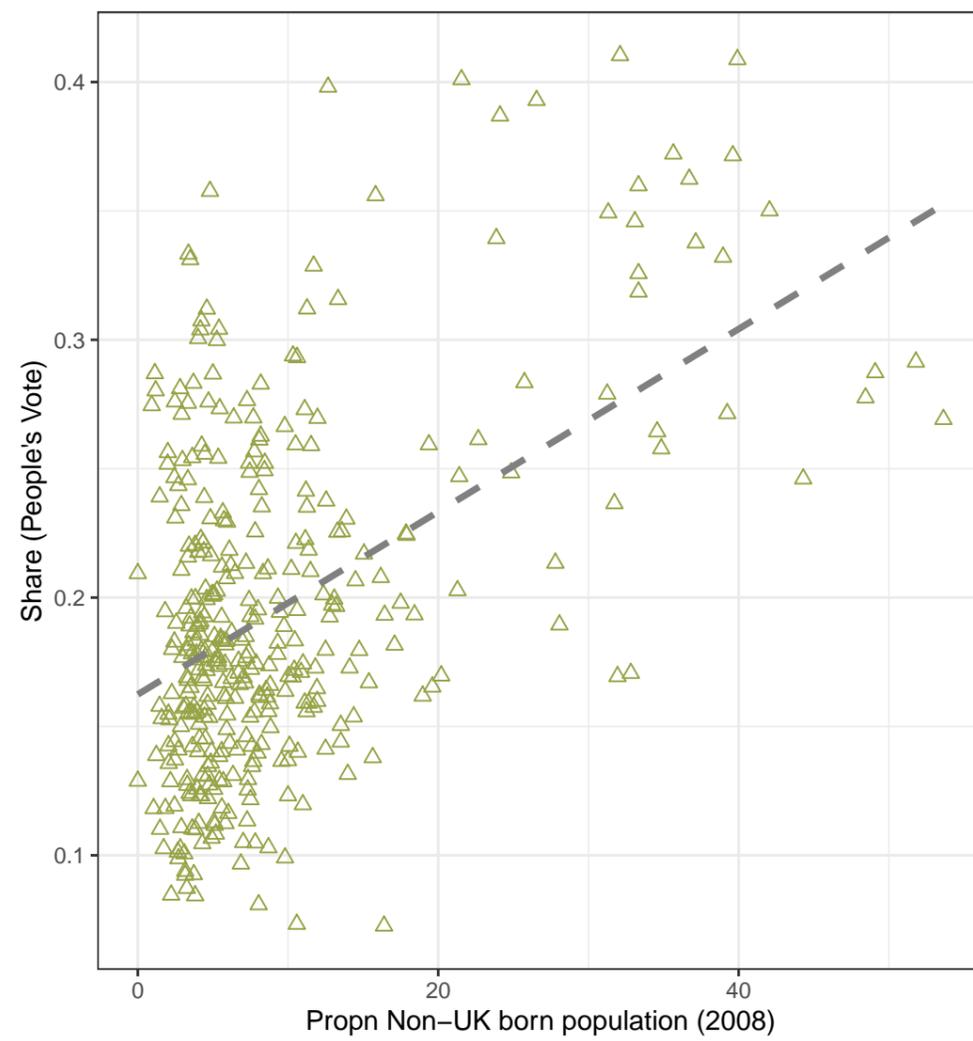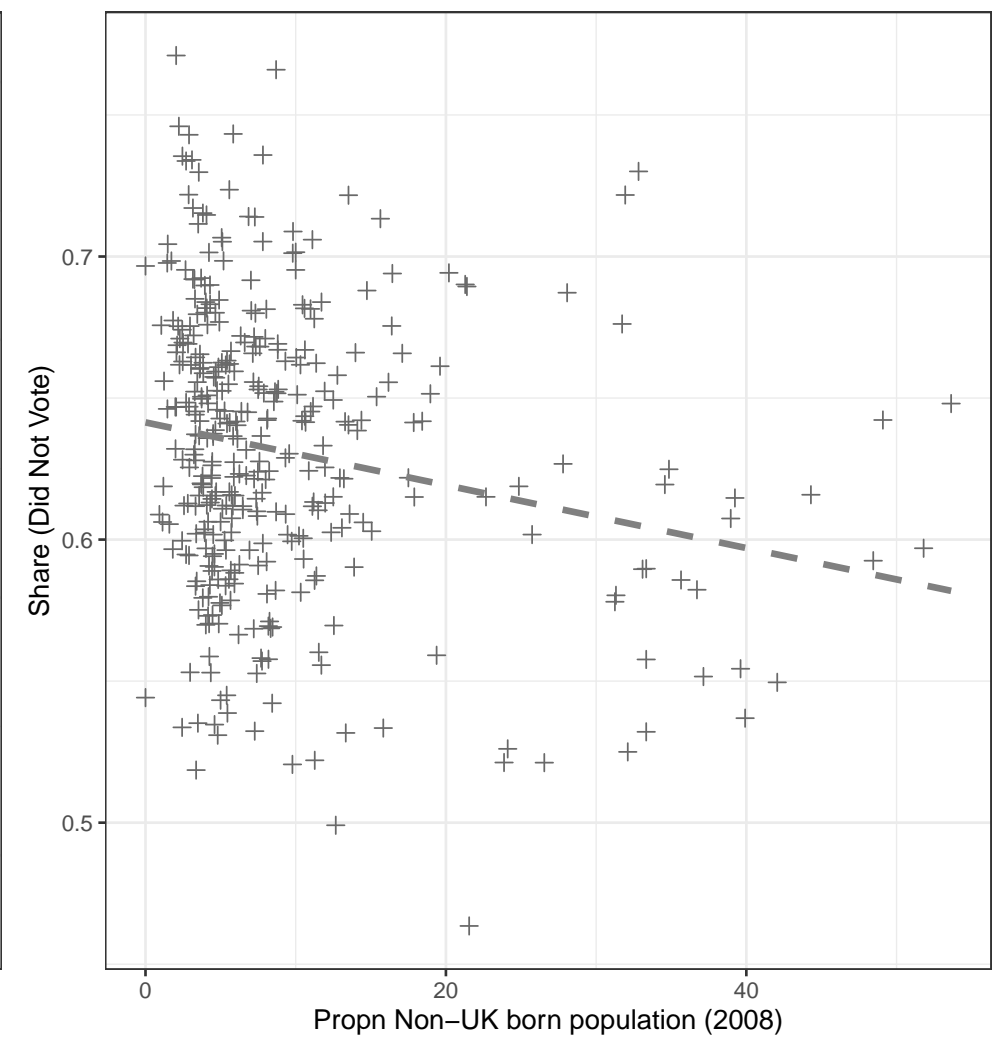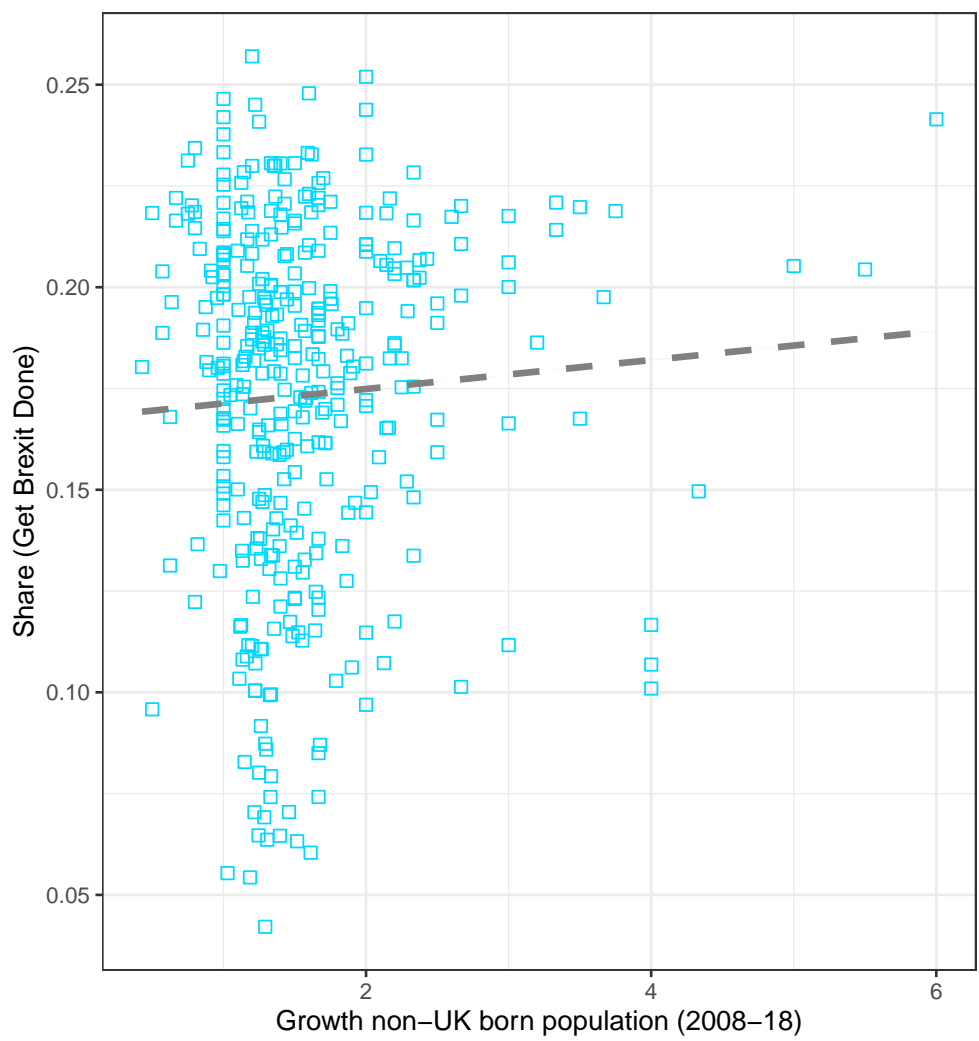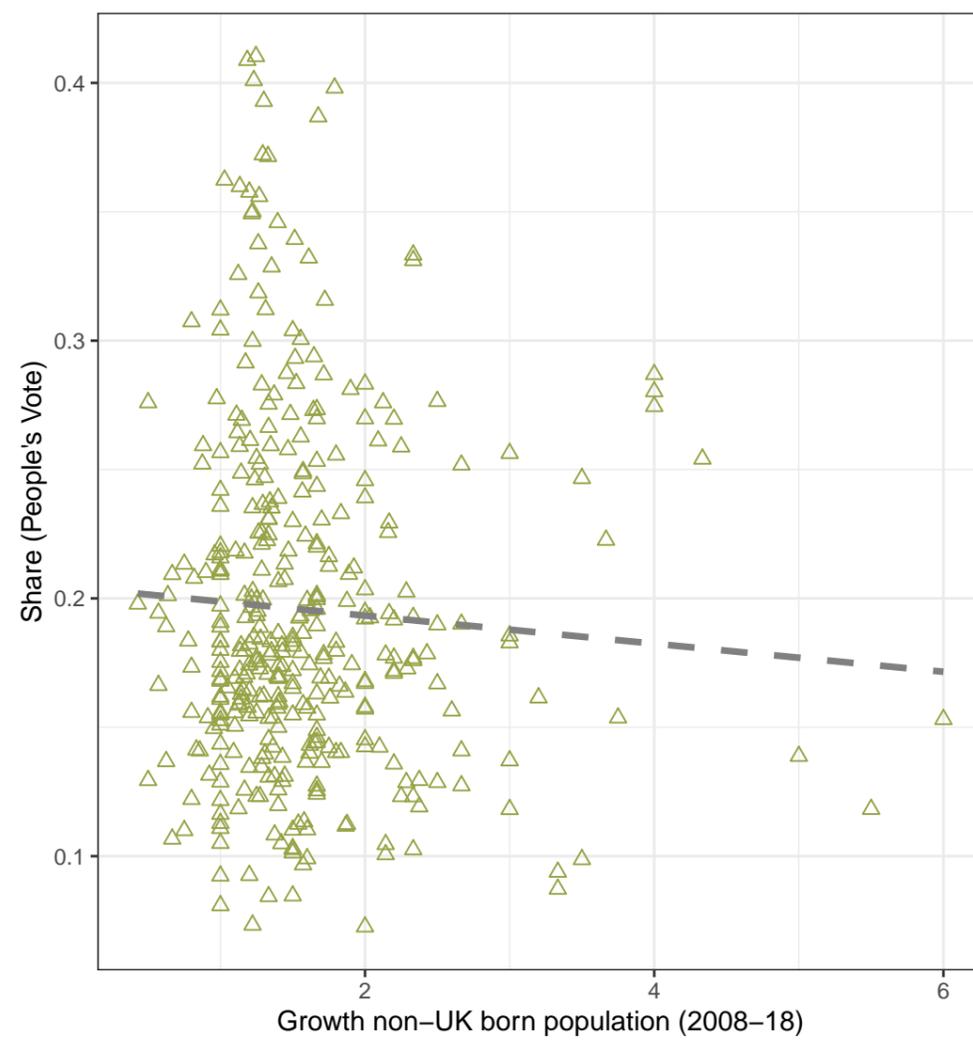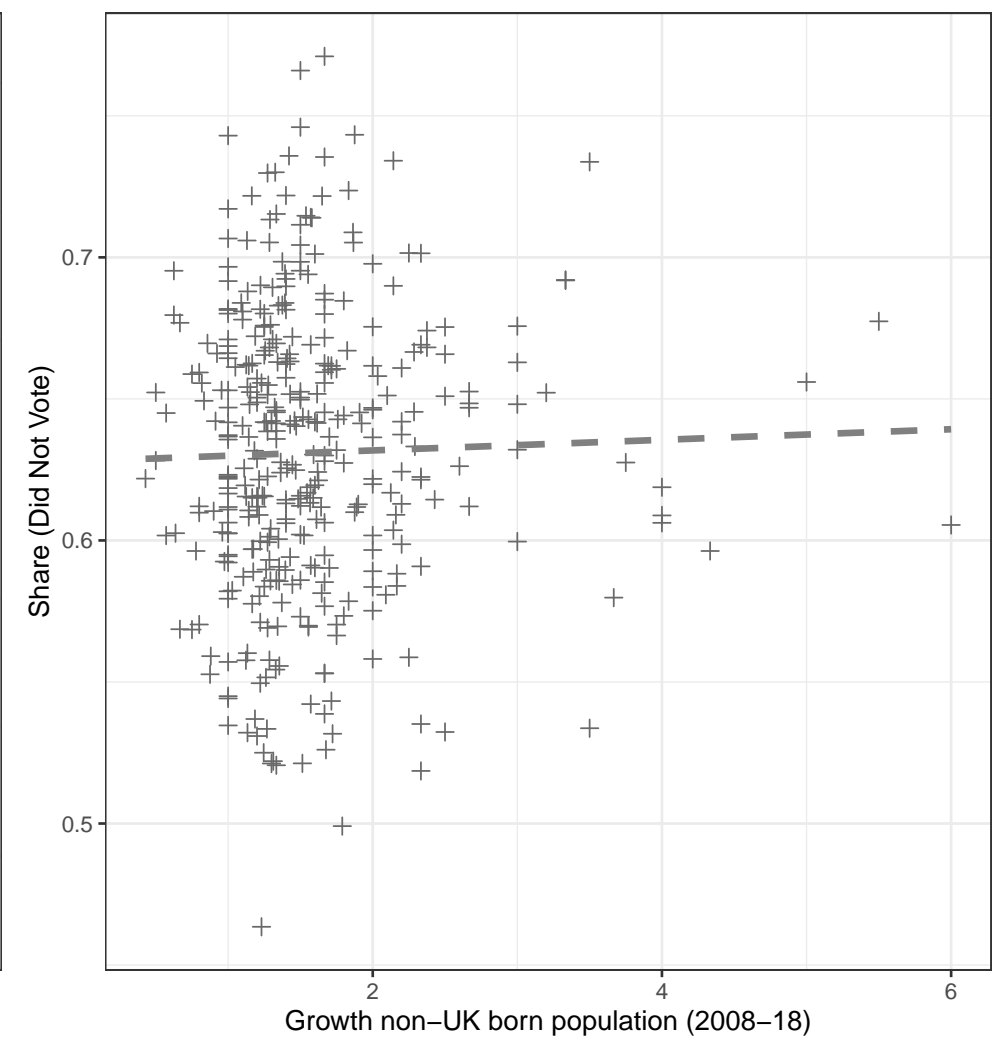

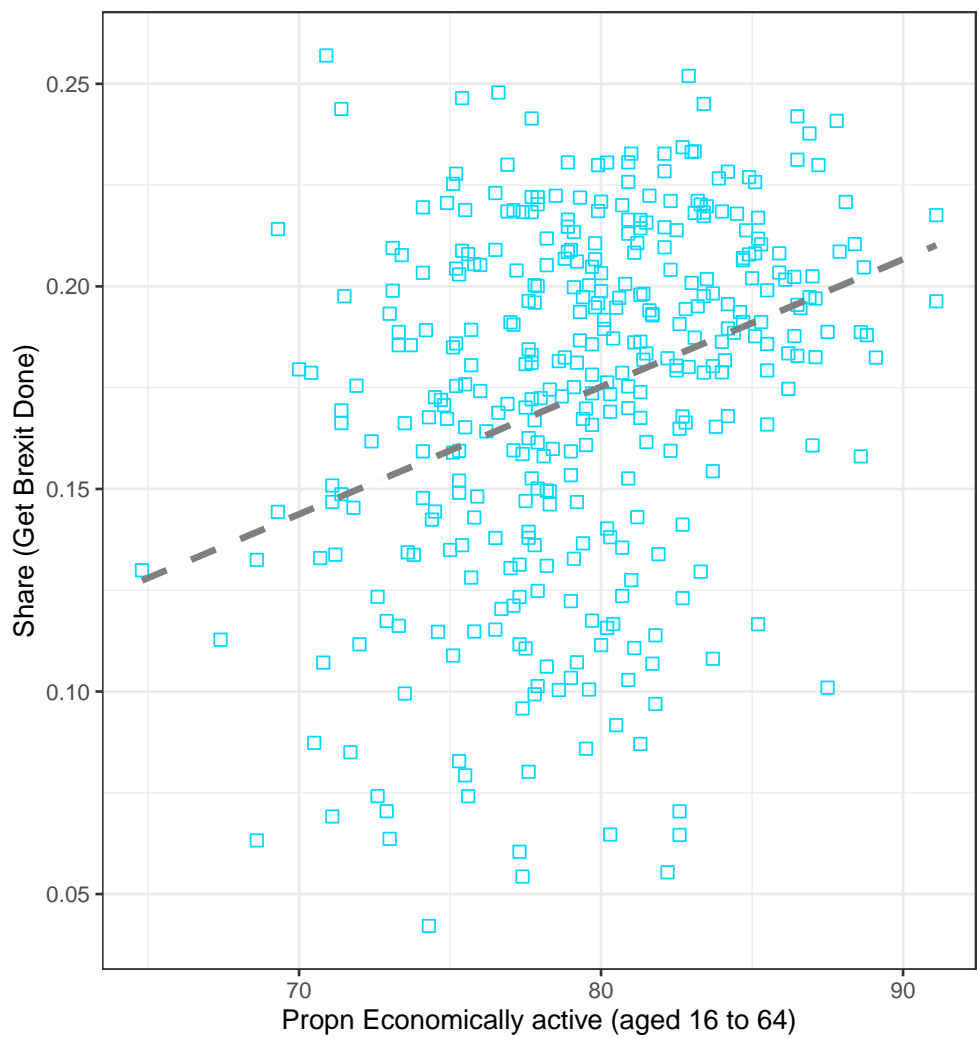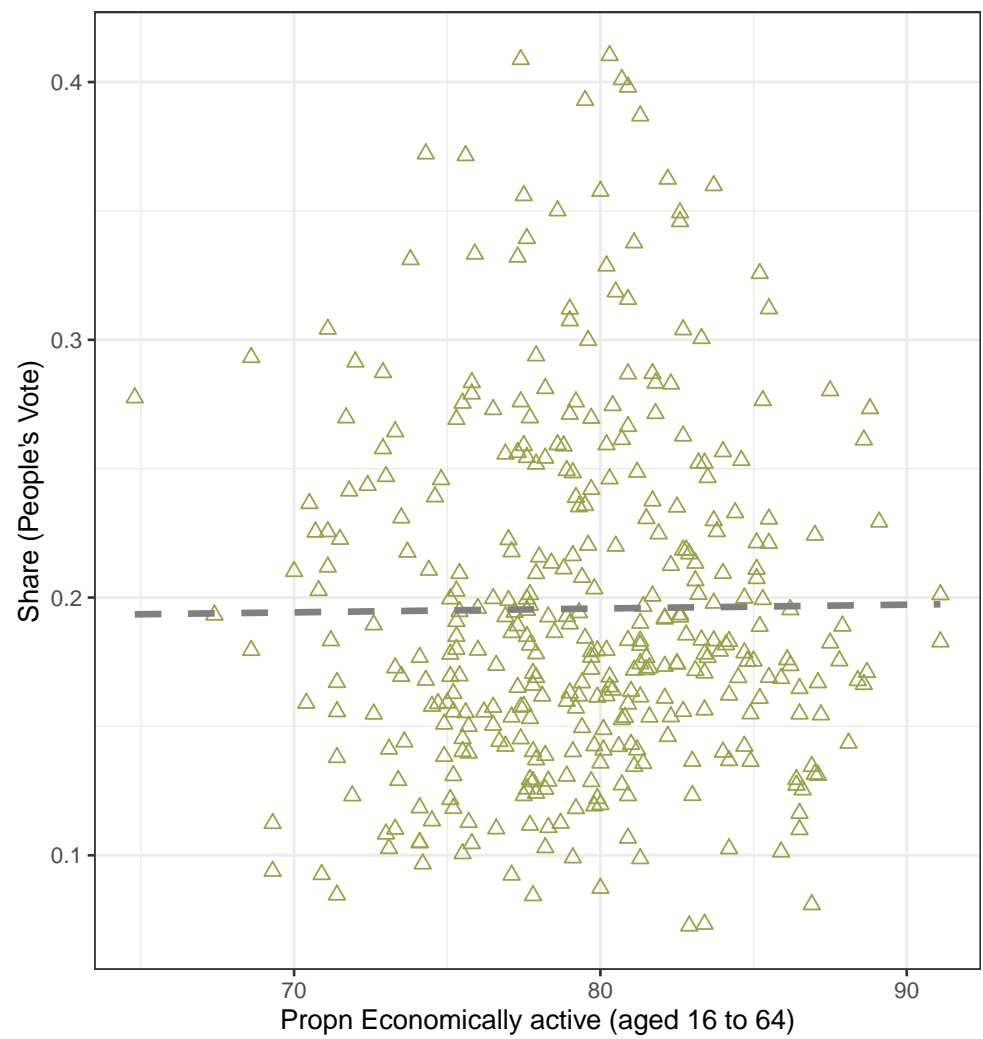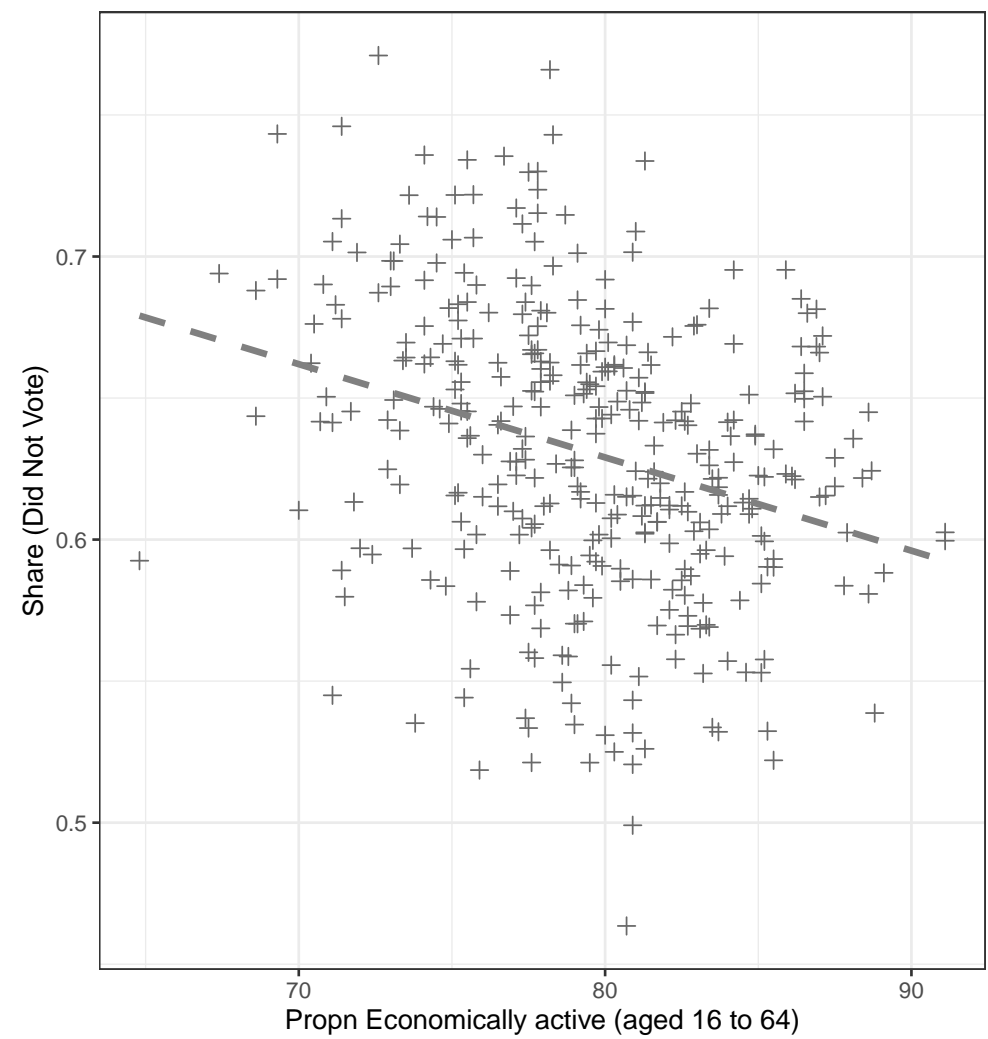
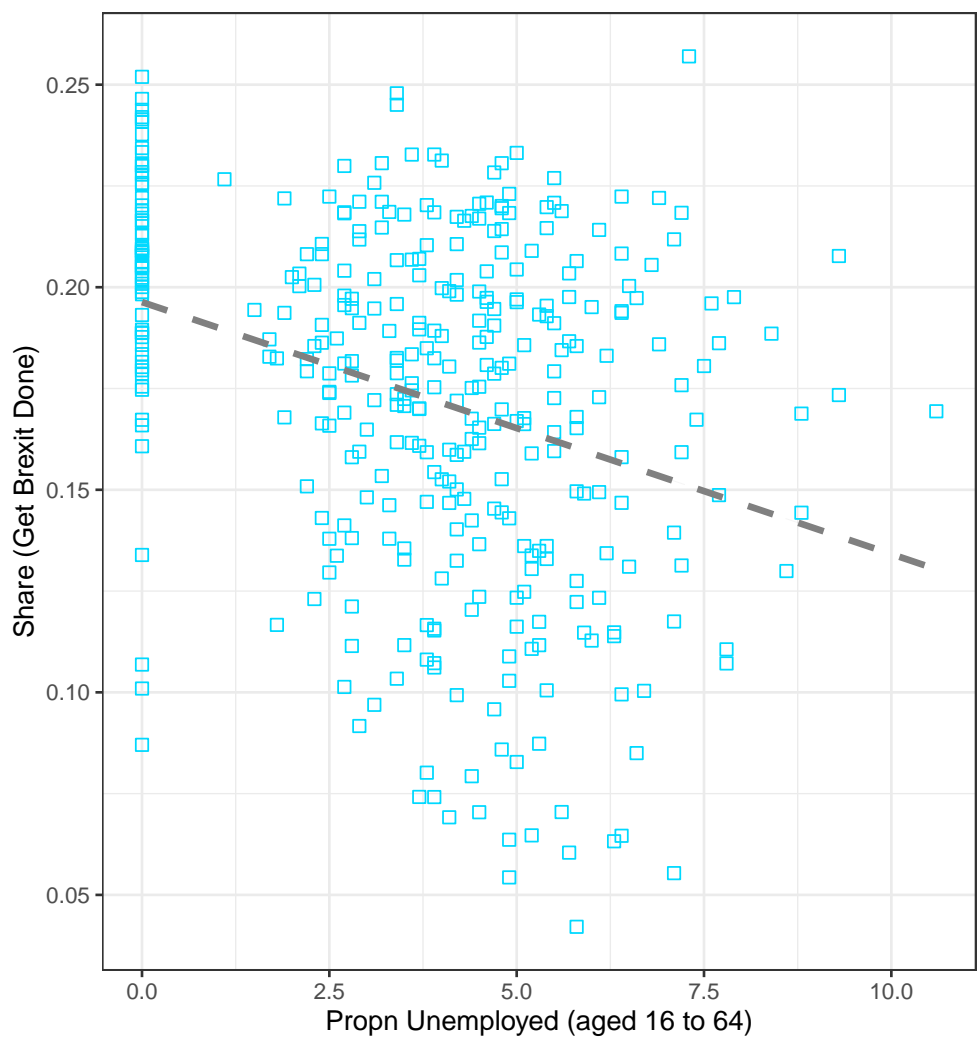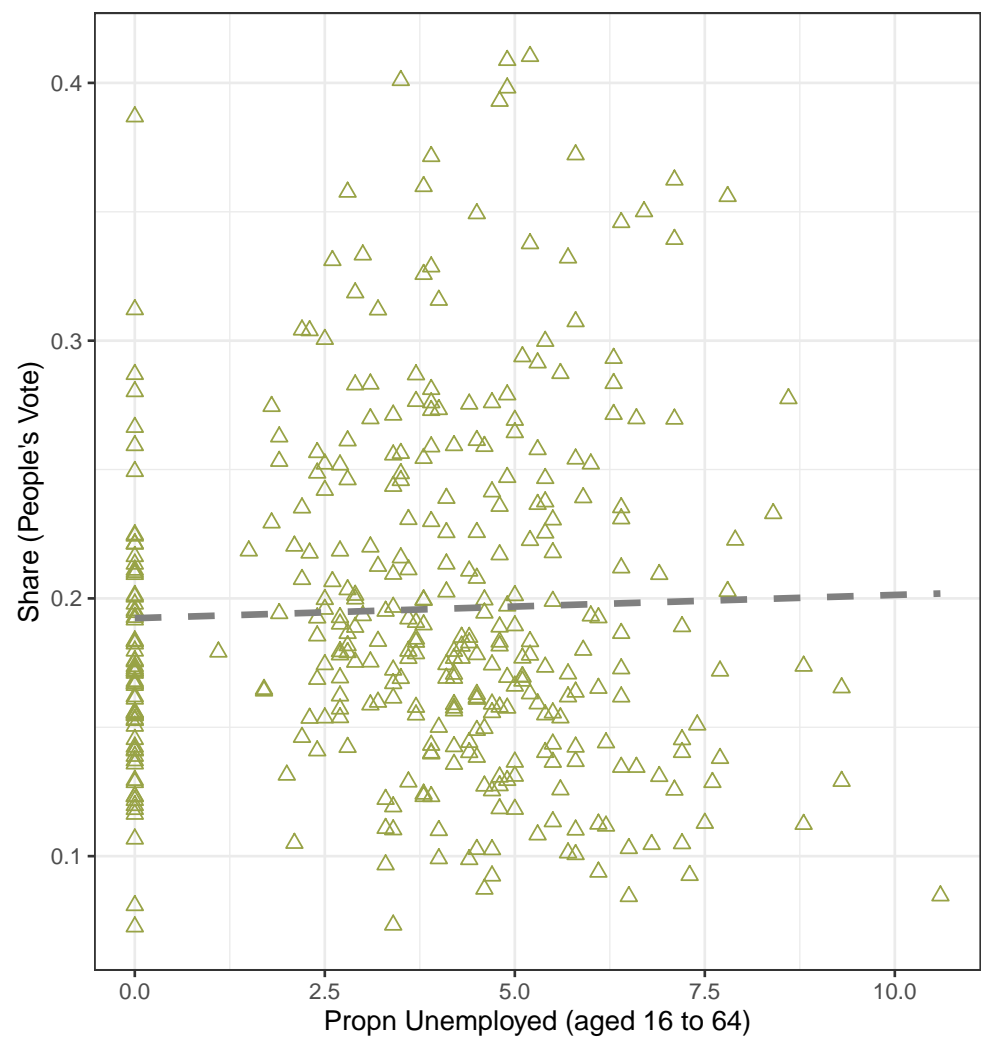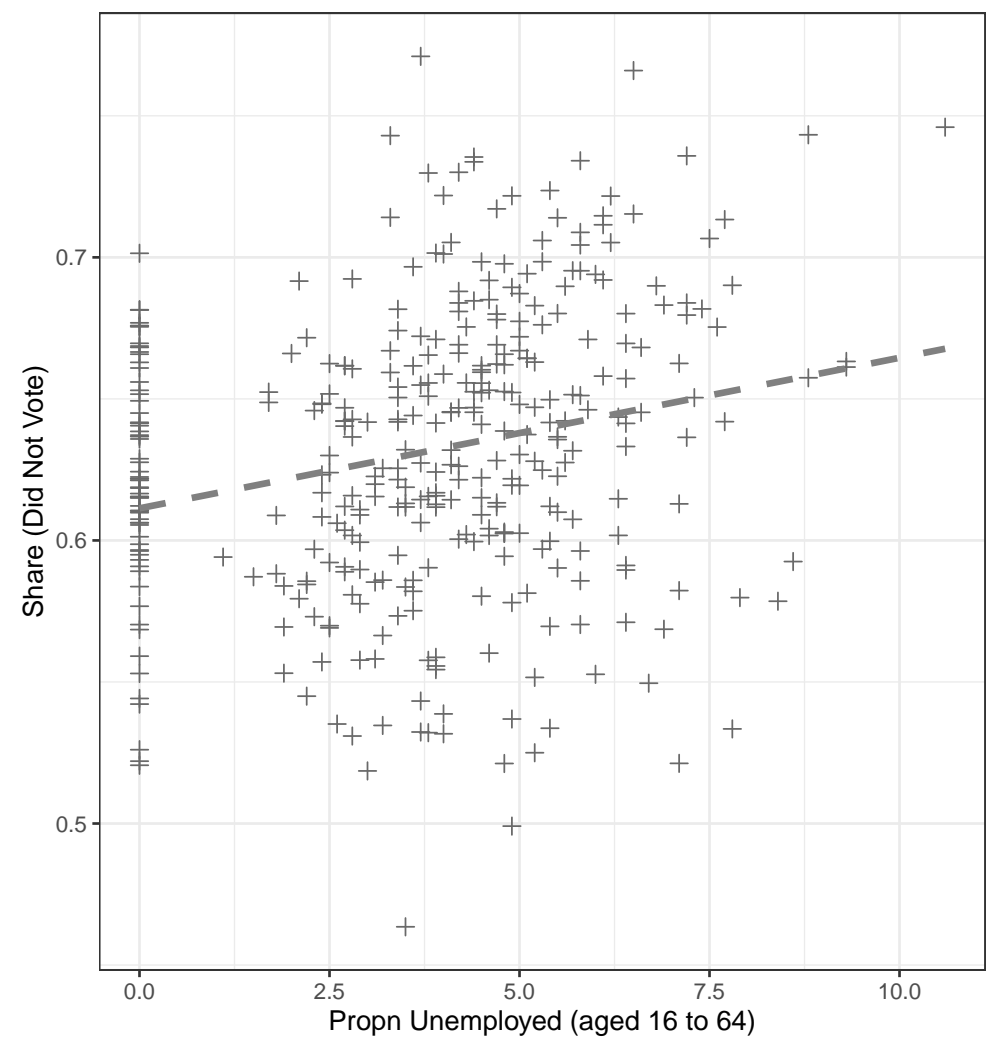

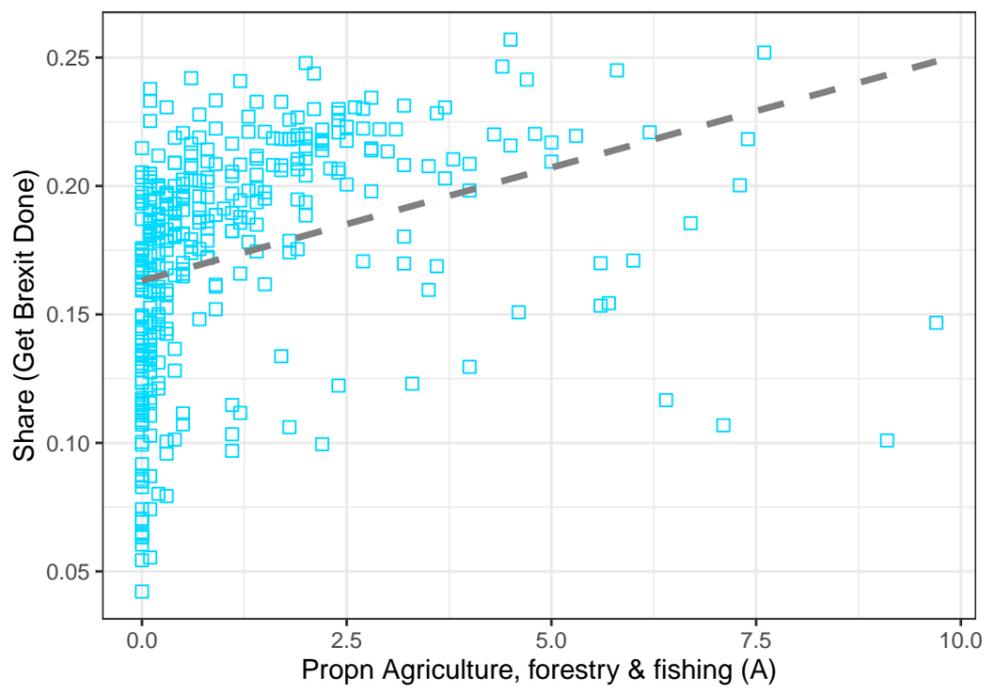 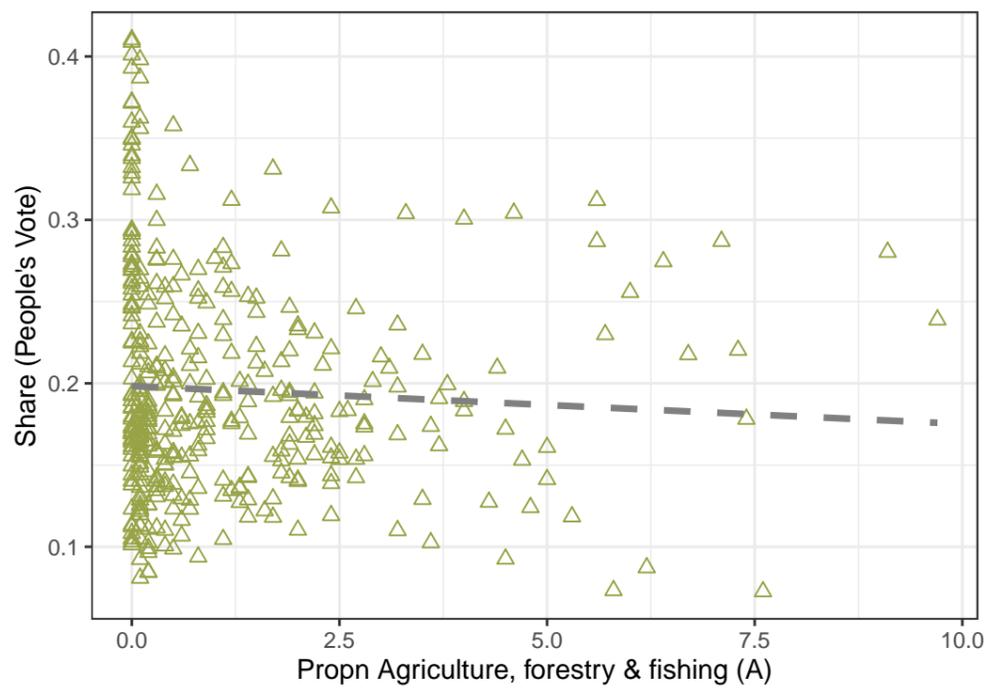 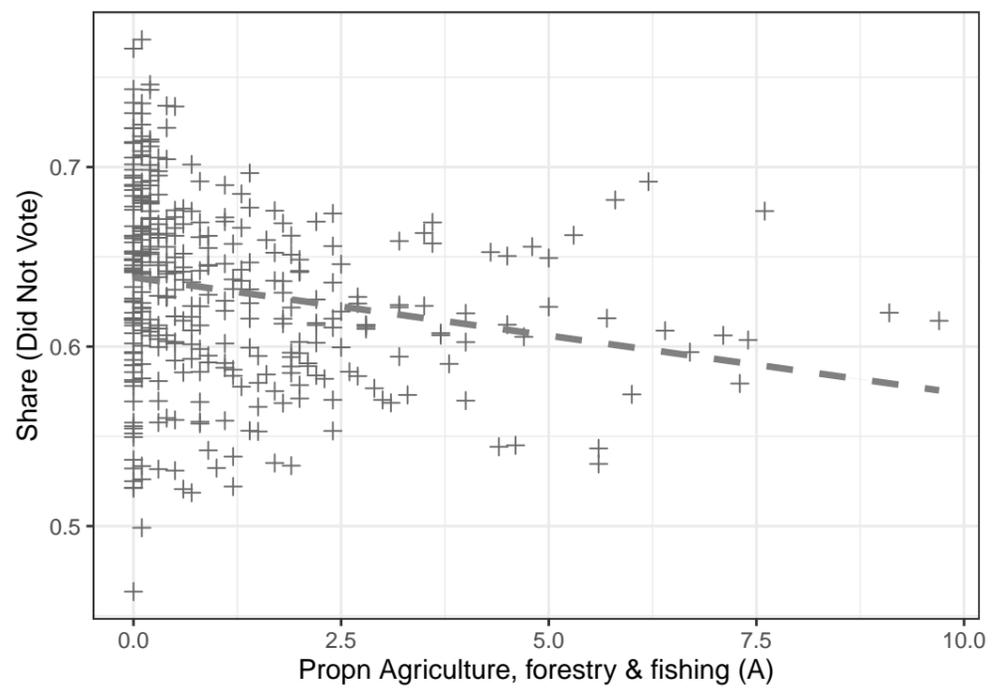
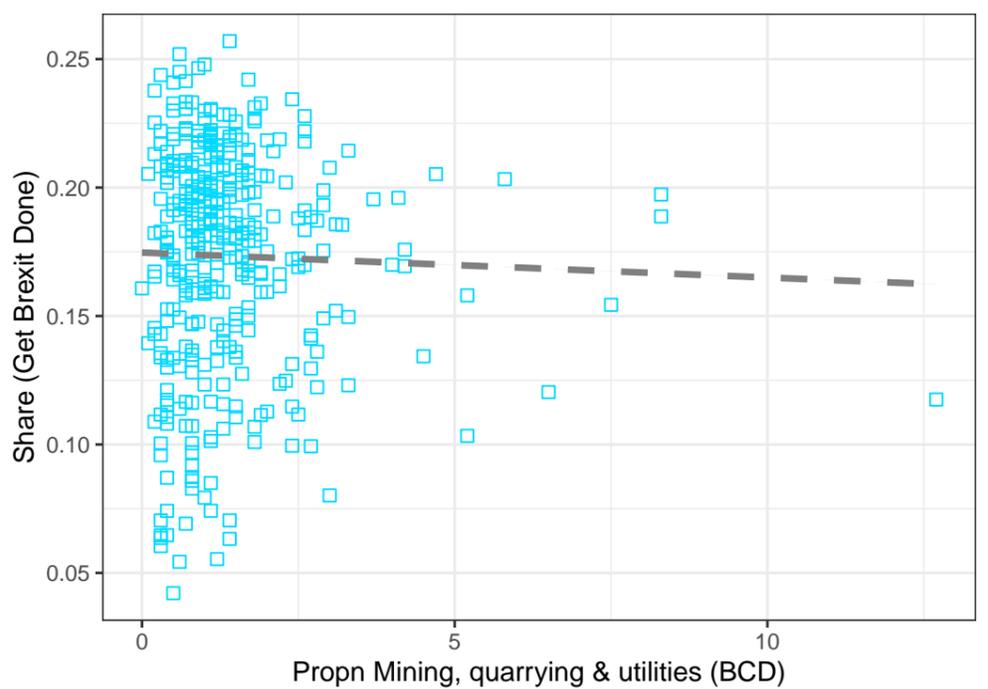 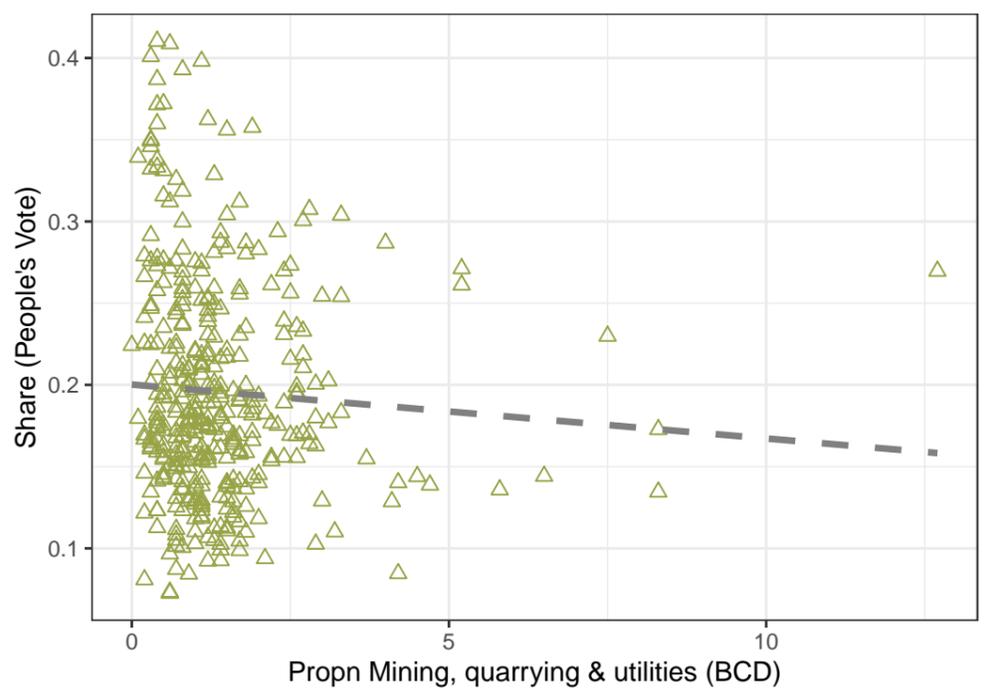 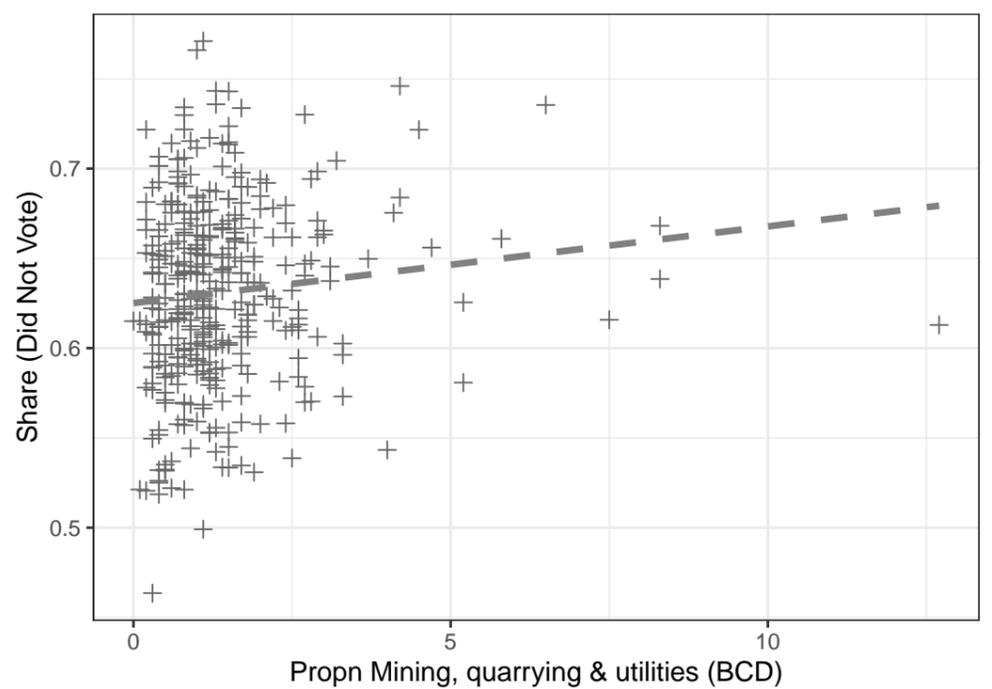
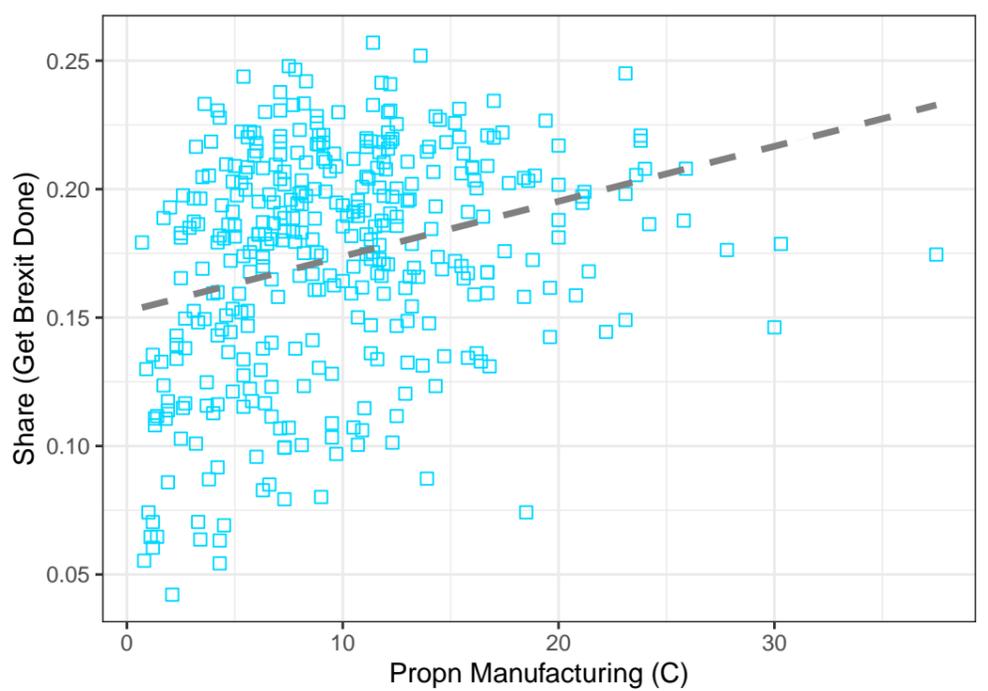 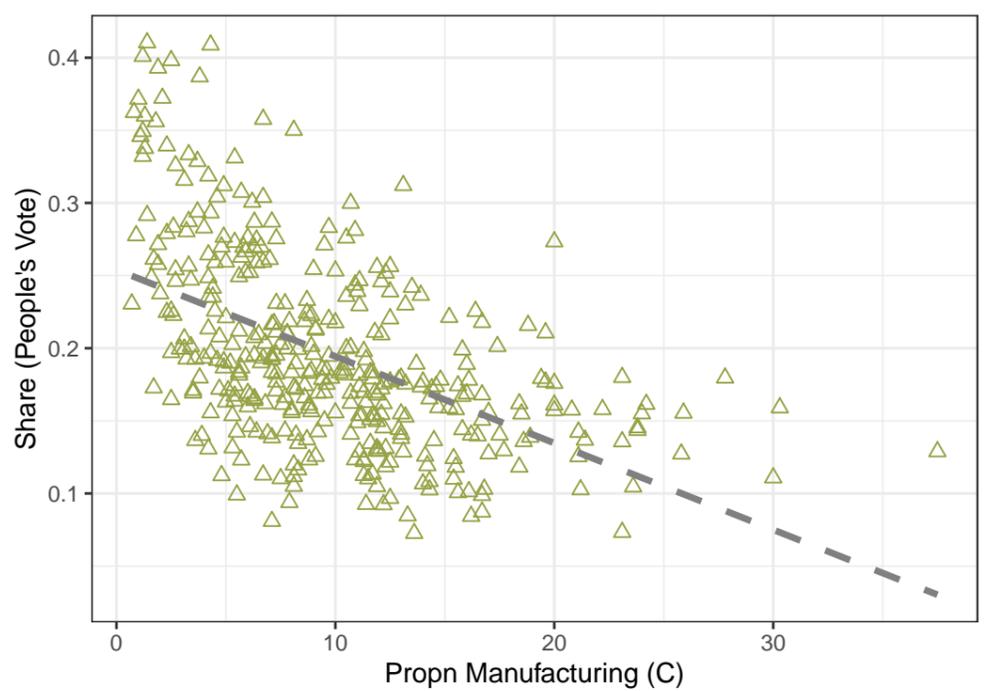 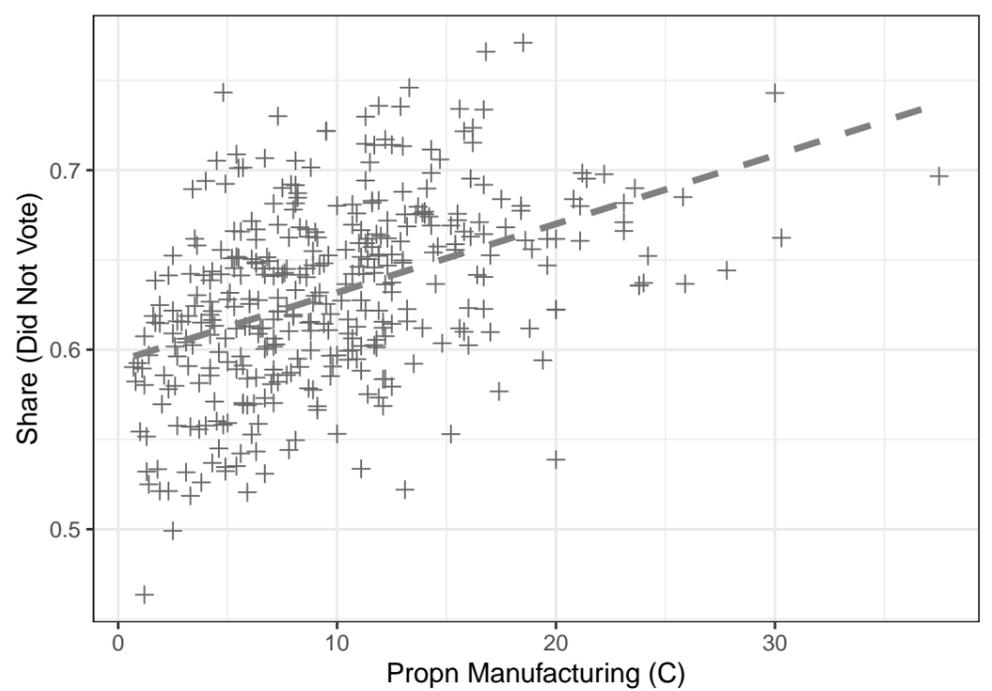

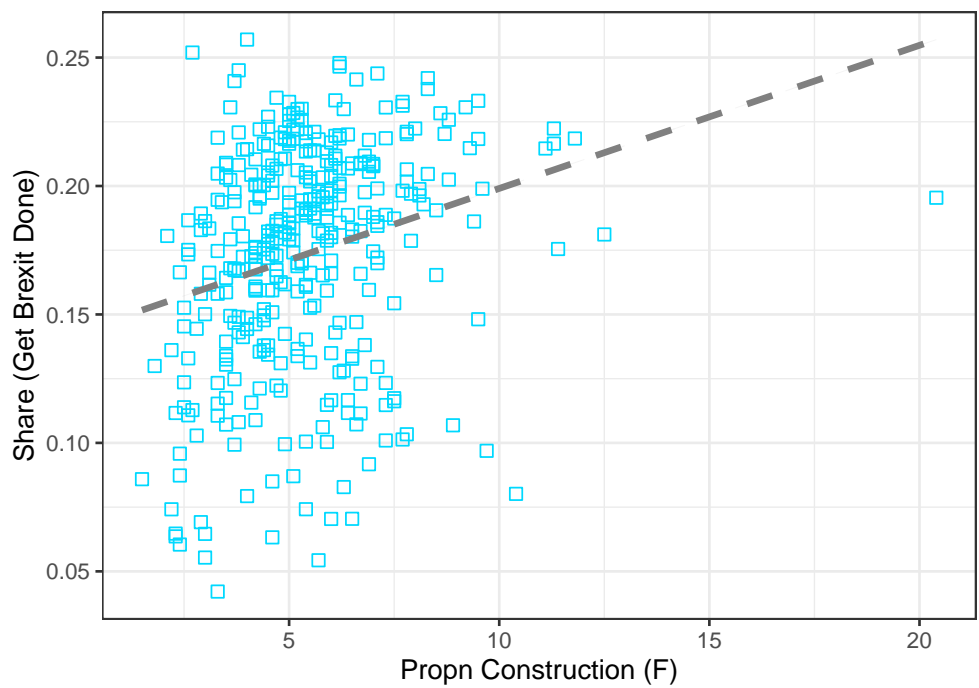 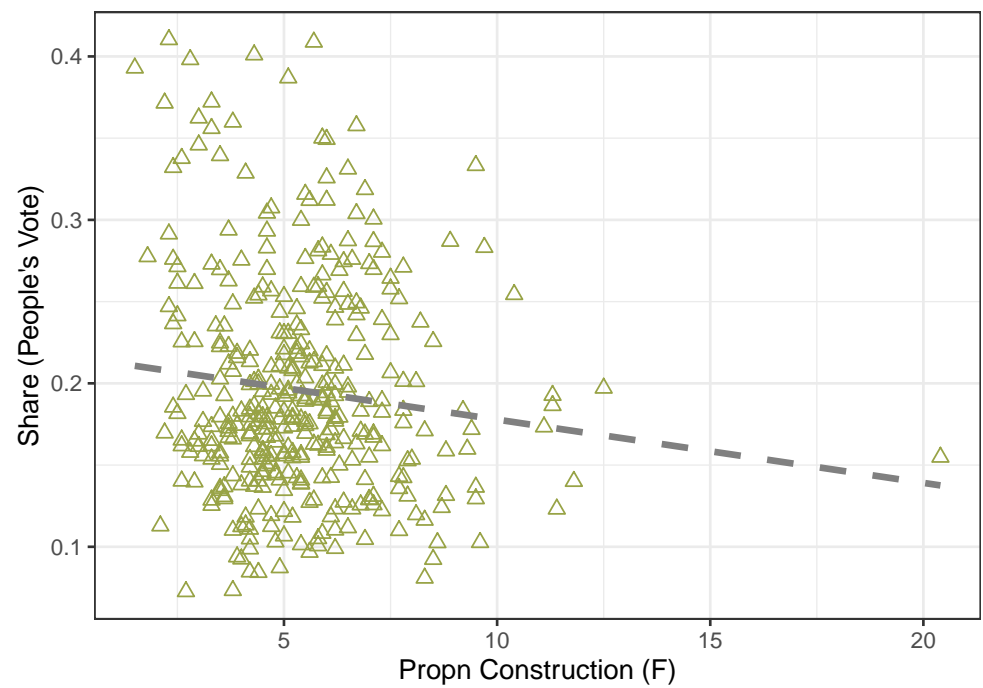 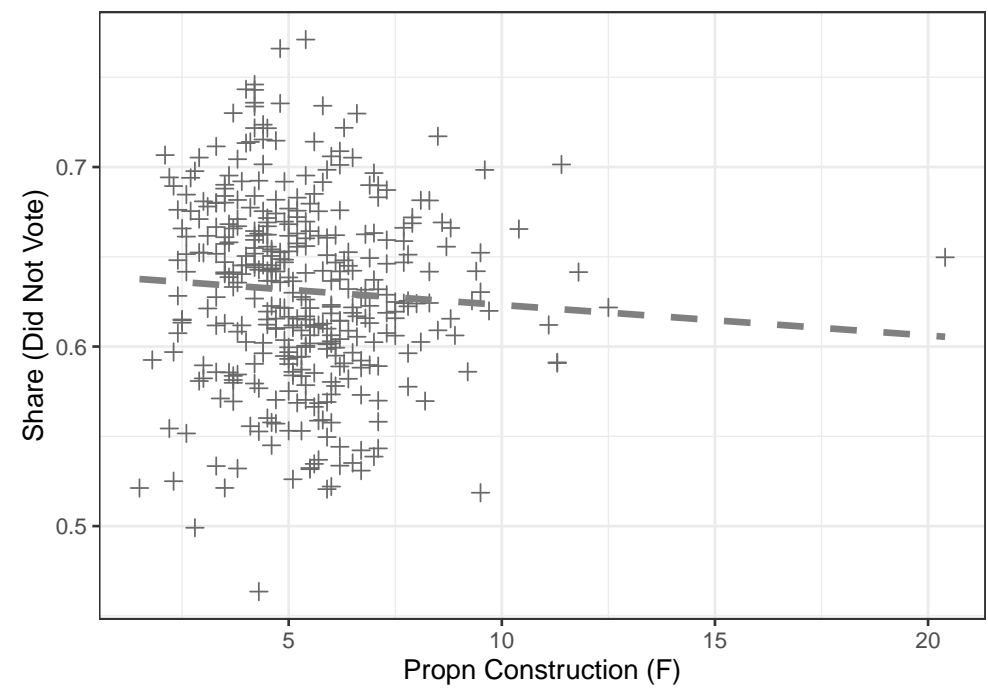
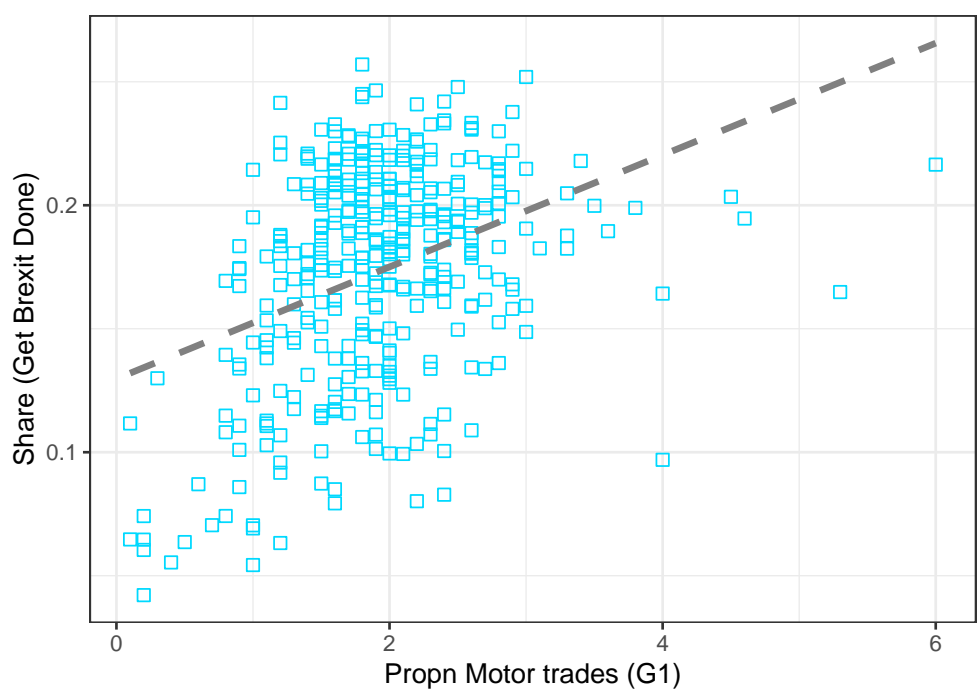 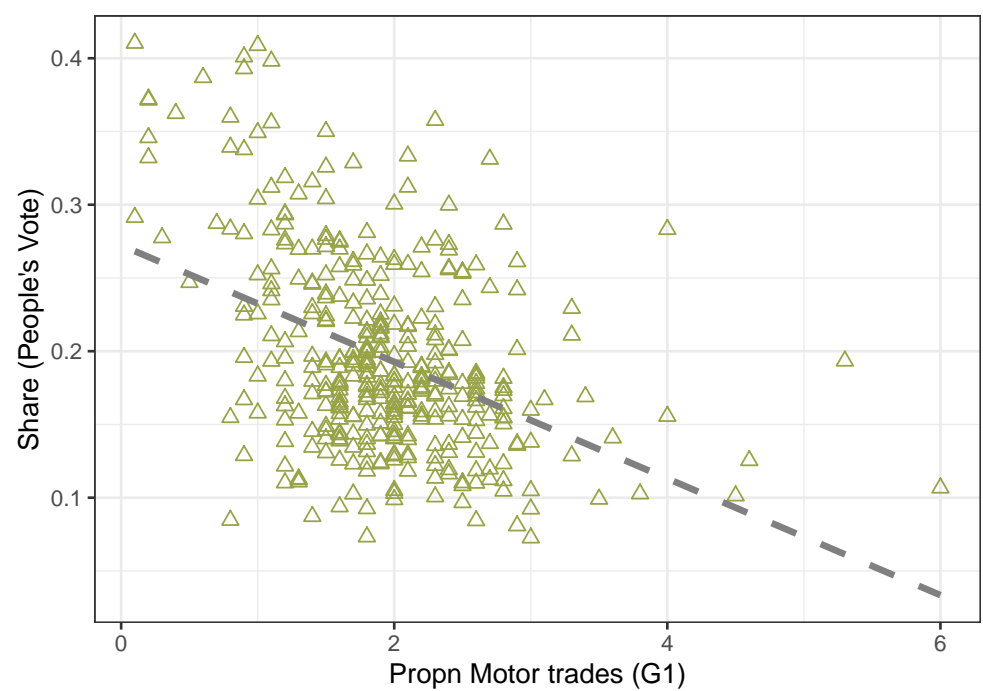 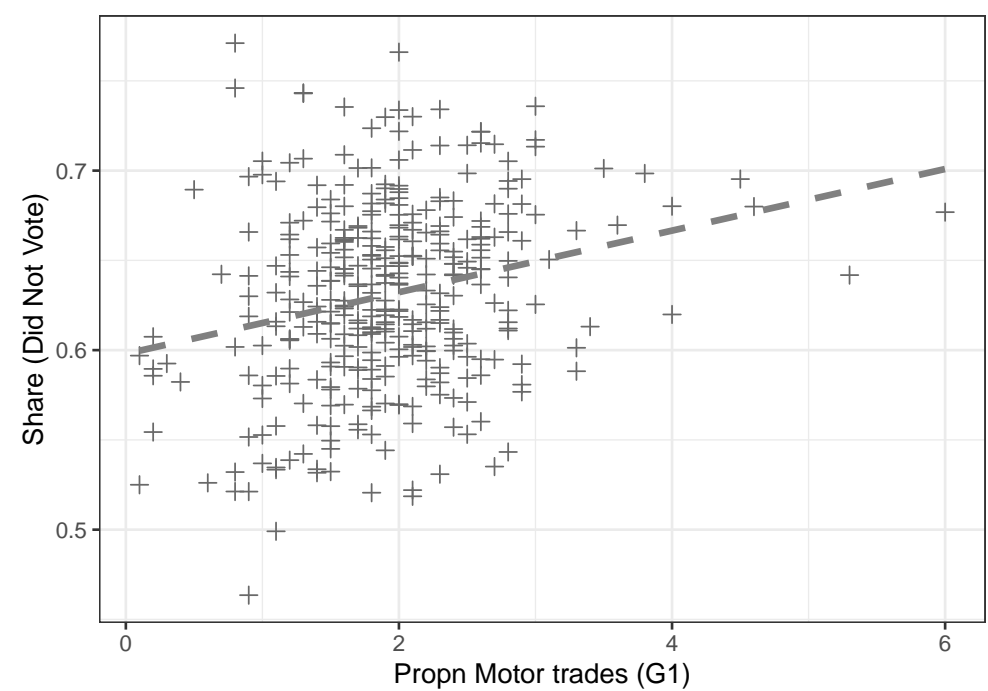
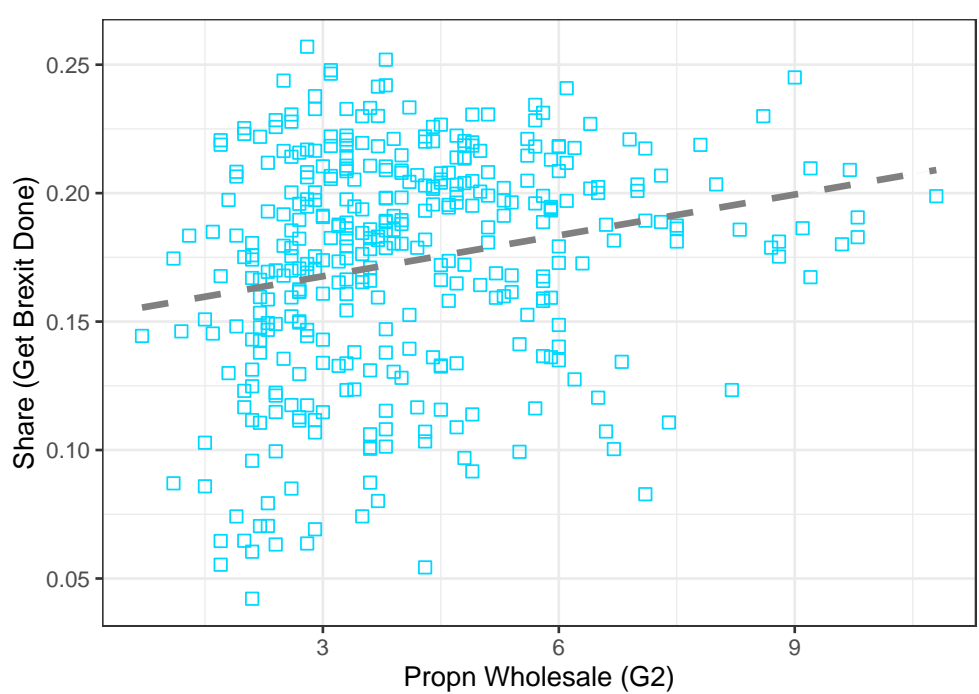 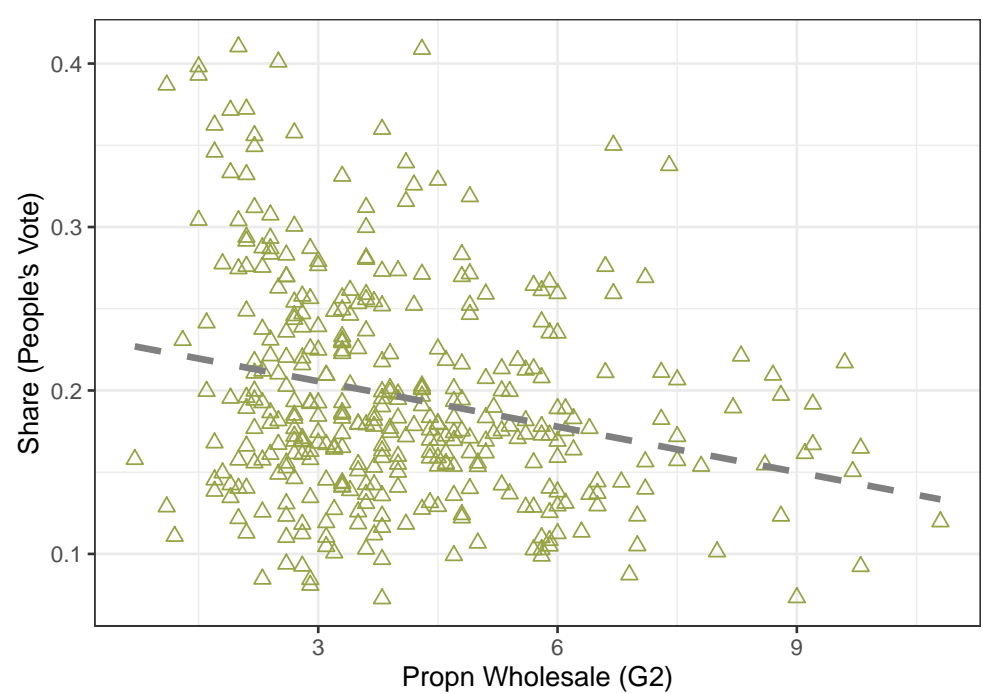 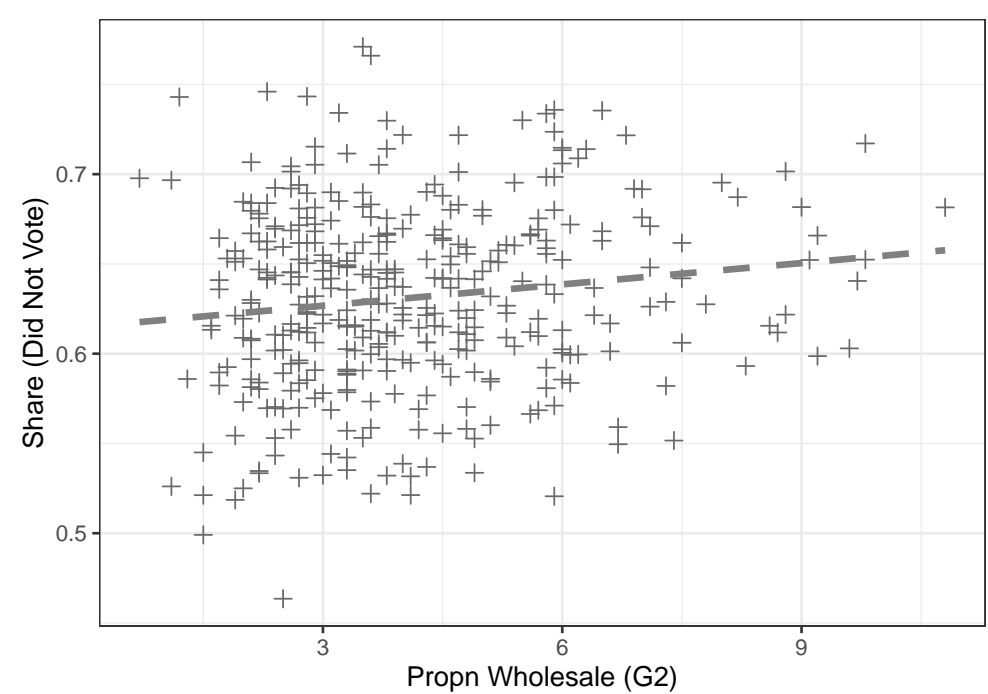

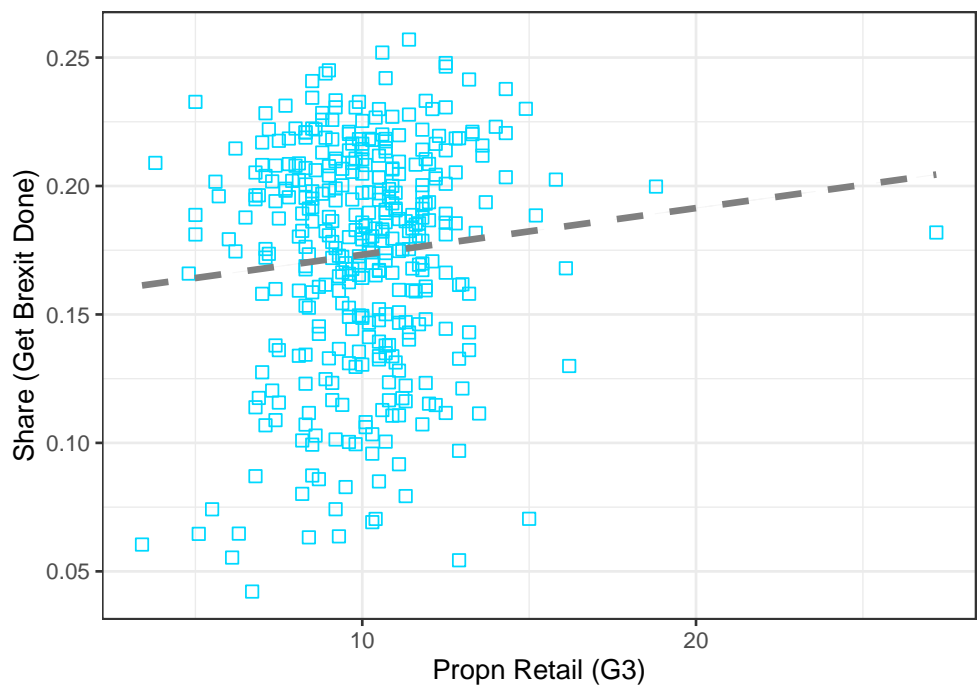 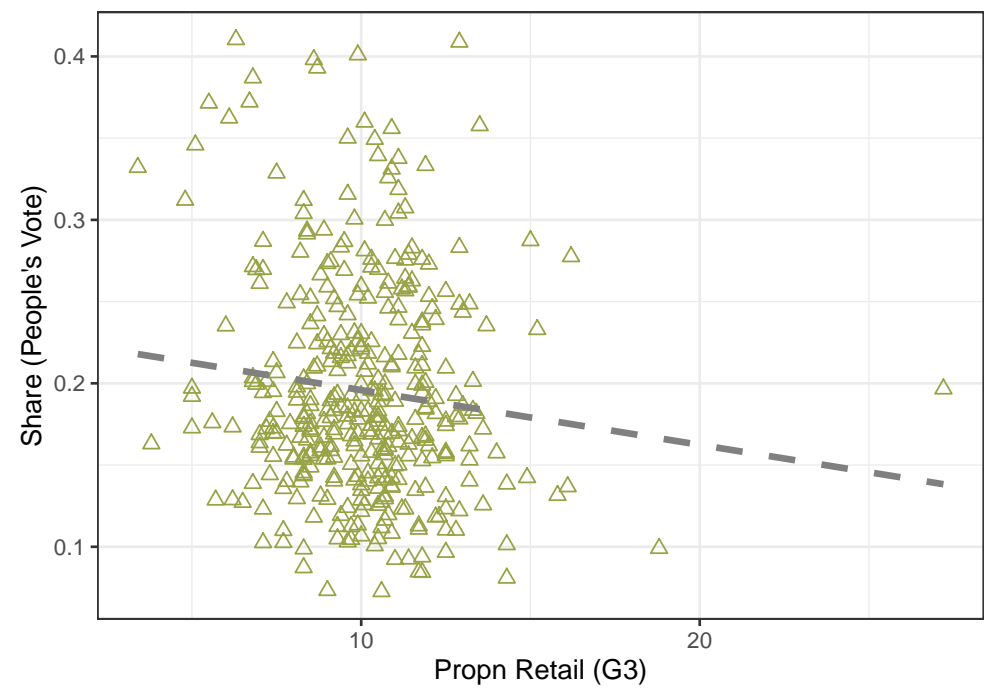 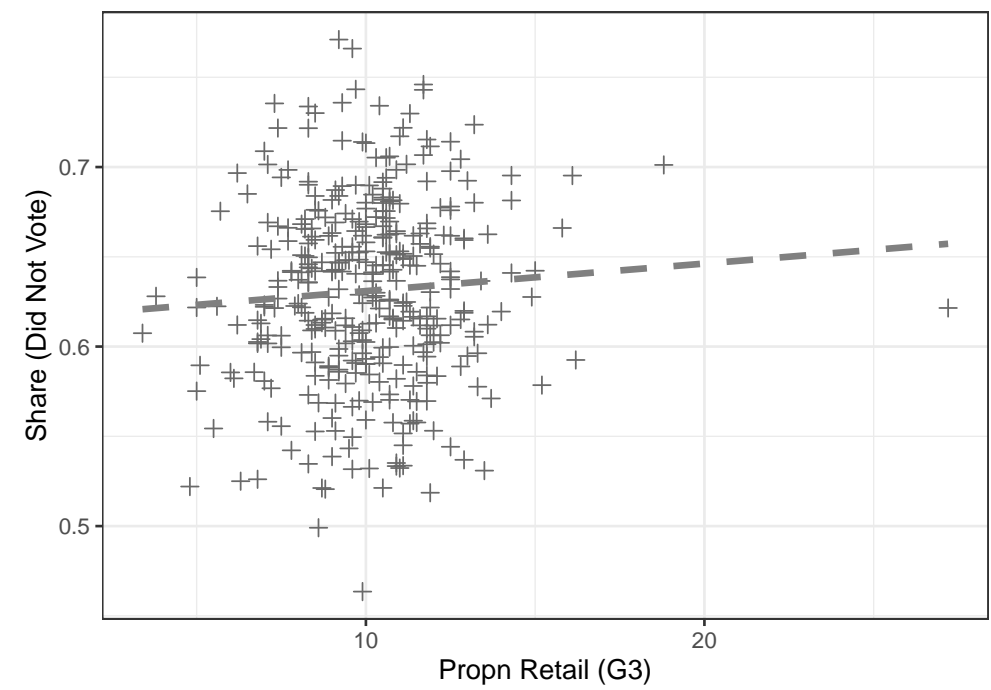
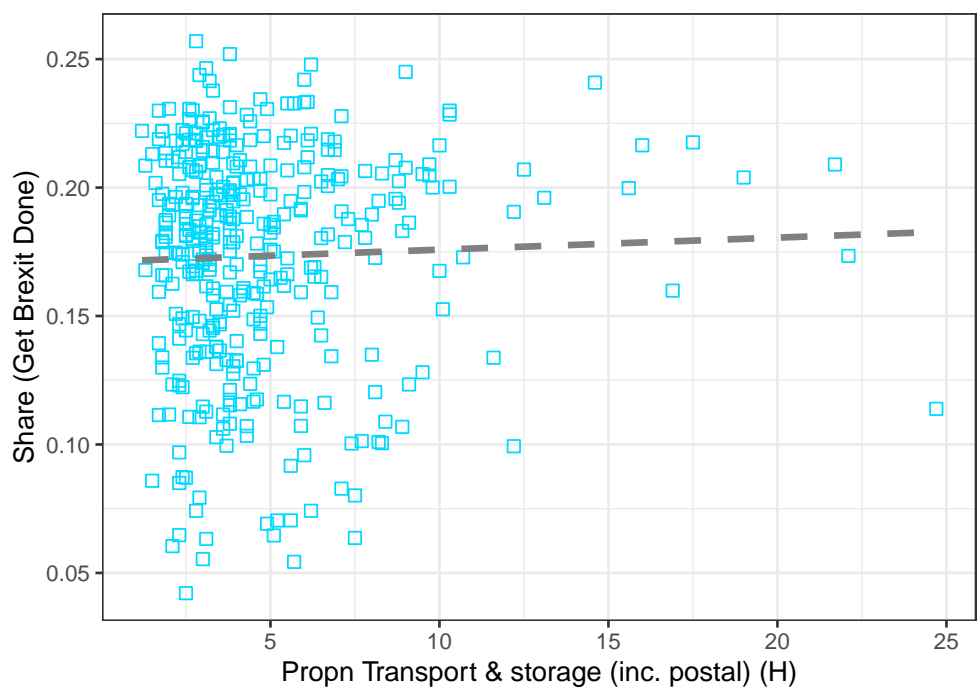 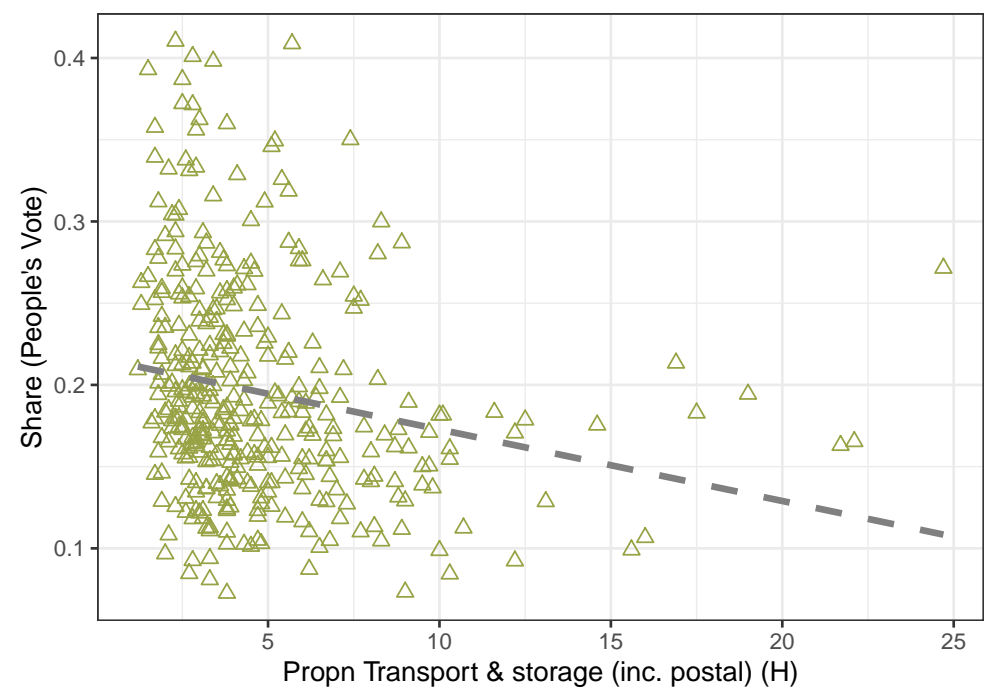 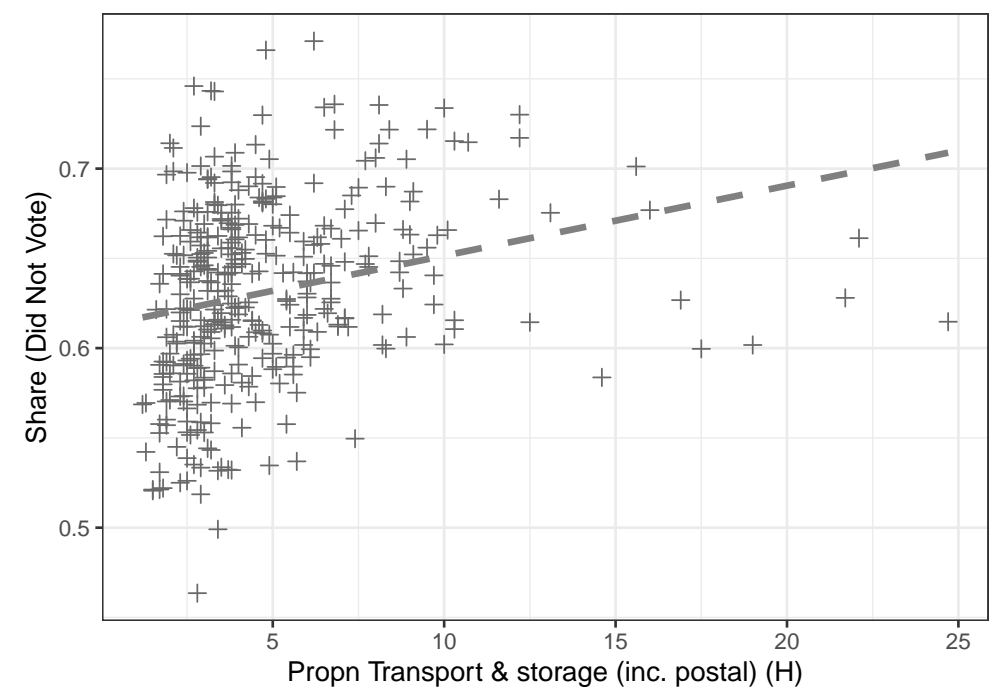
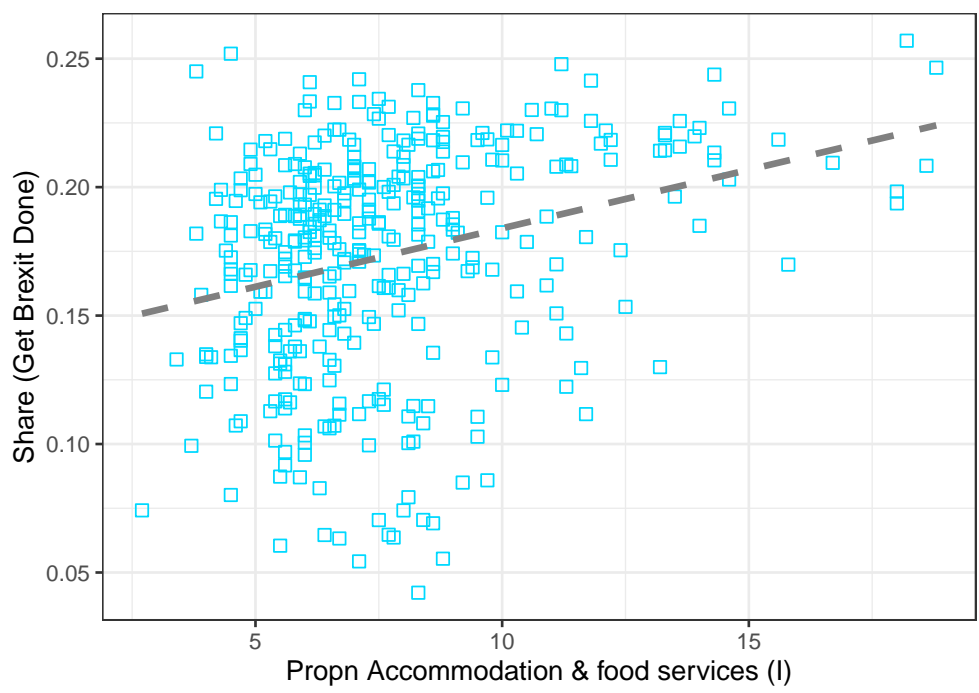 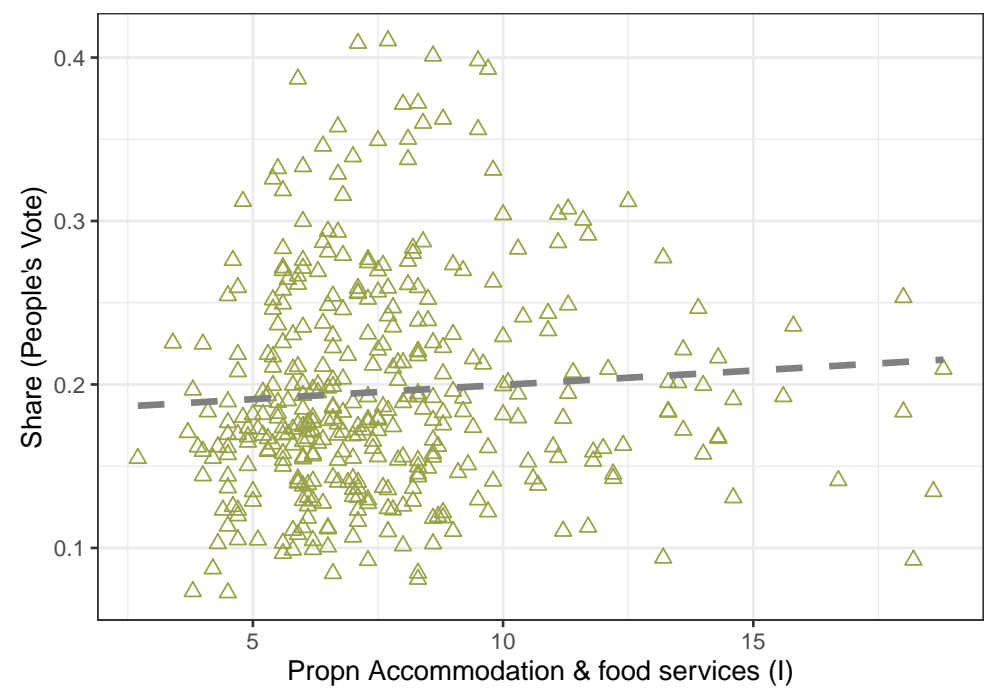 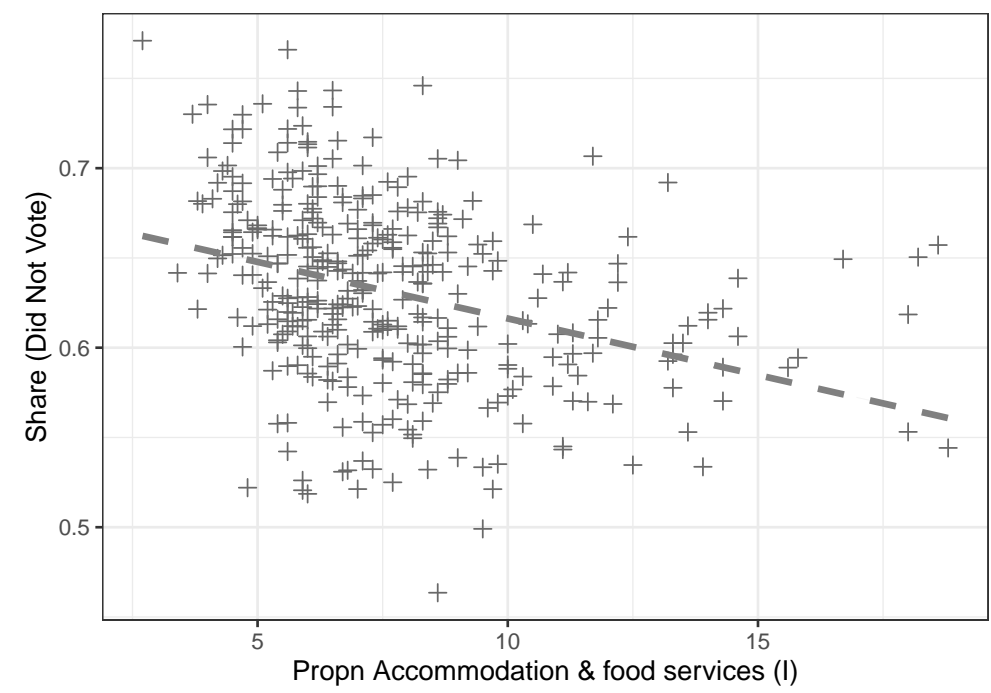

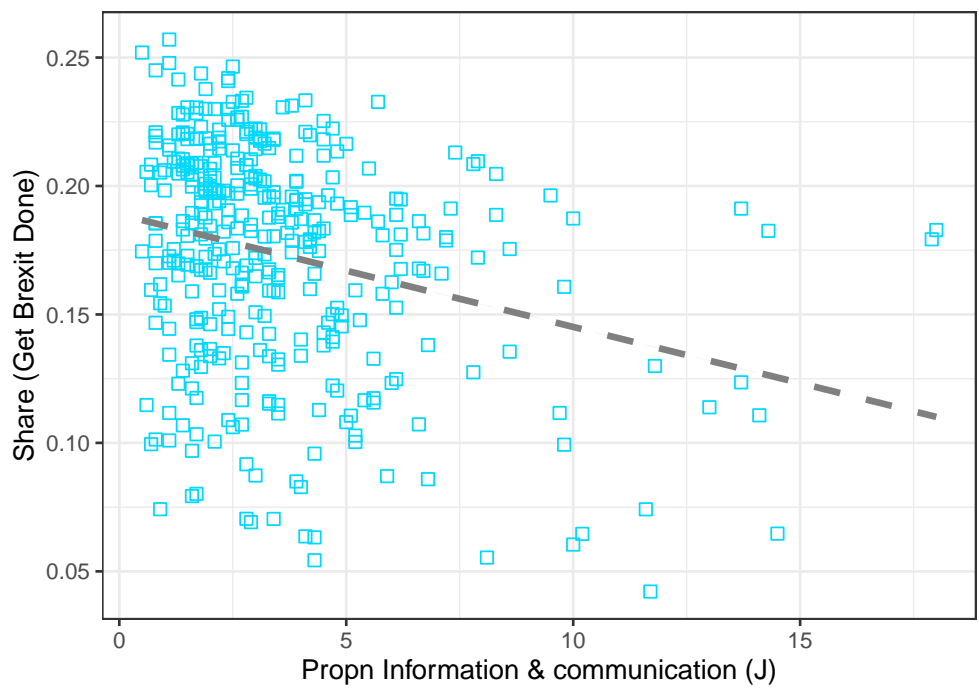 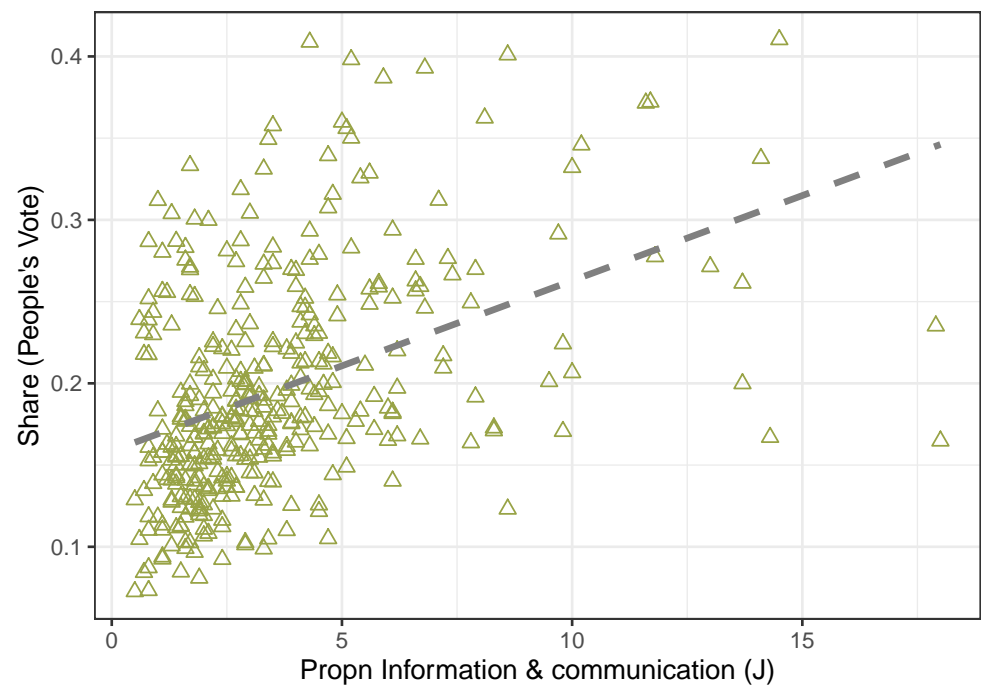 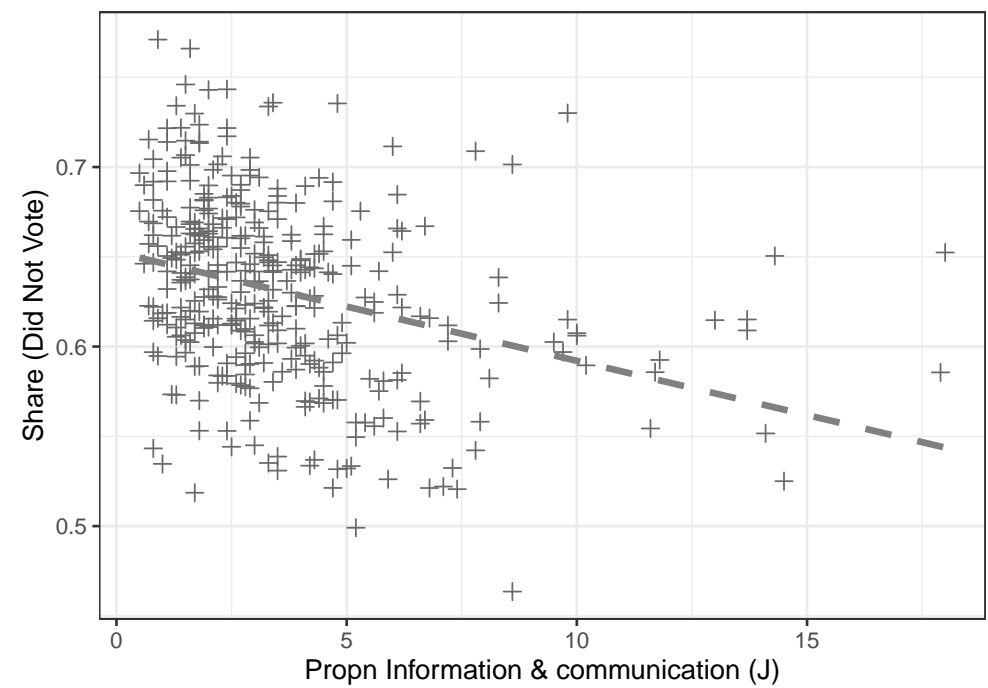
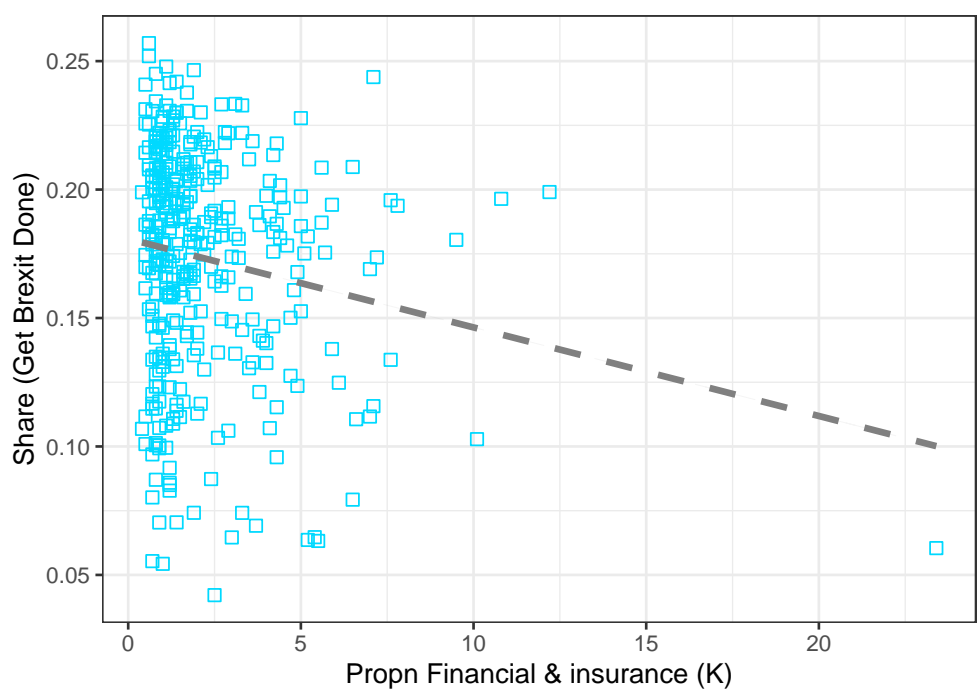 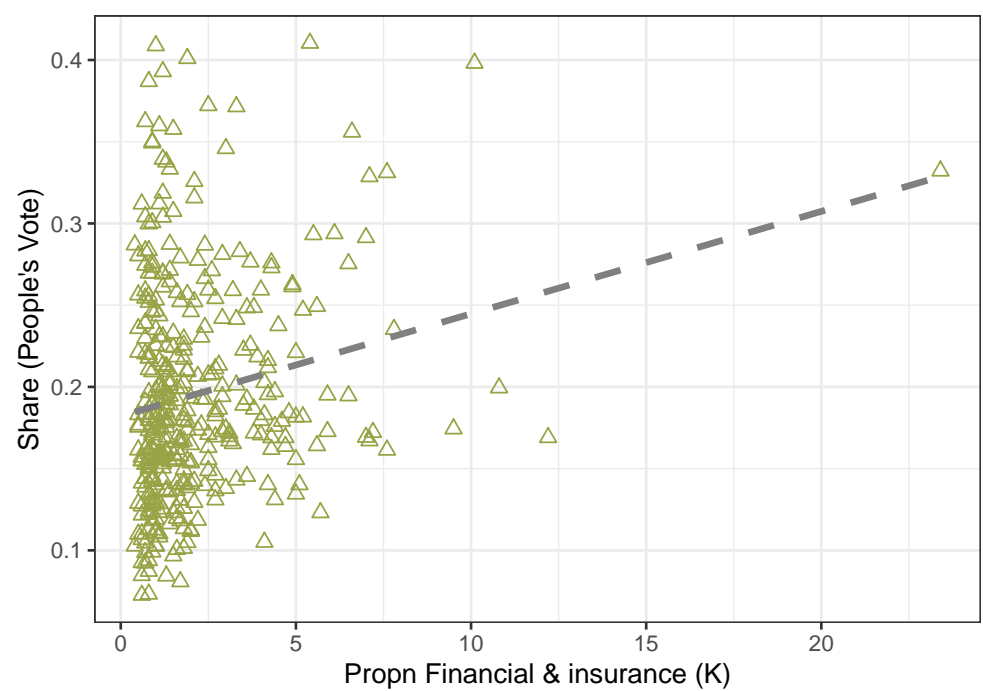 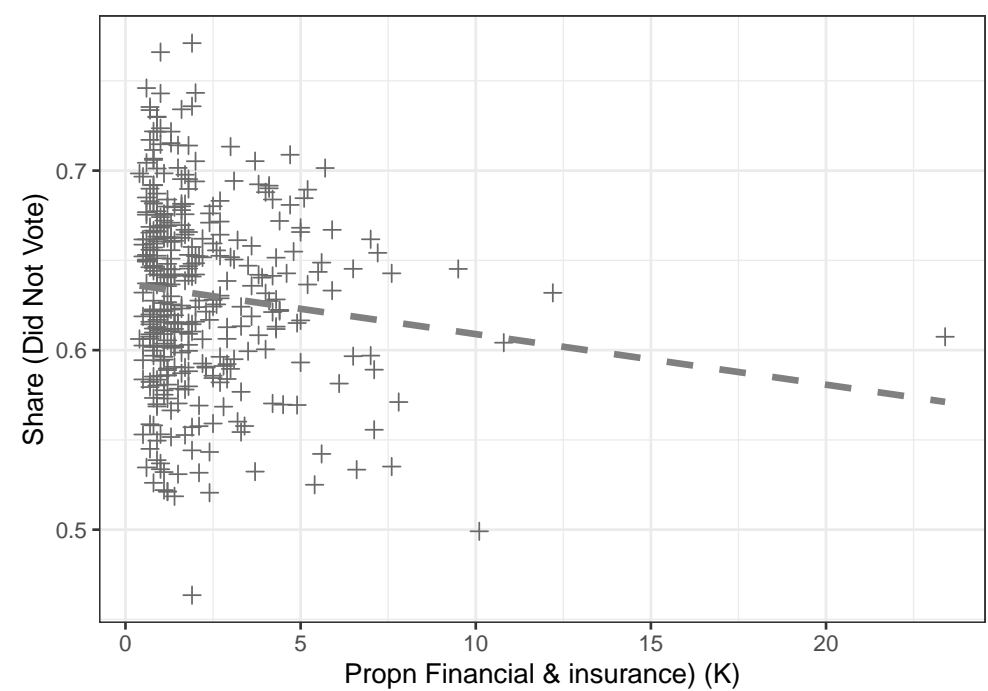
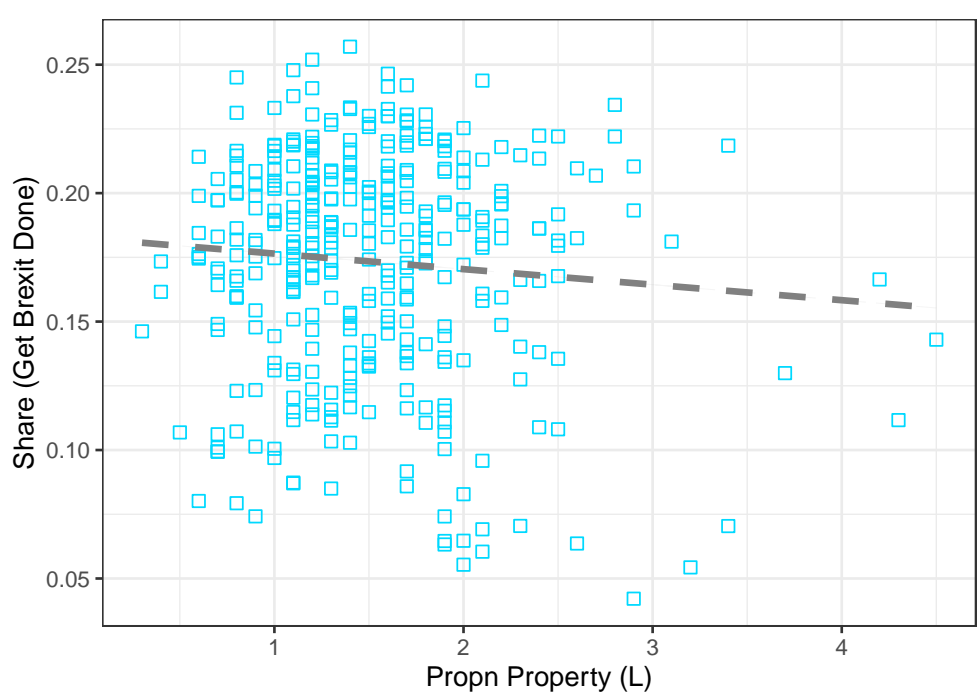 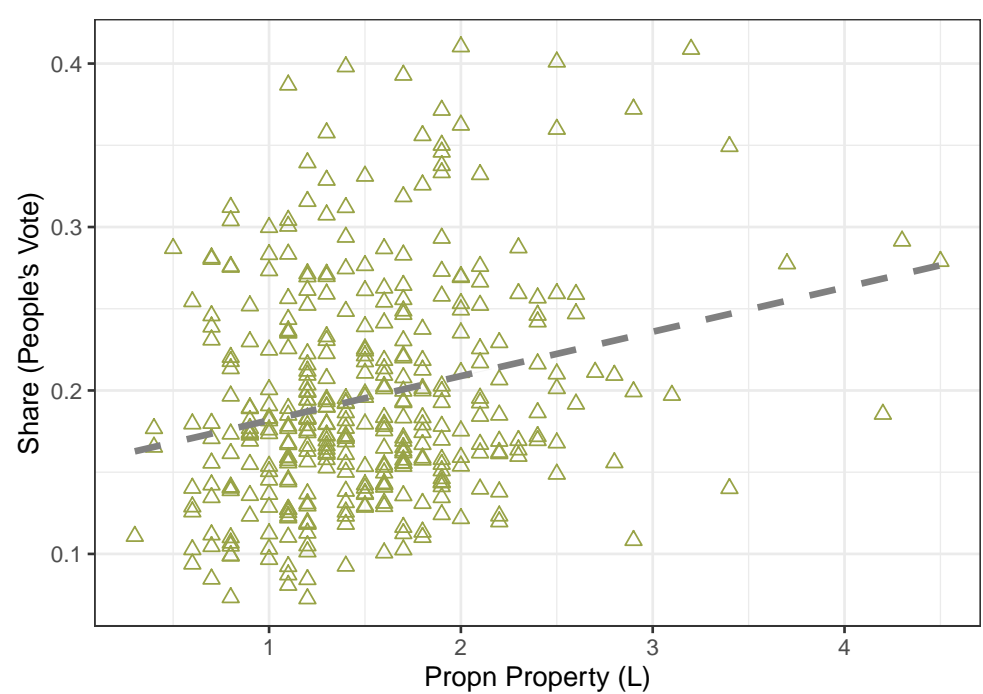 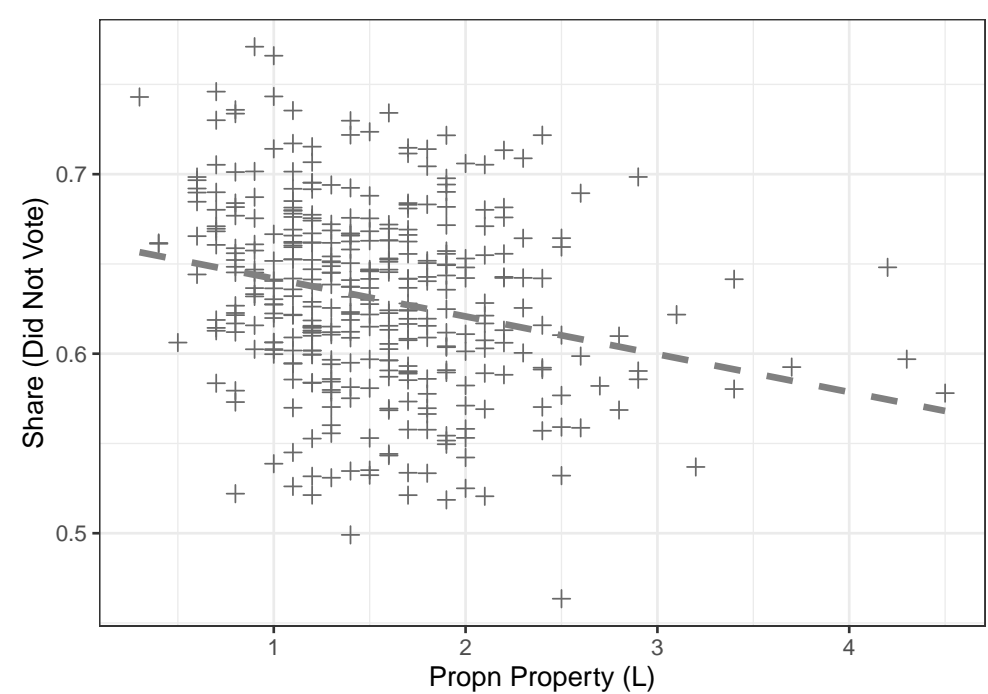

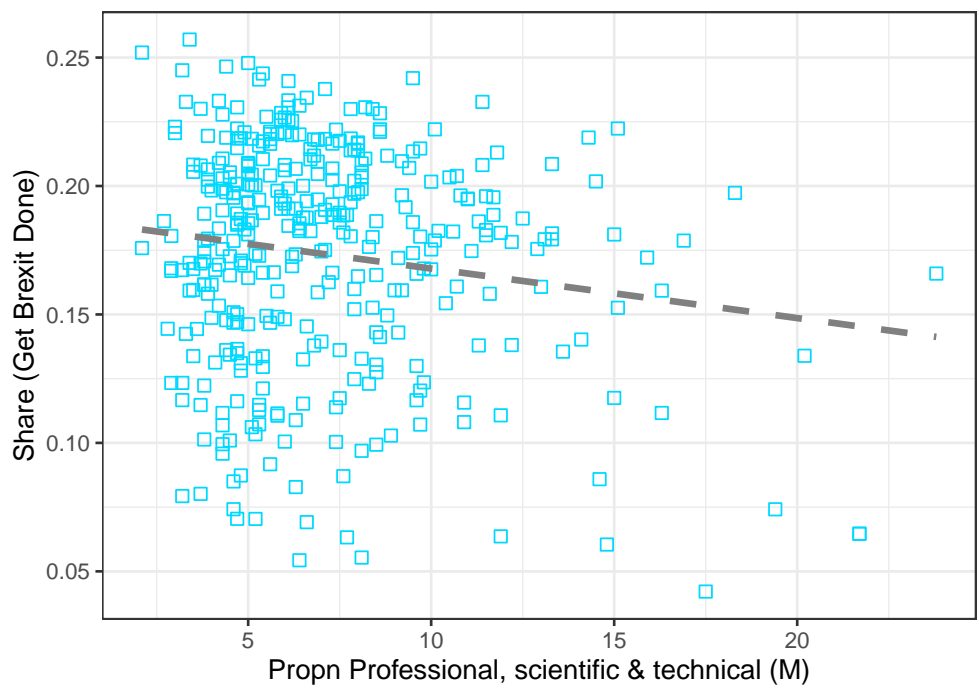 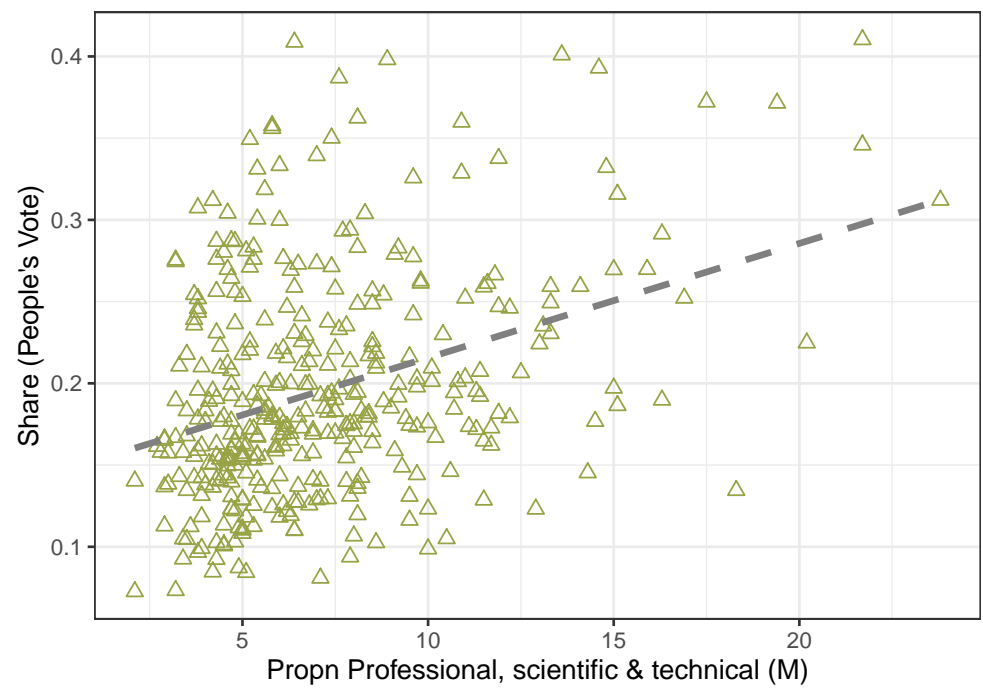 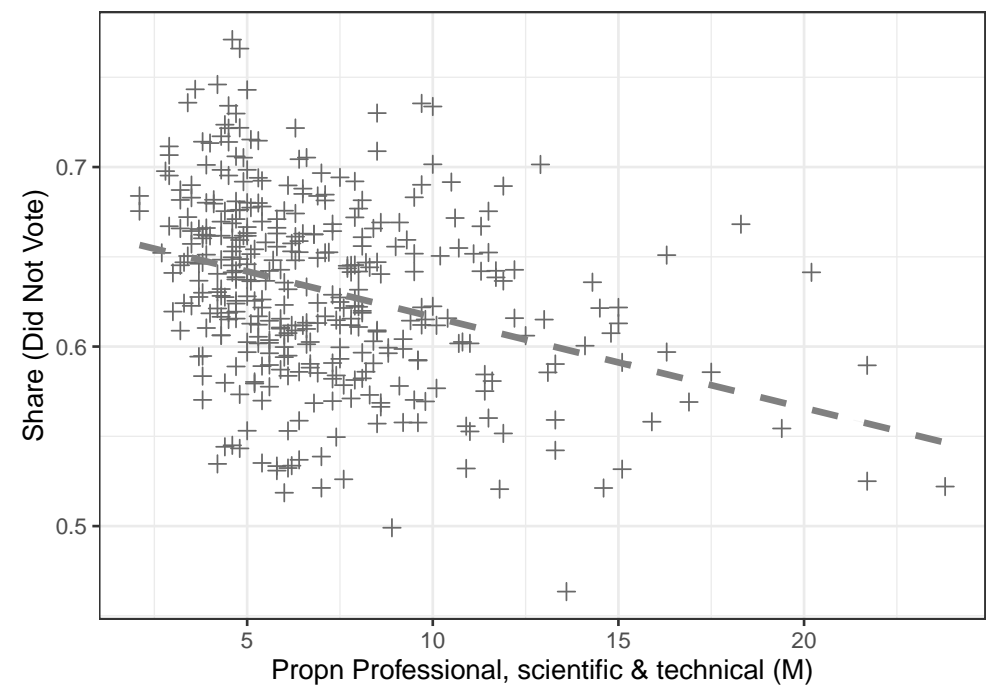
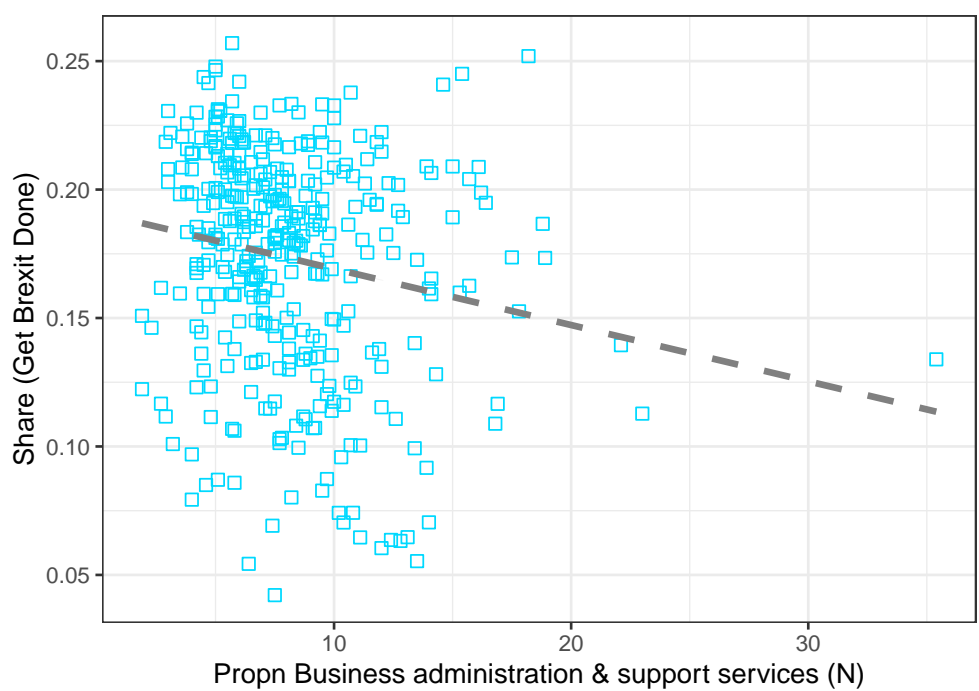 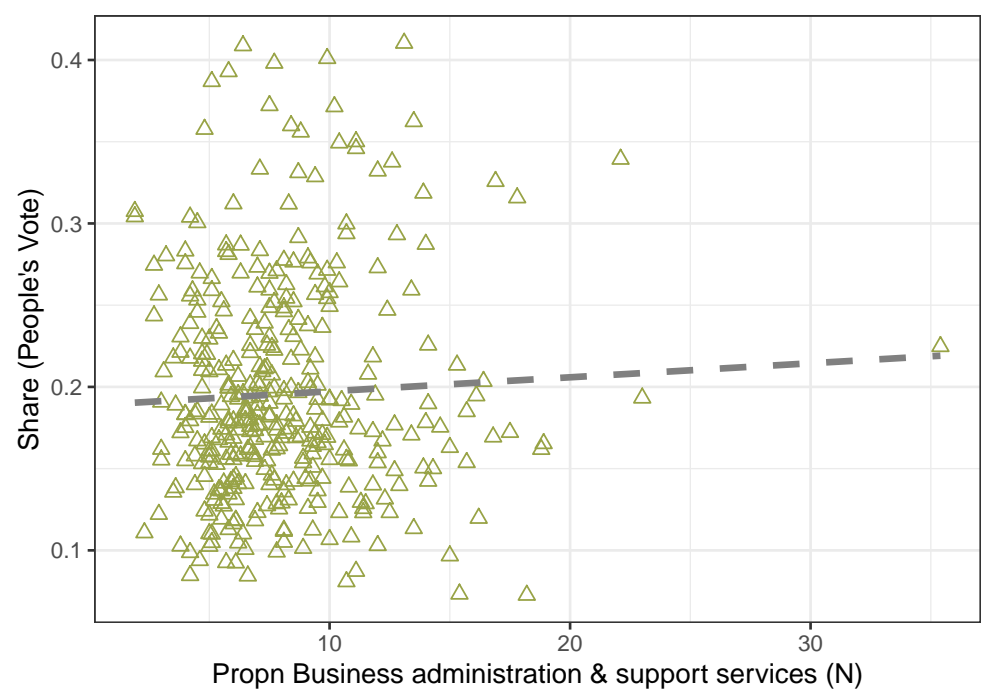 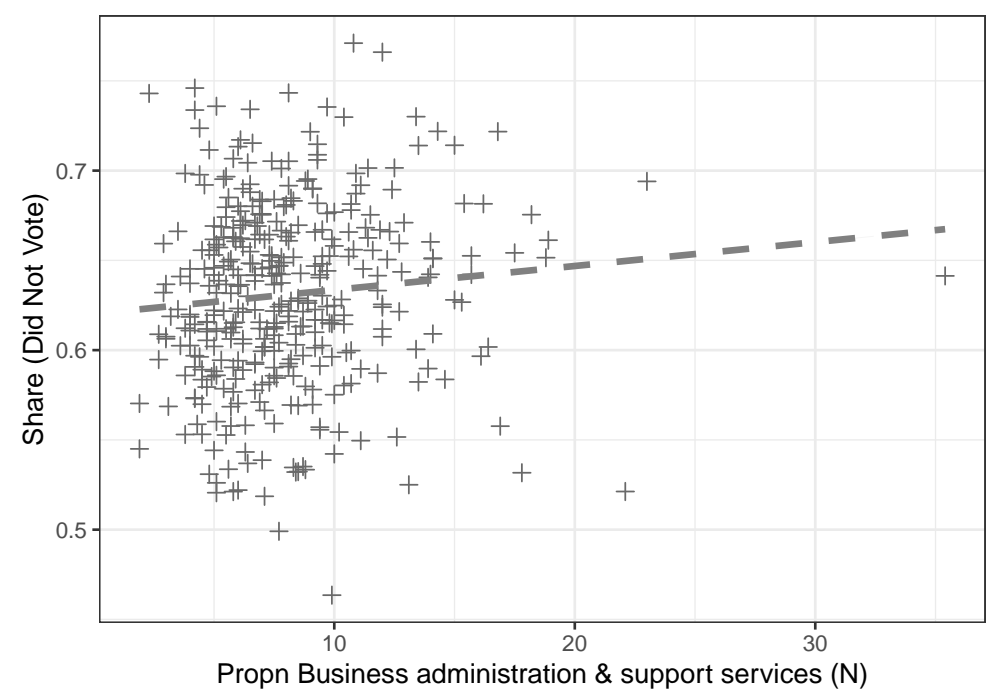
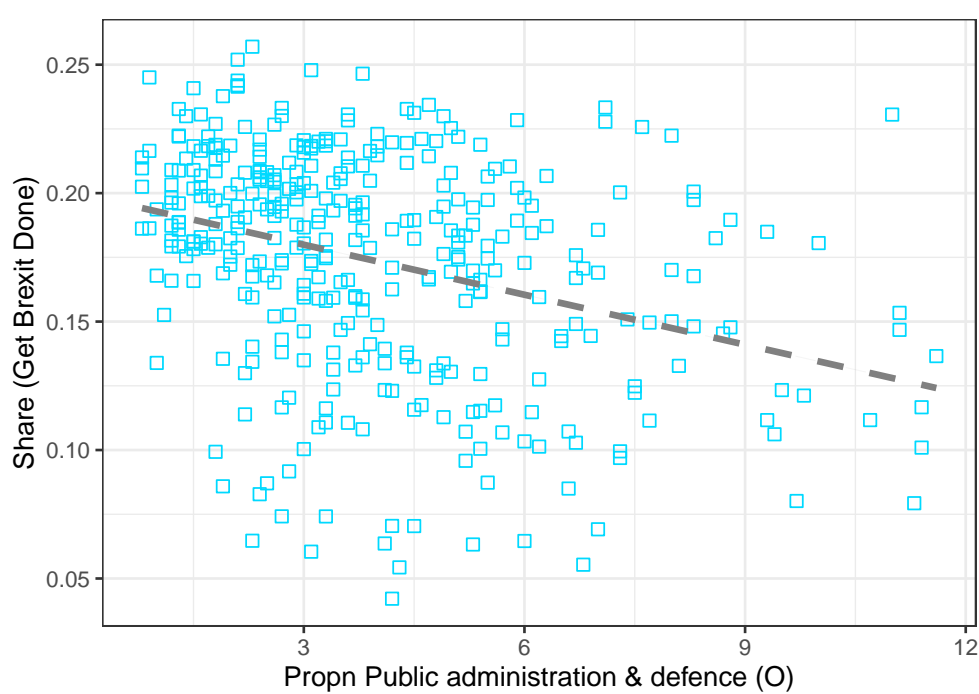 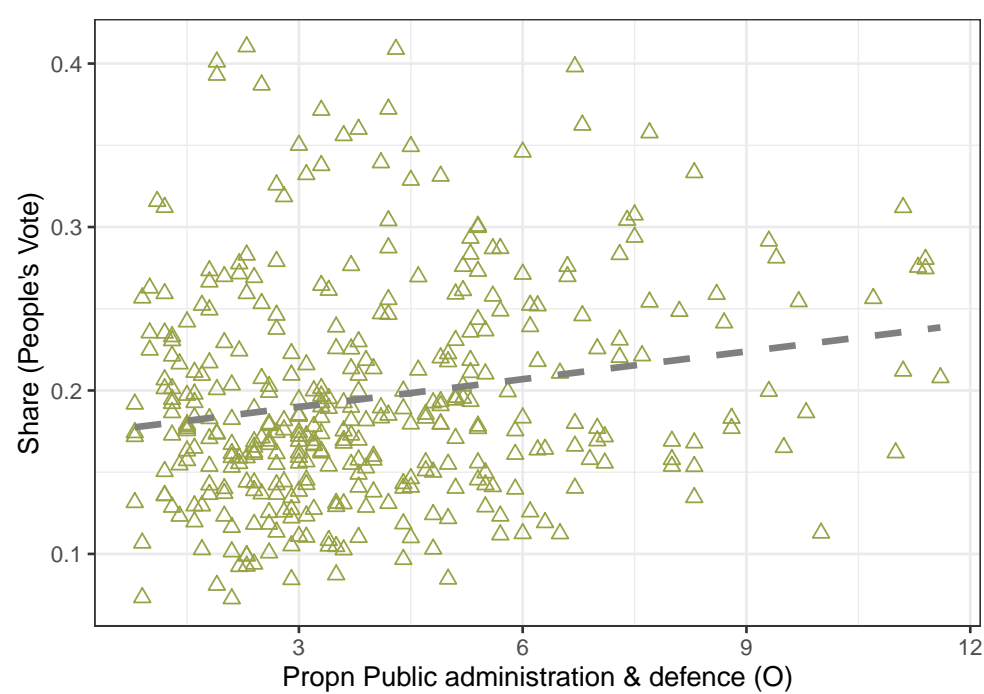 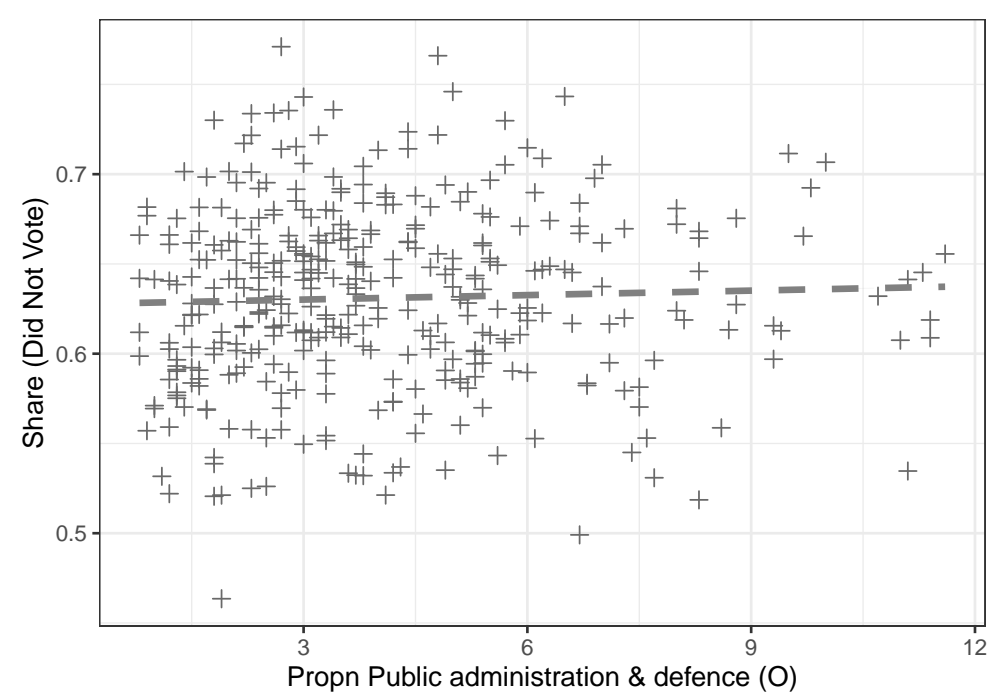

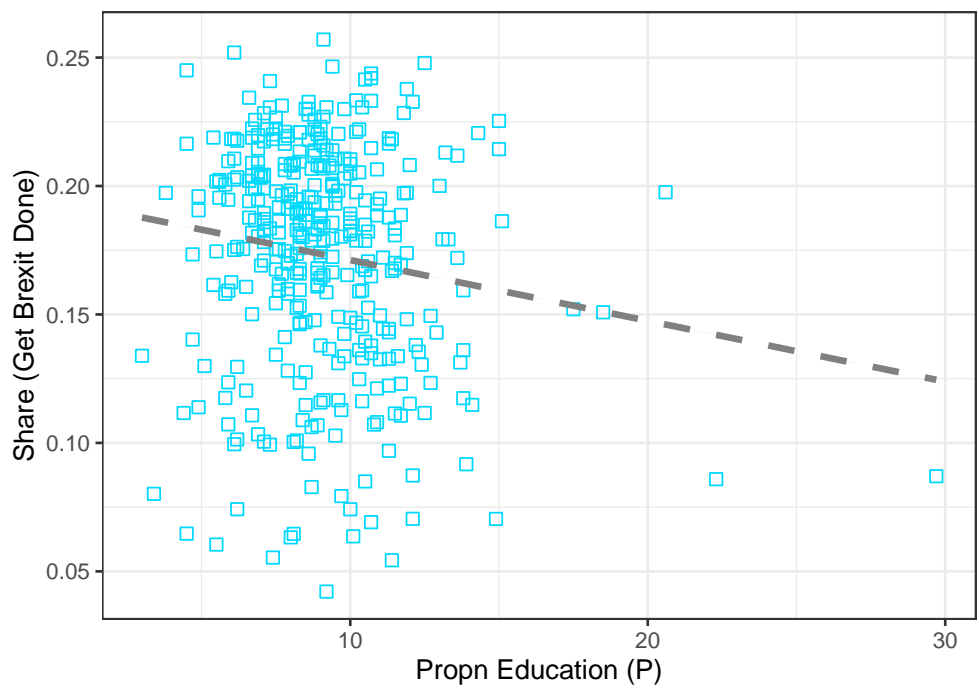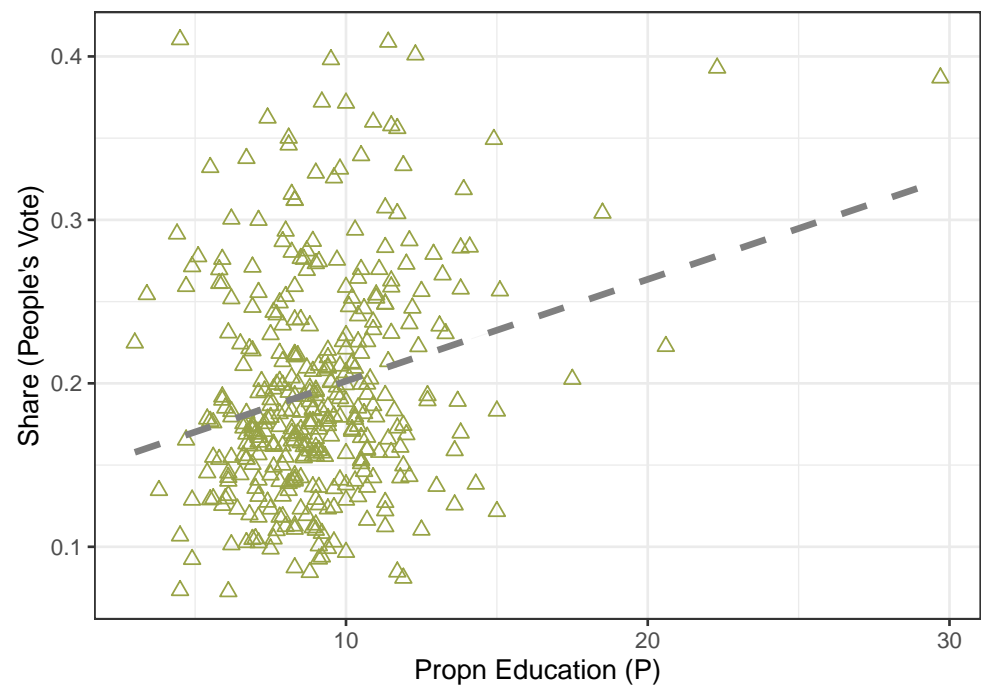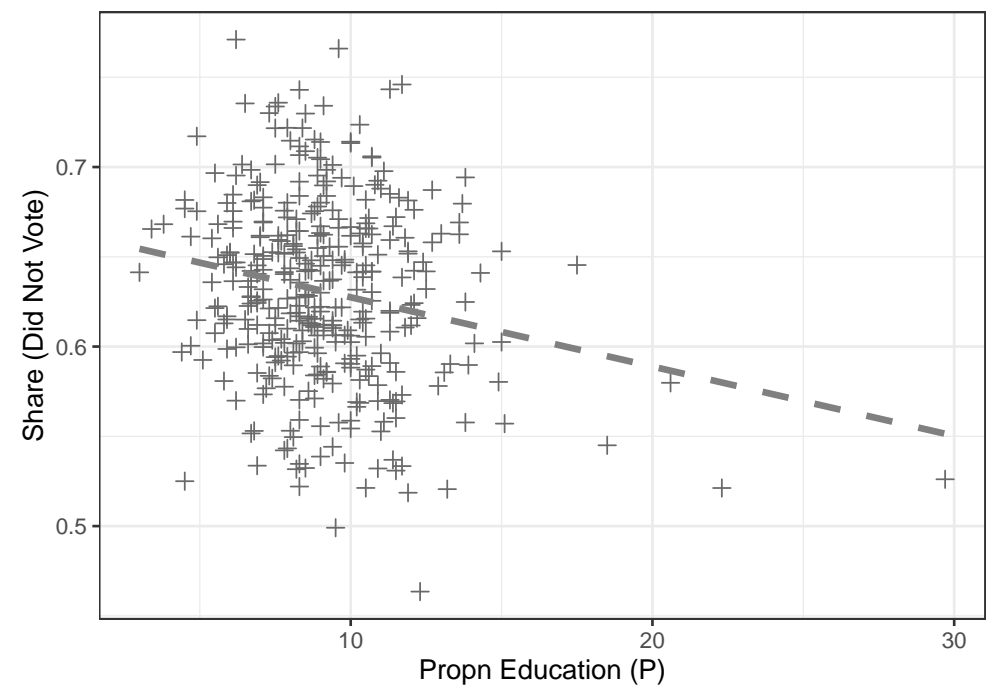
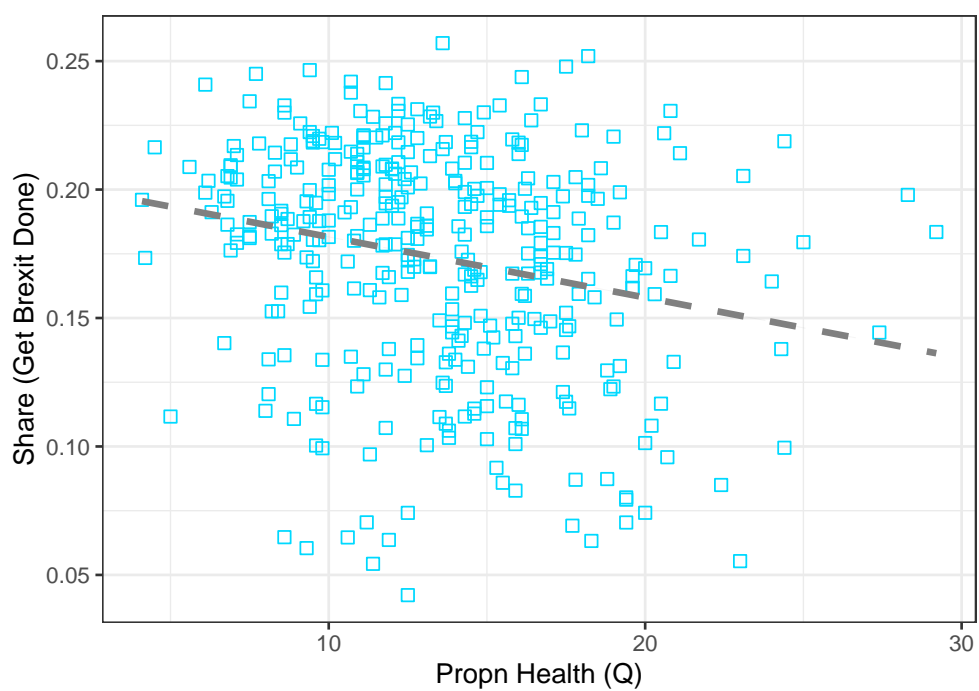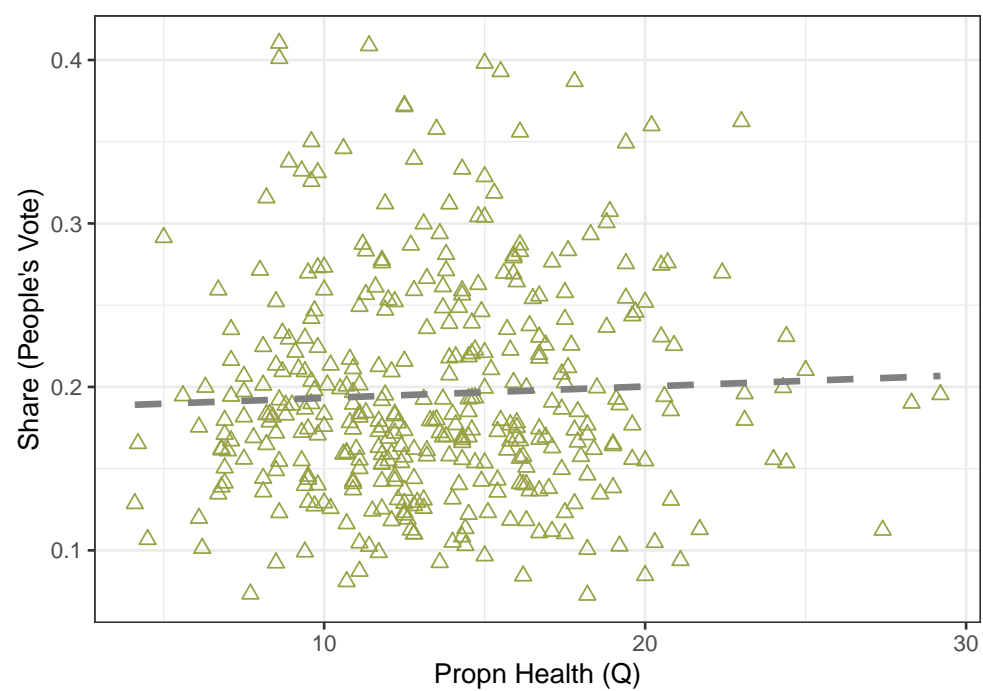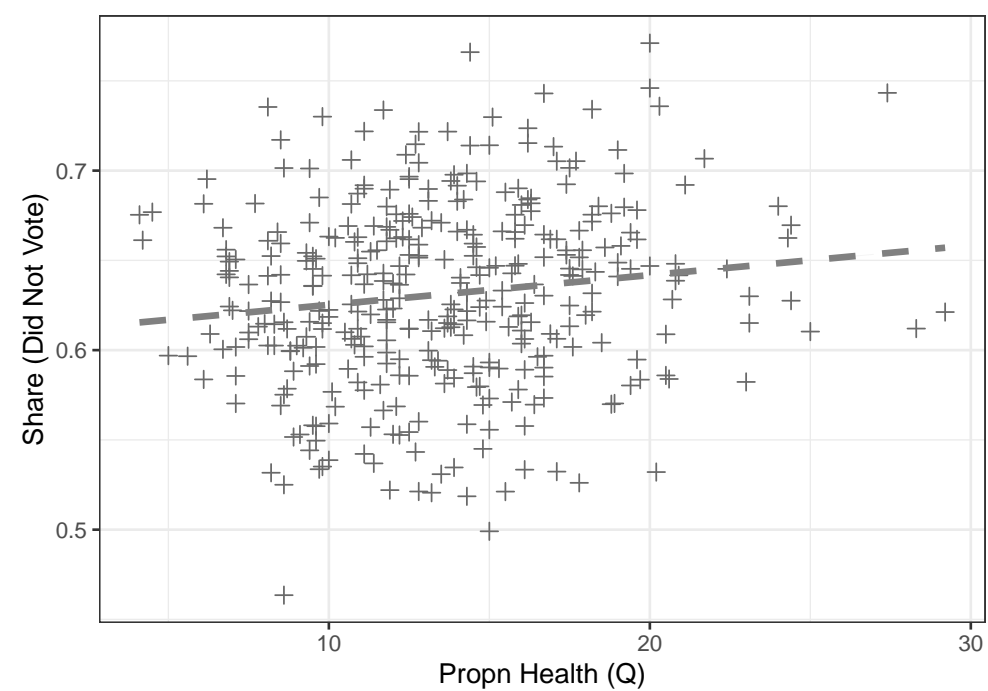
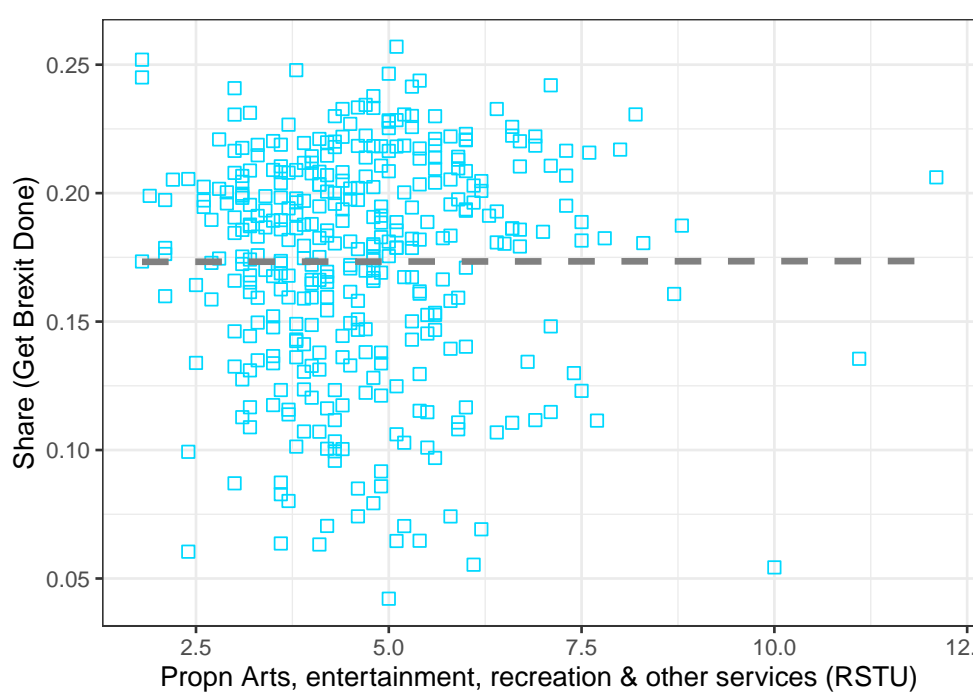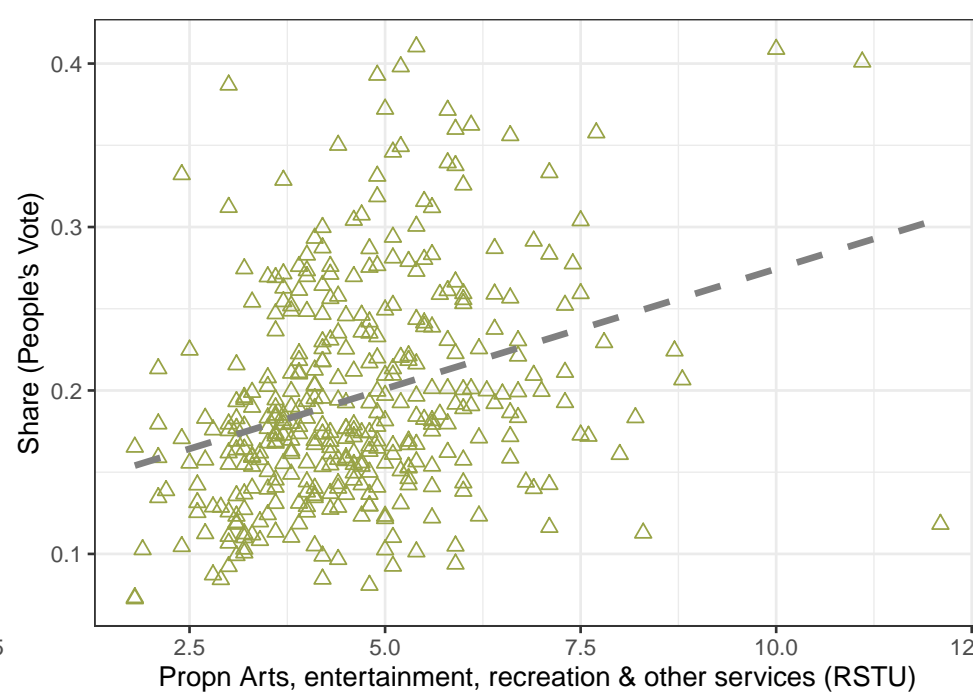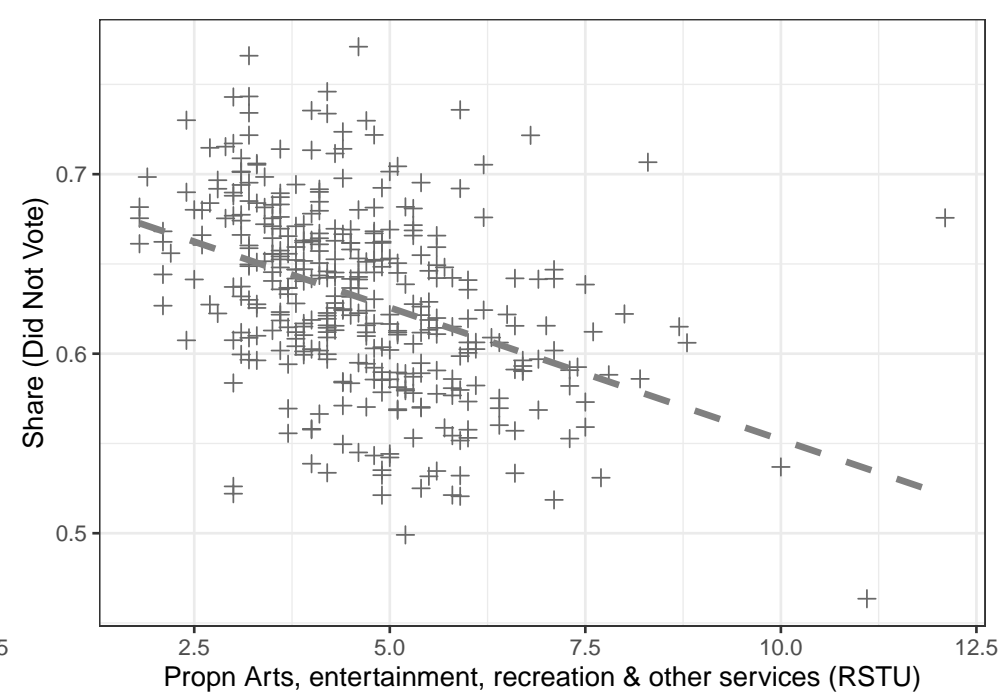

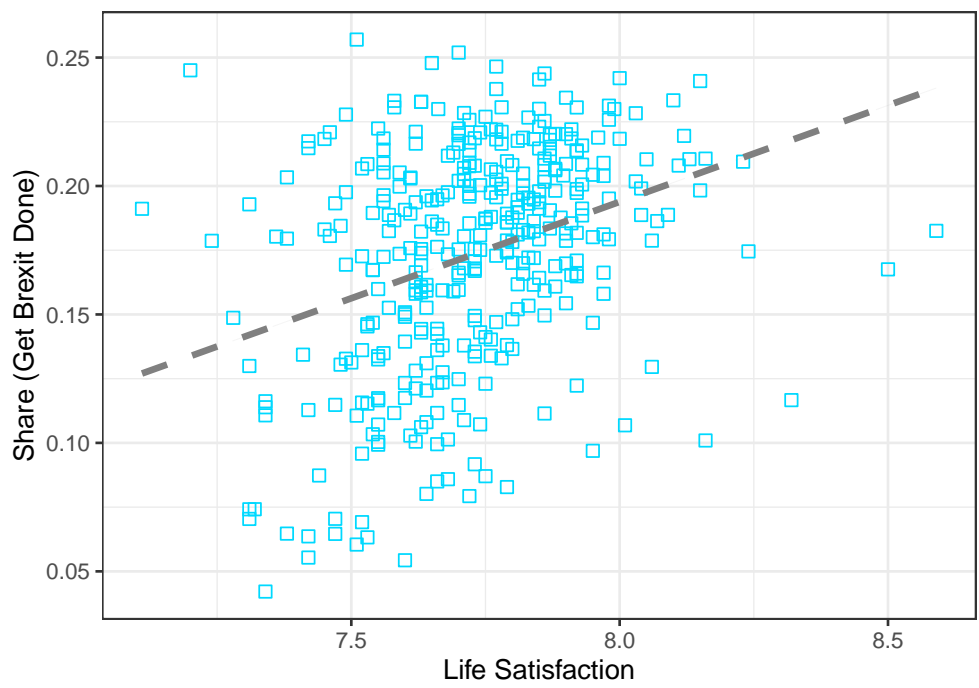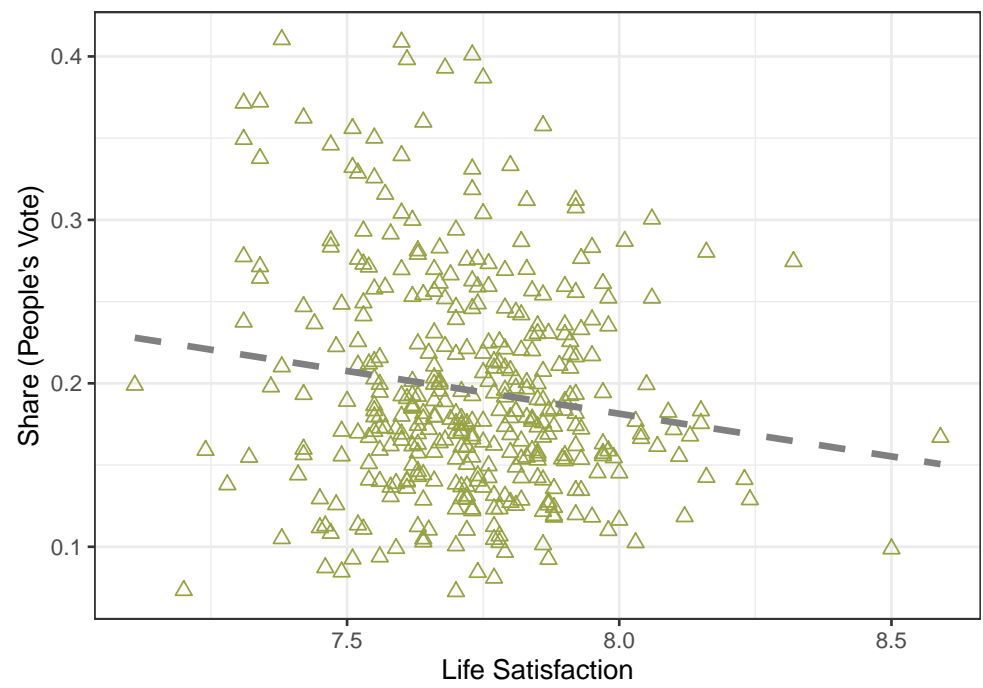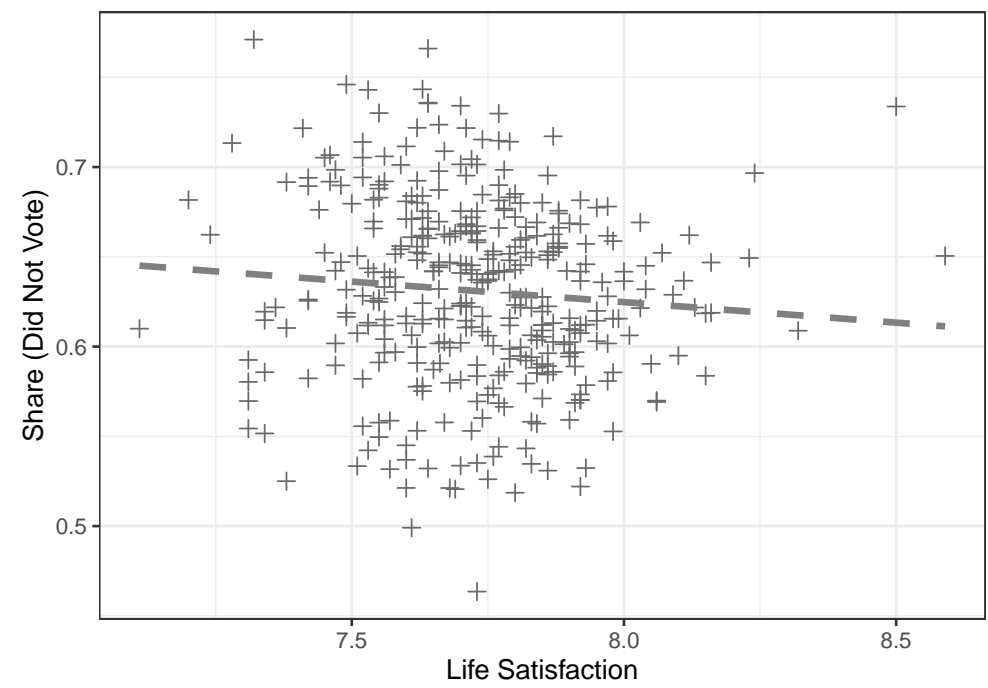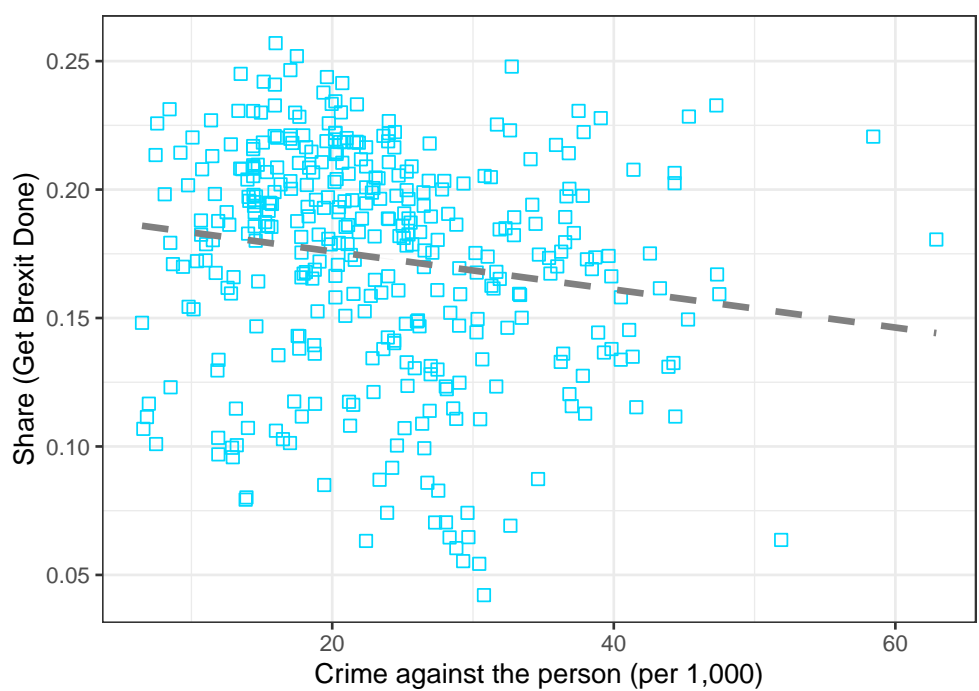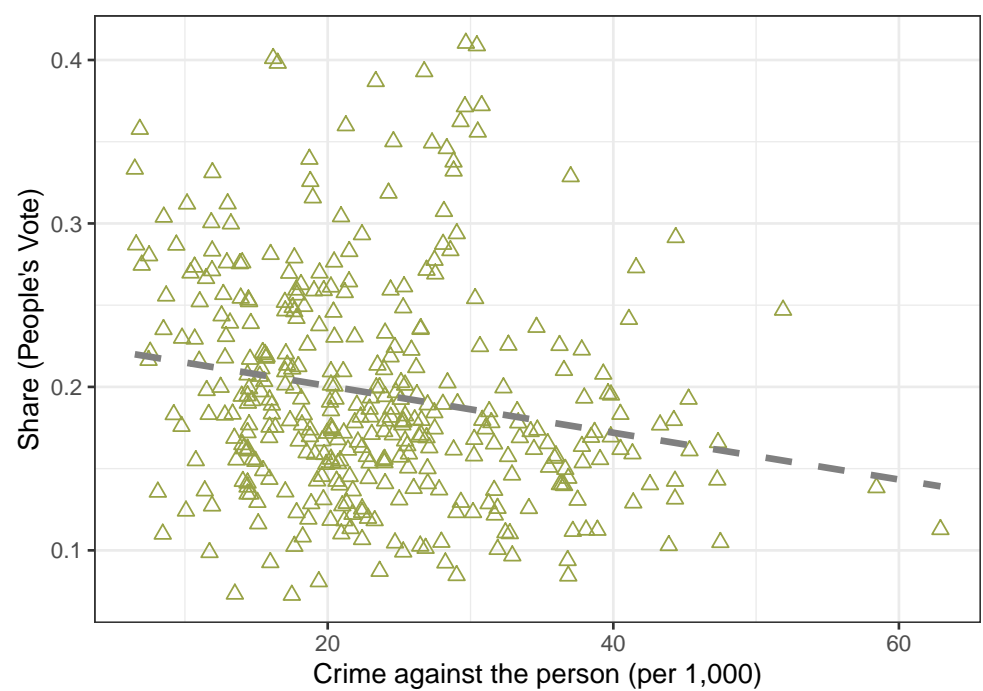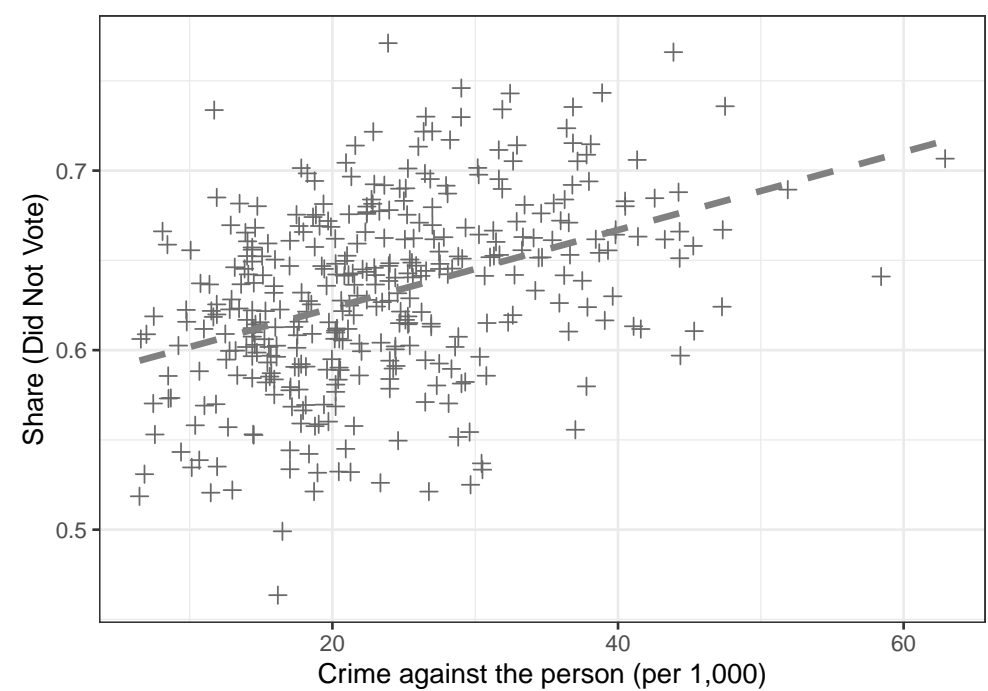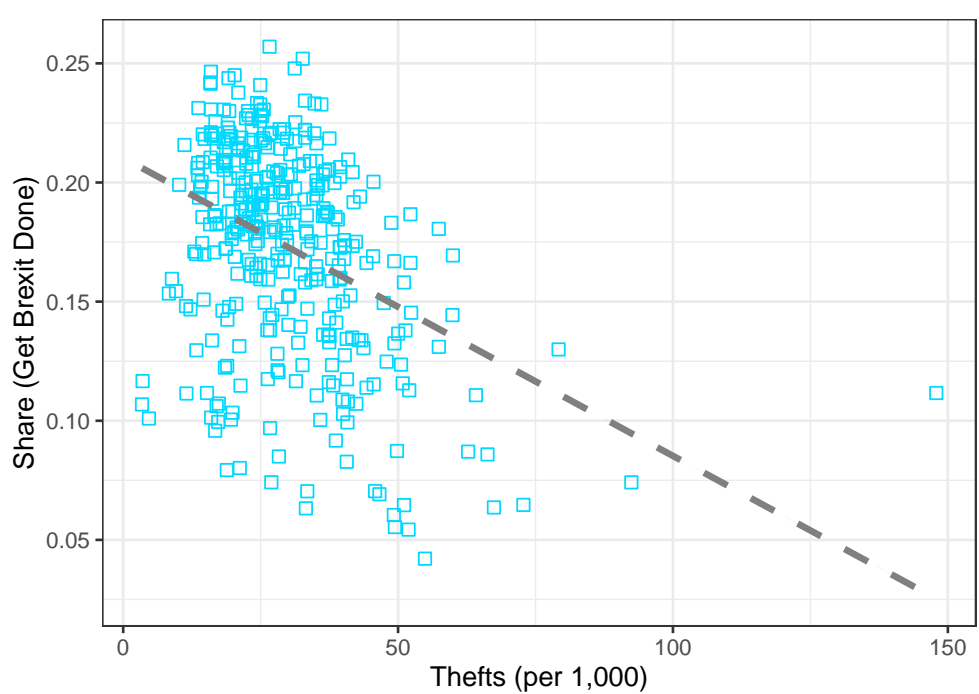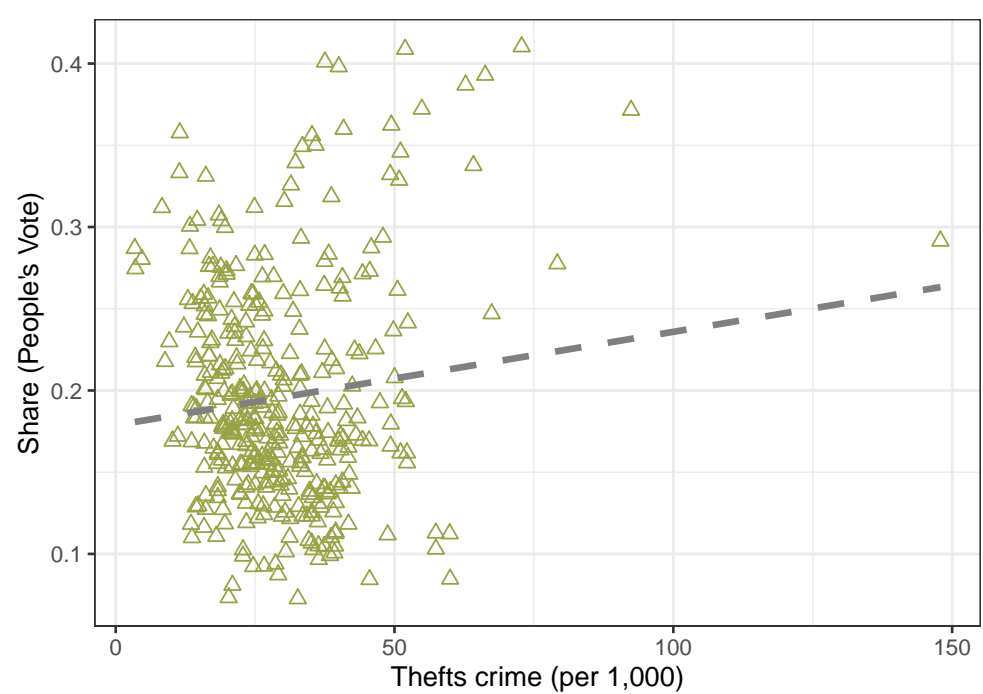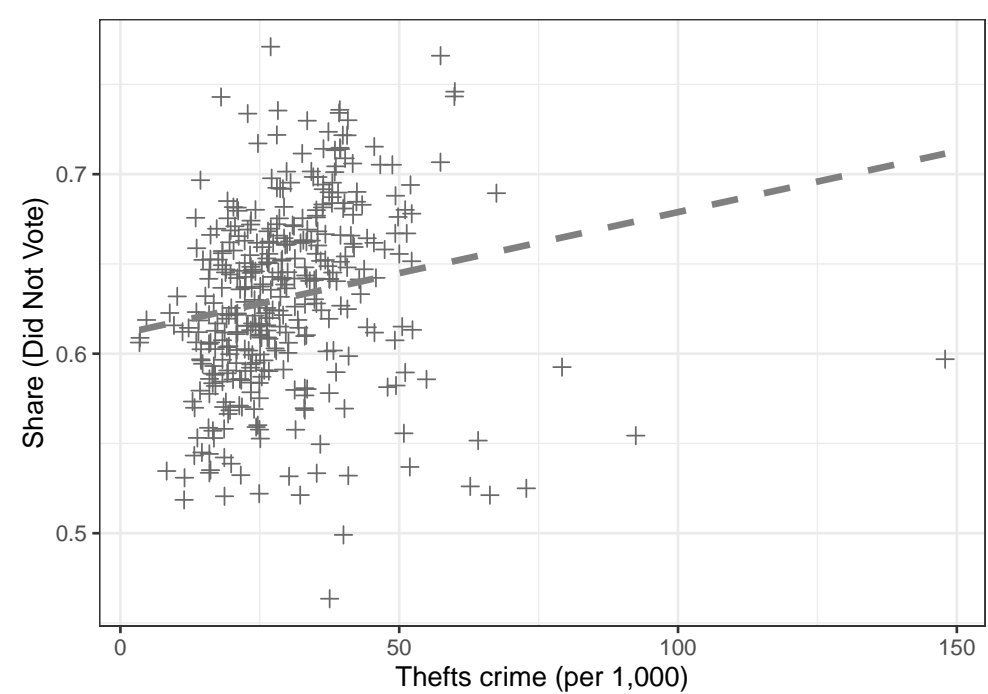

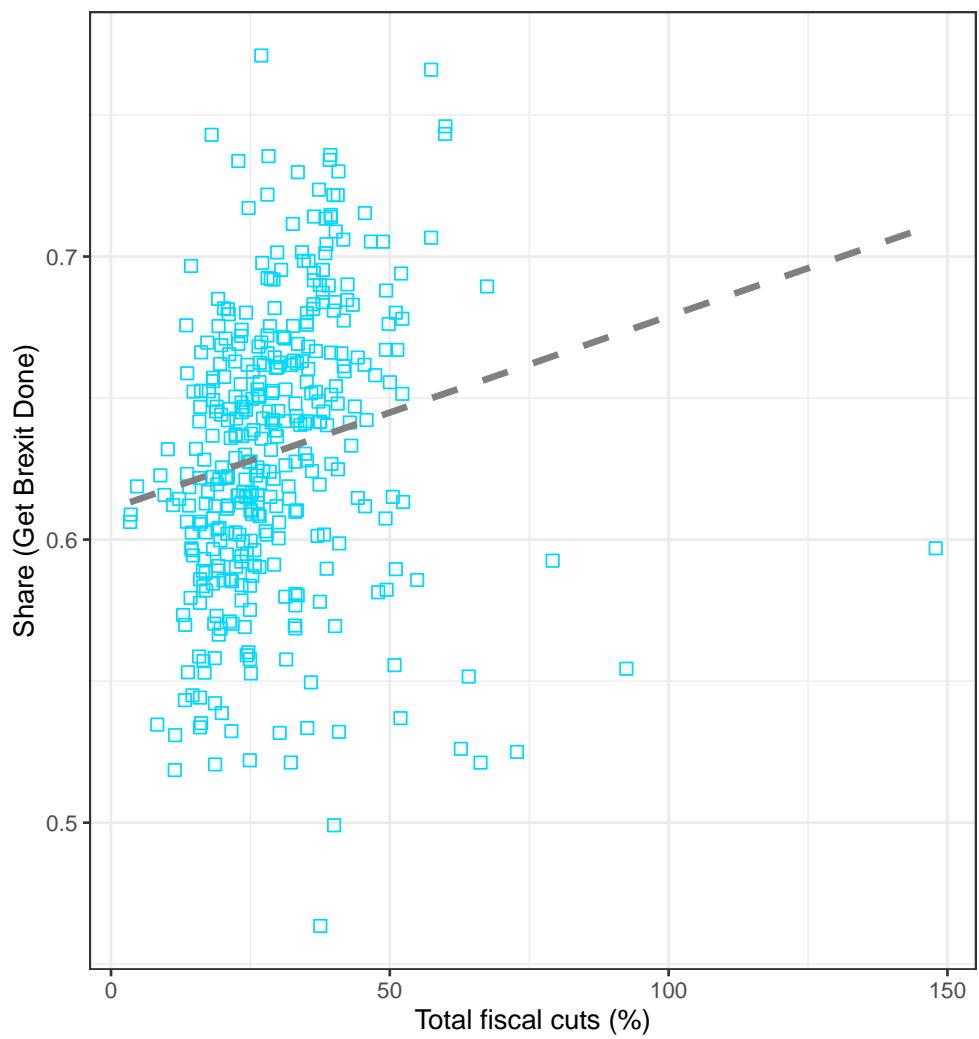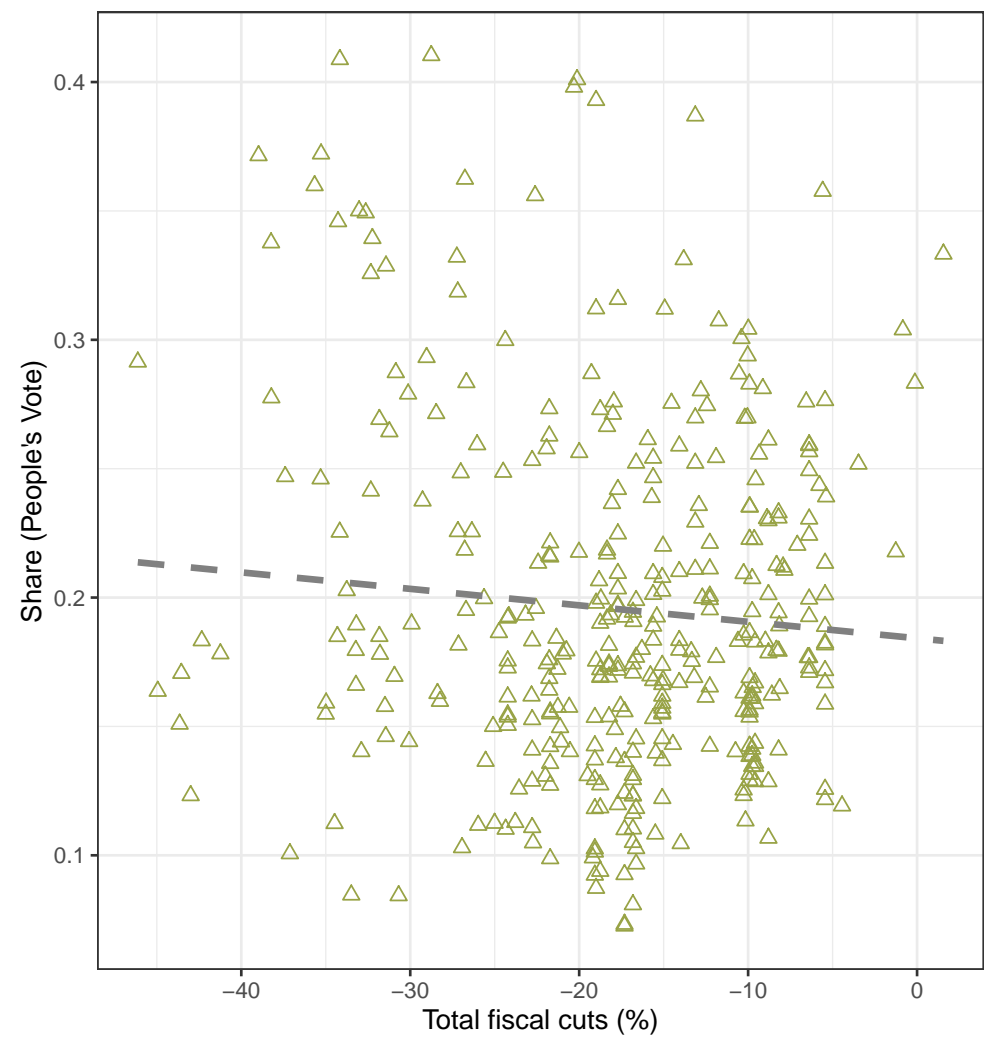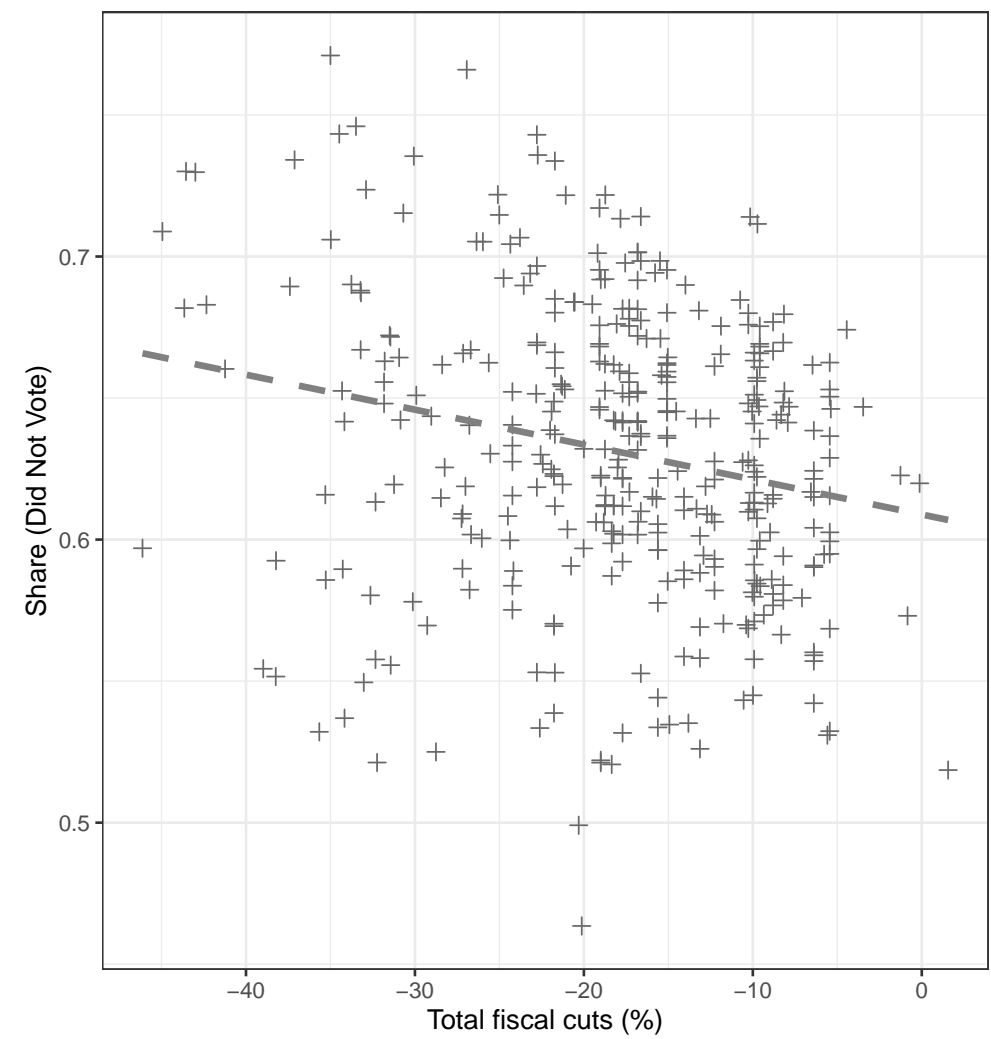
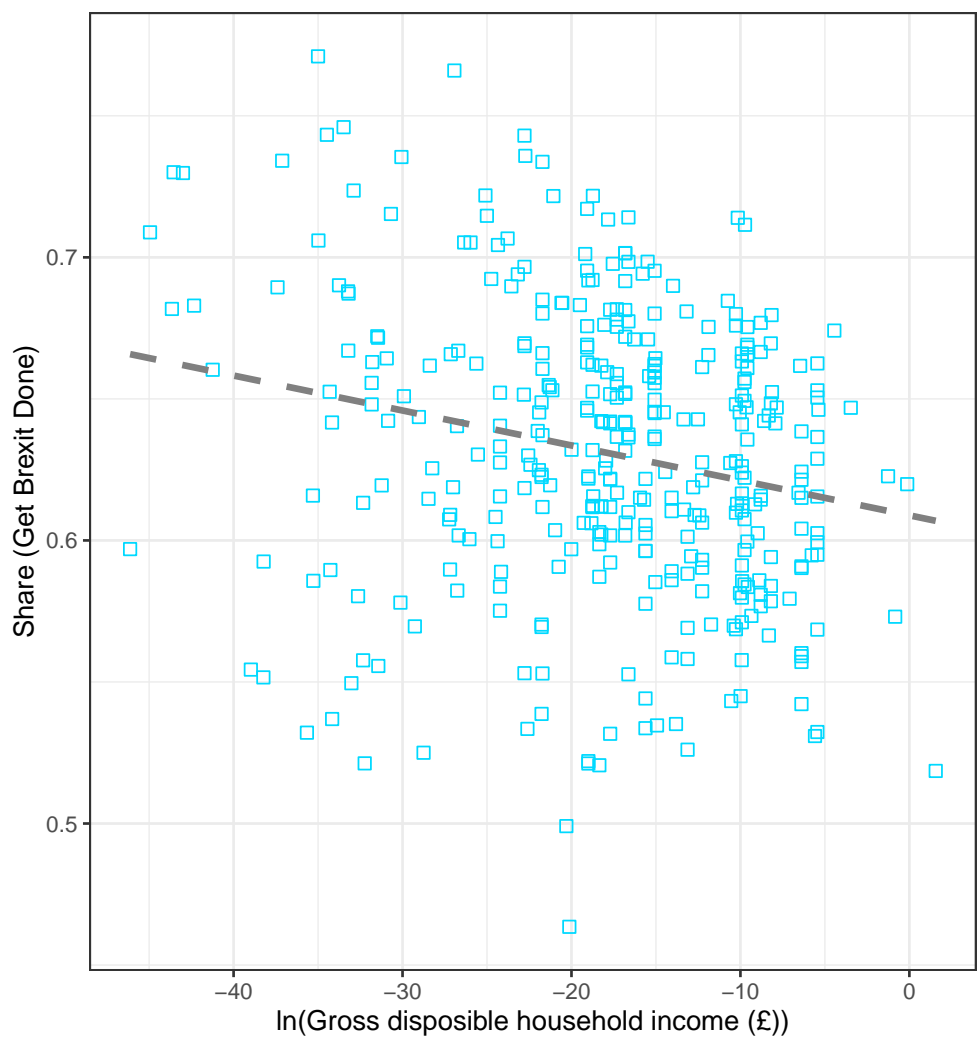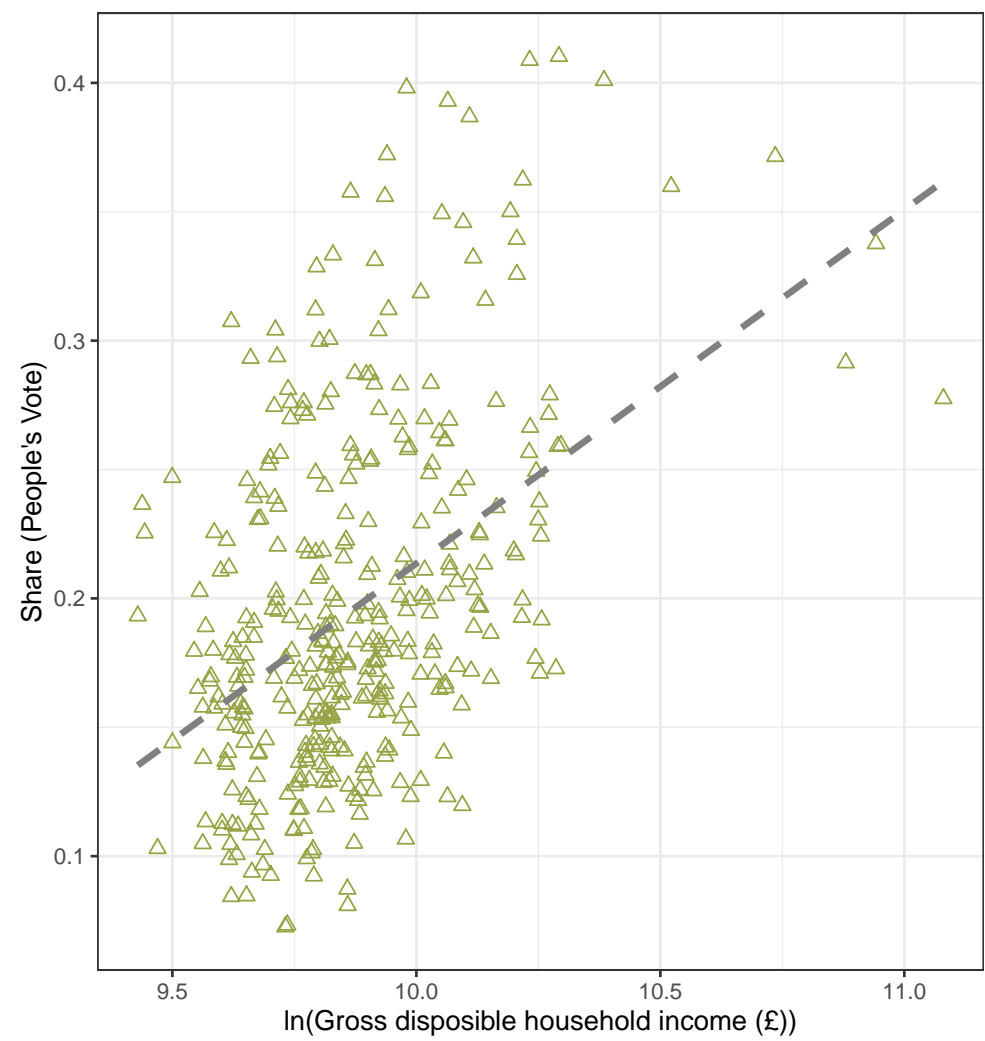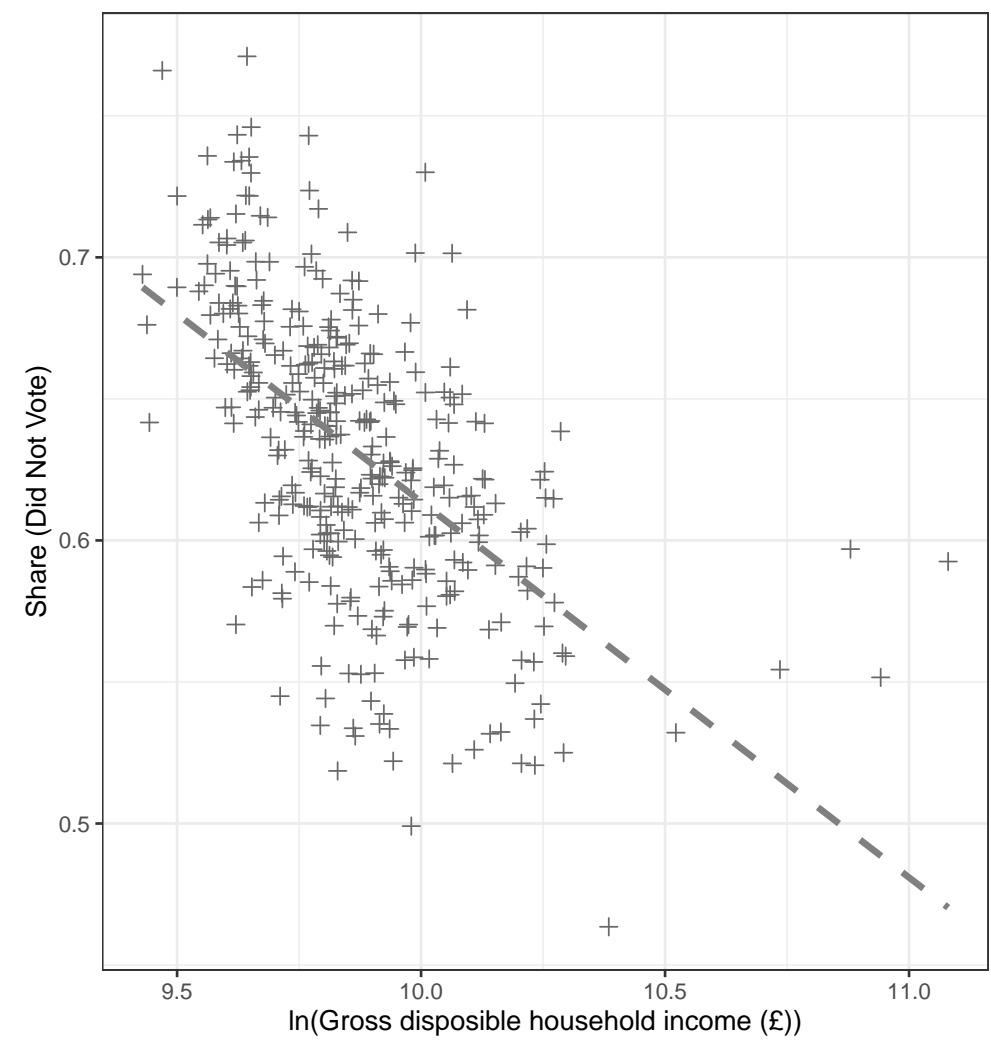

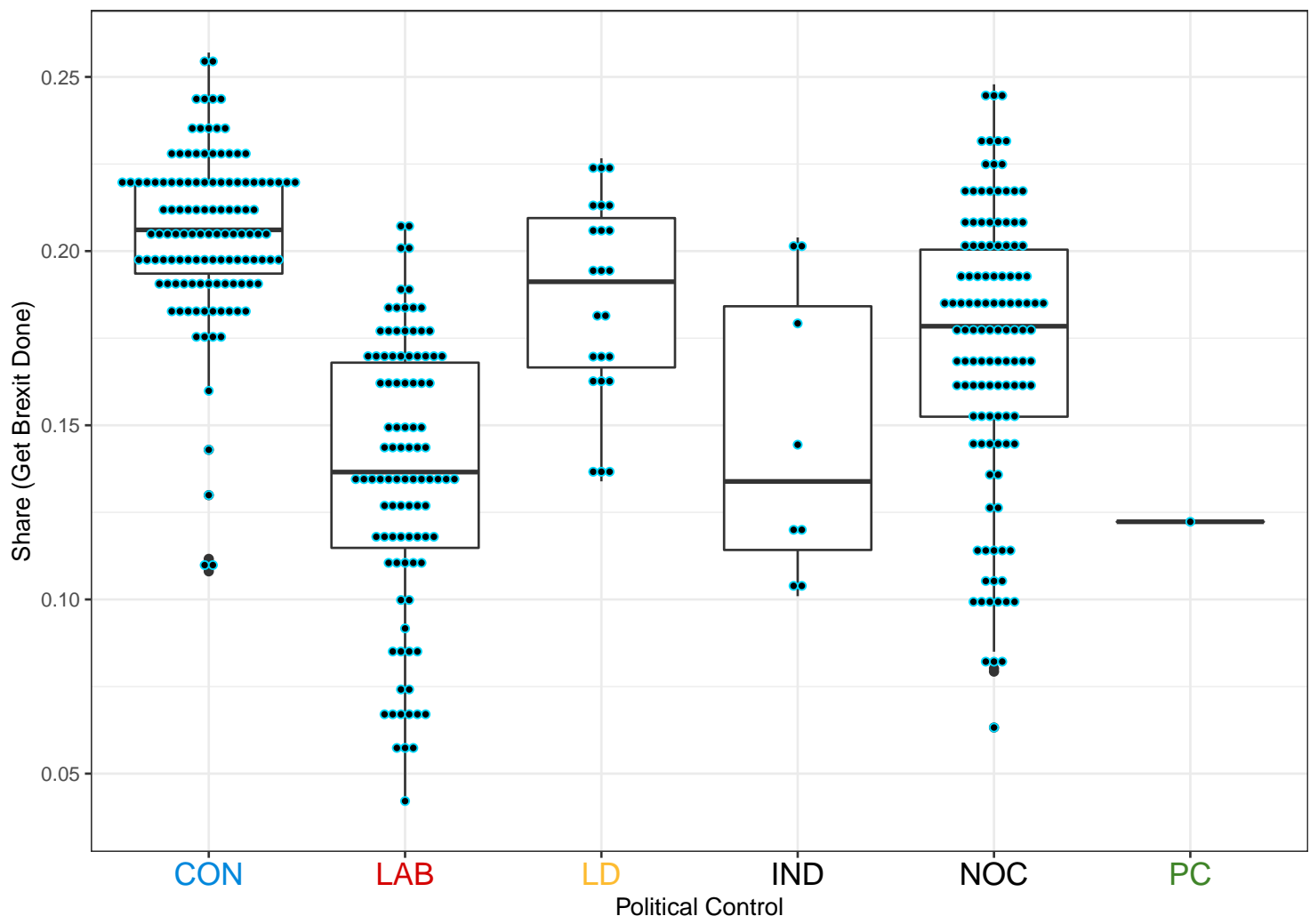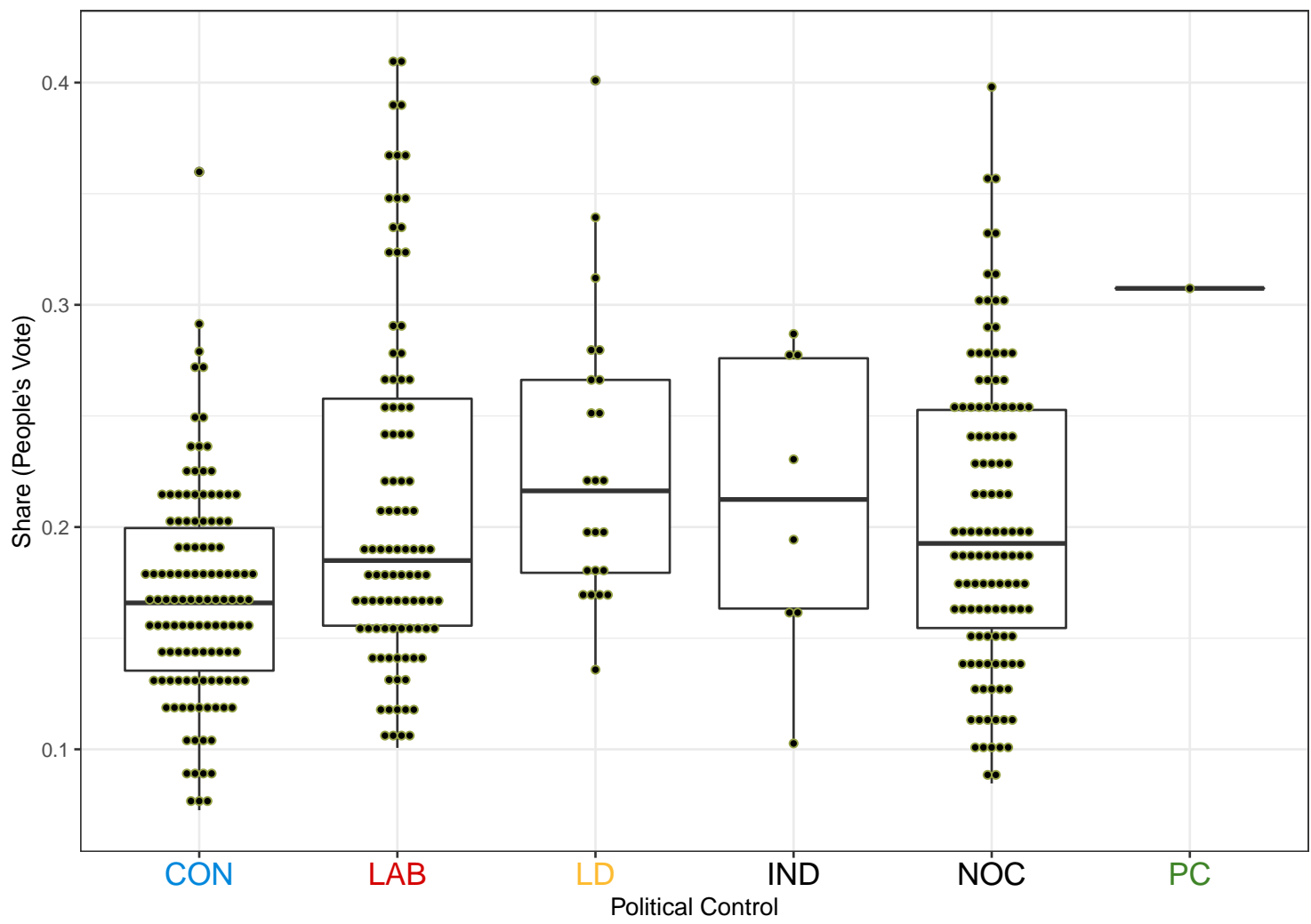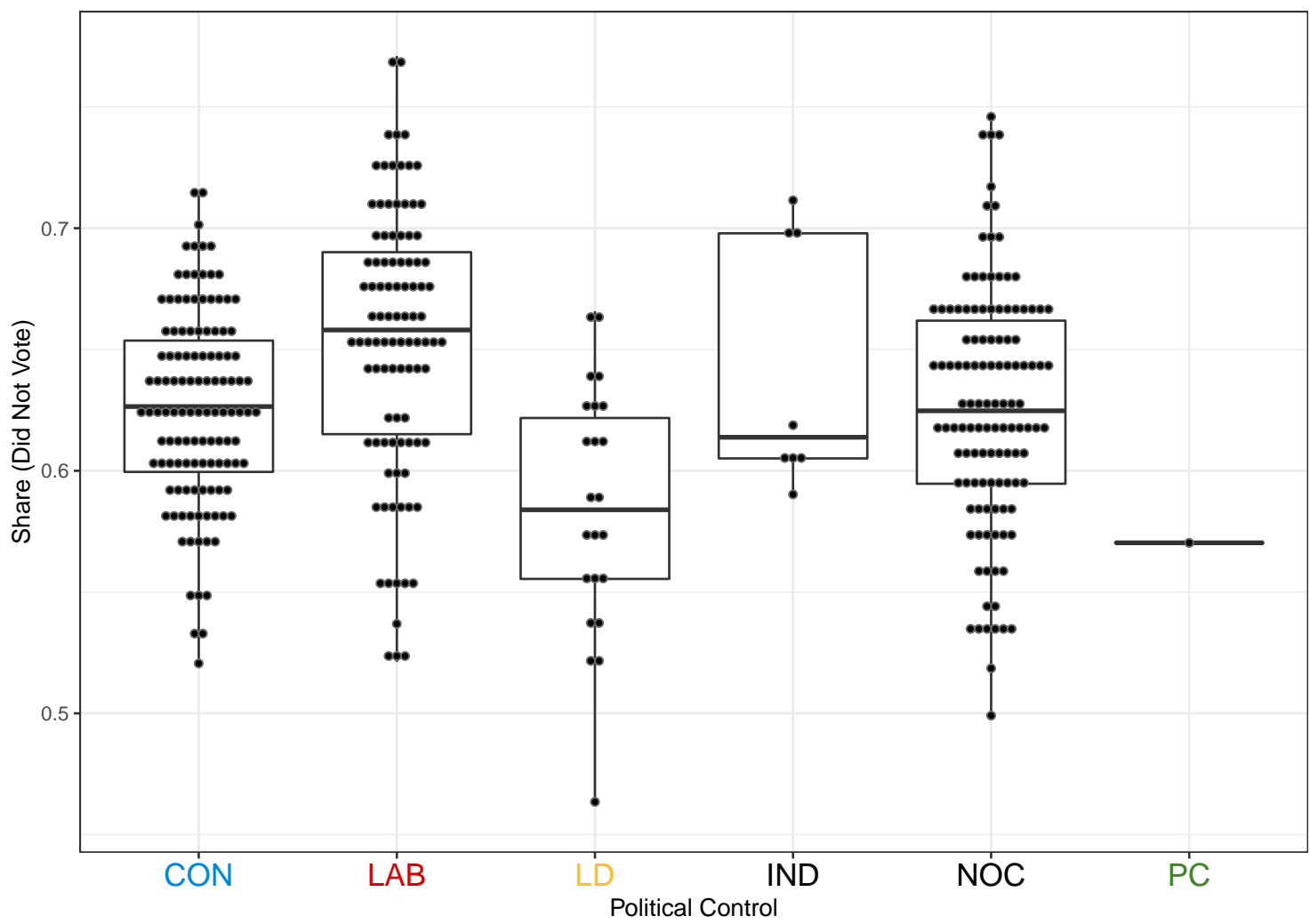

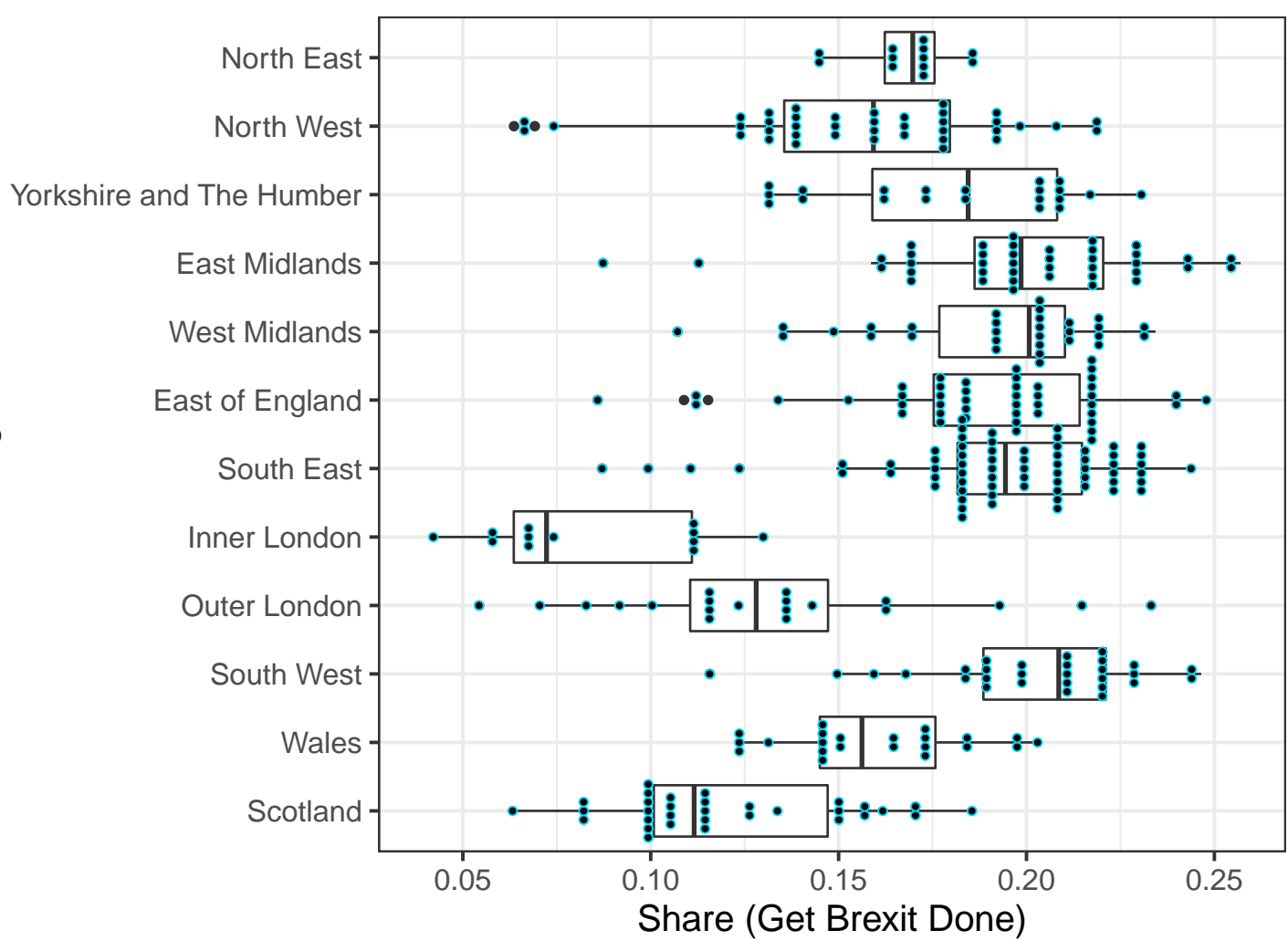
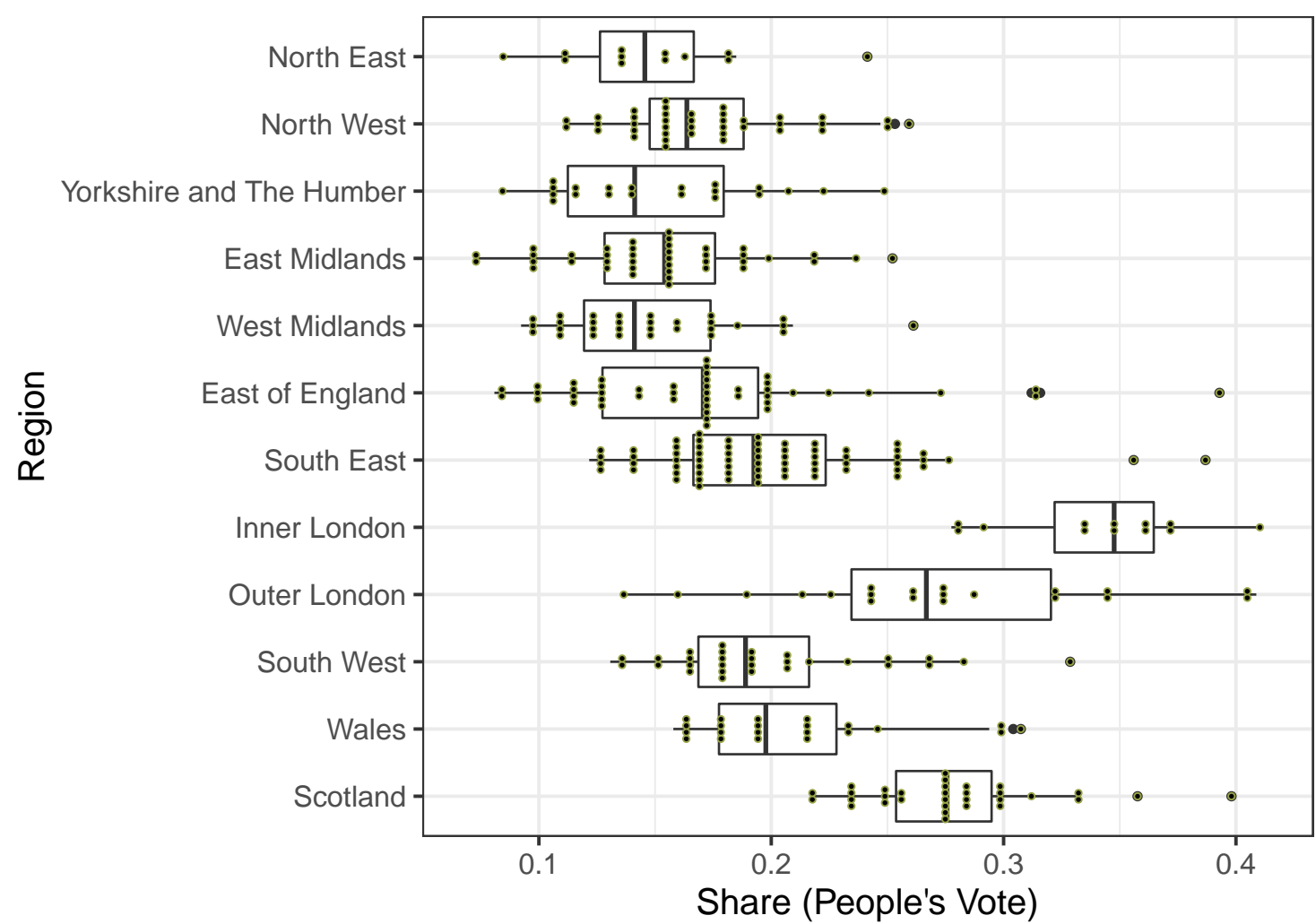
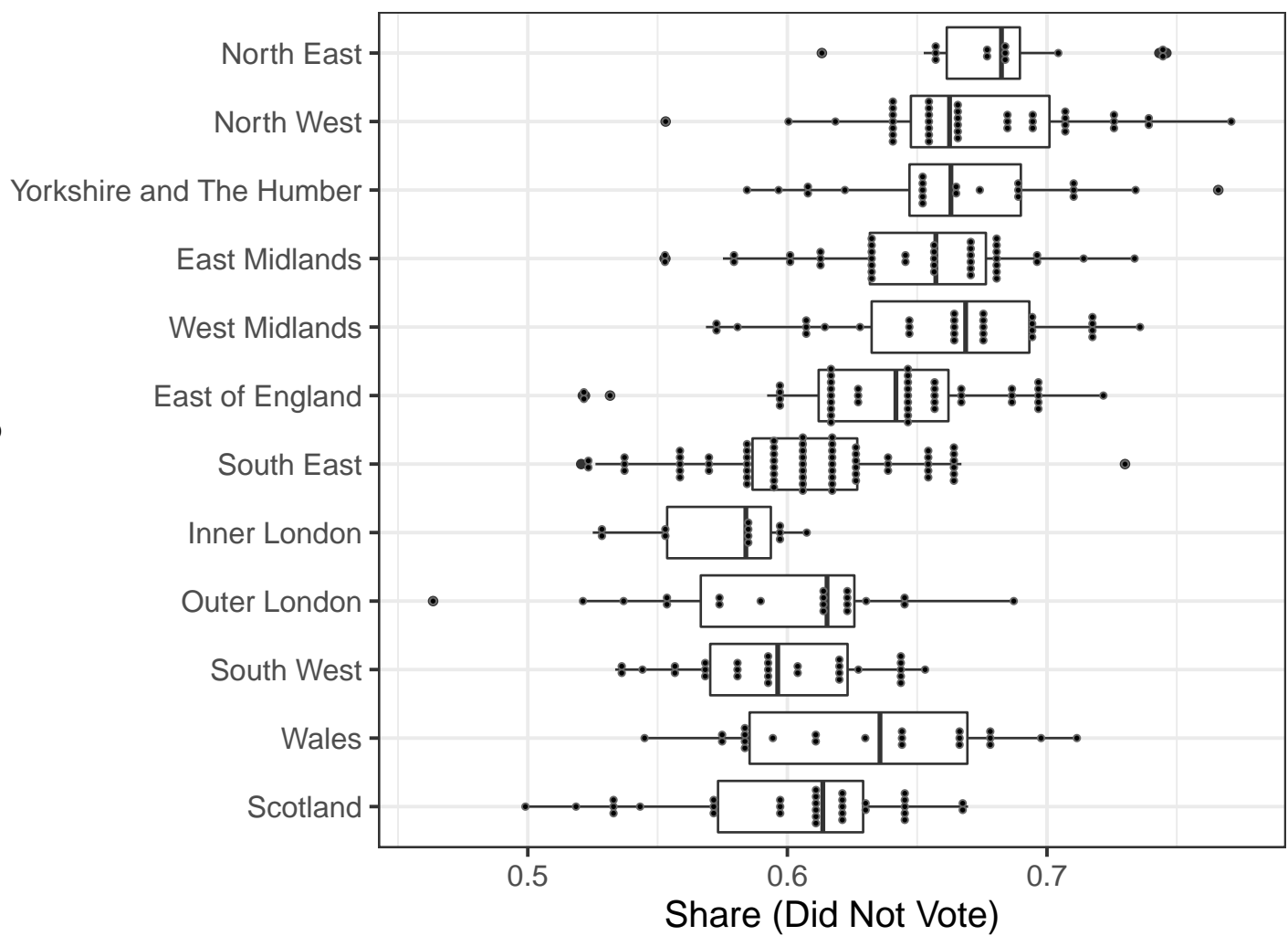

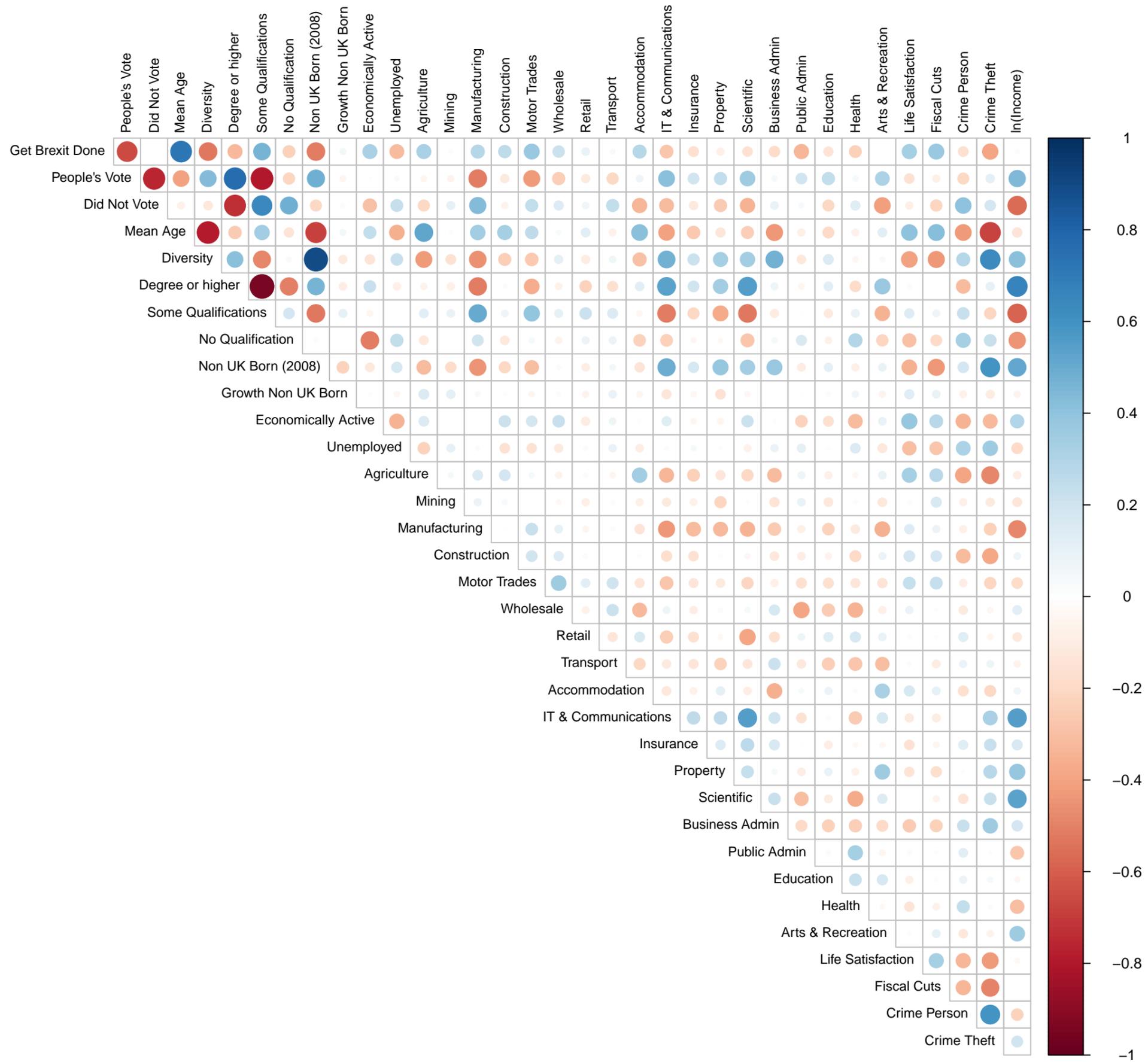

Correlation plot

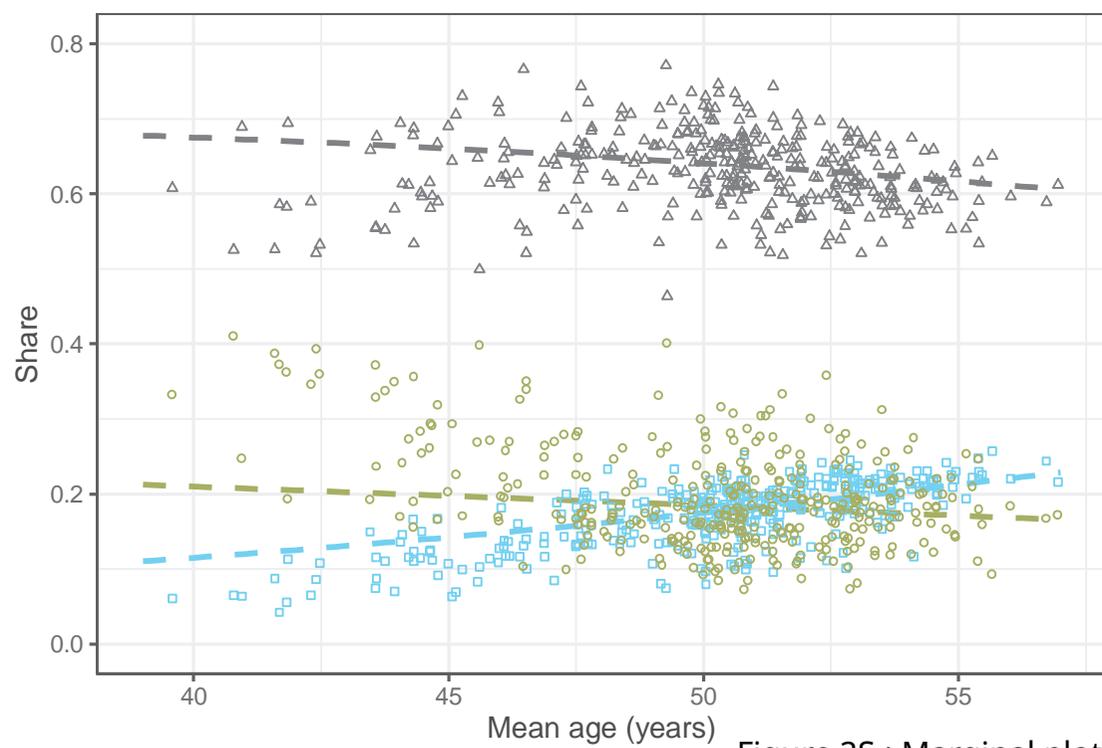
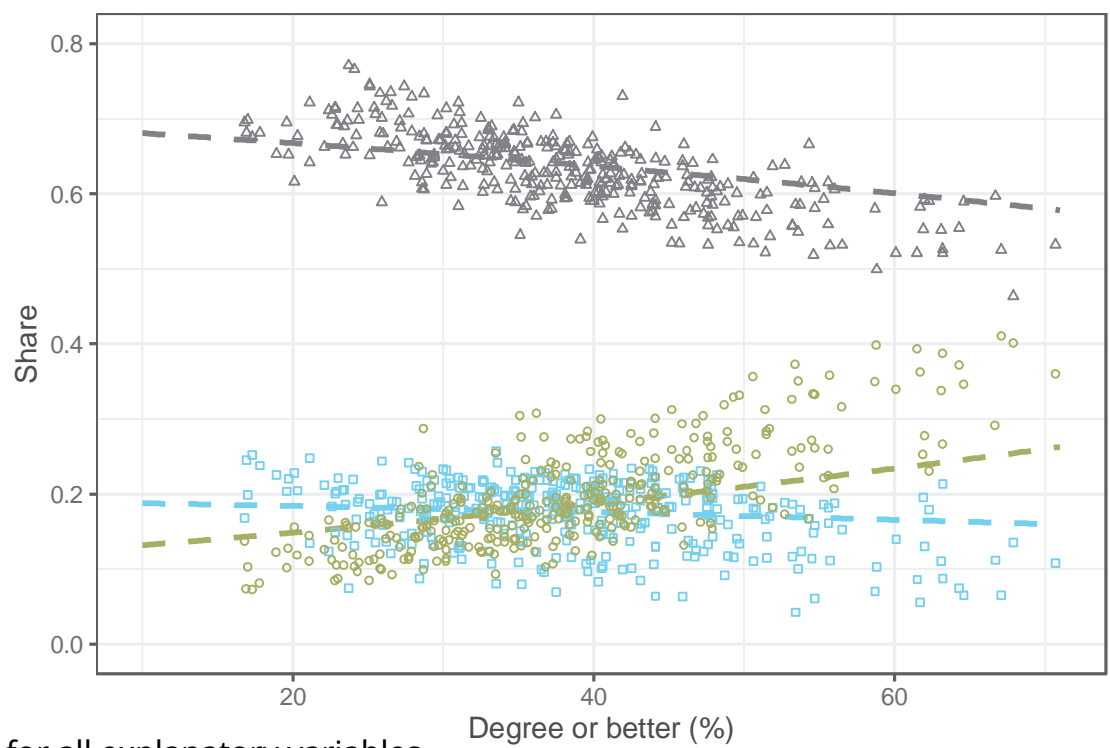
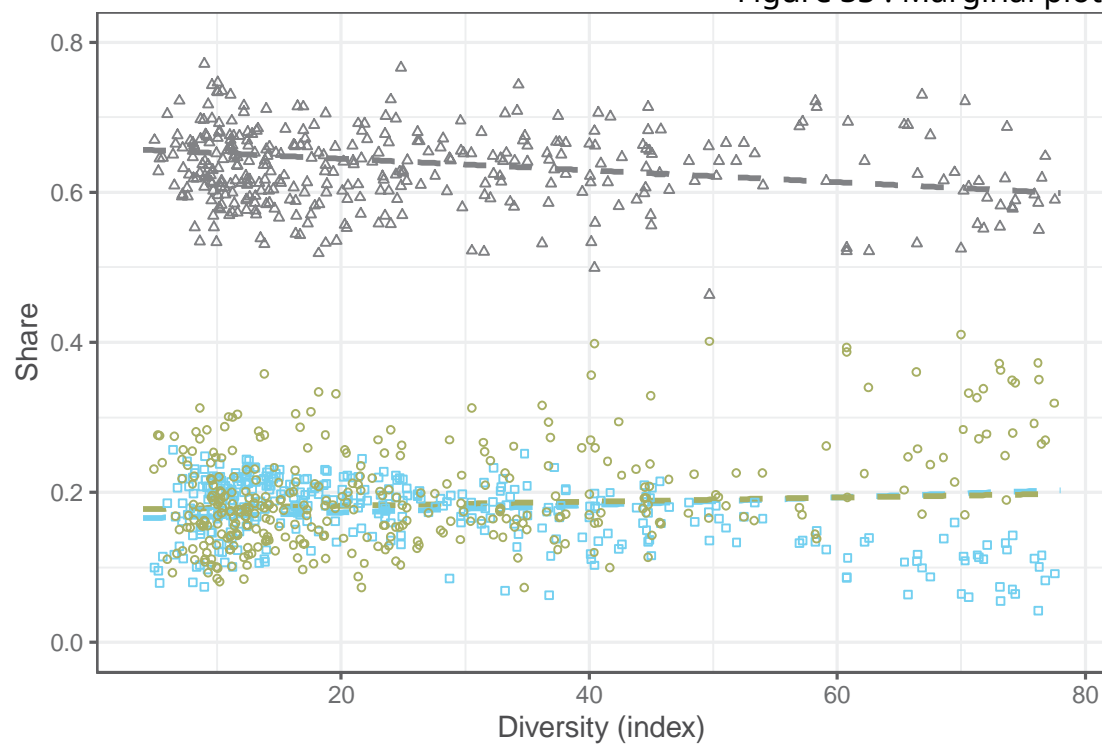
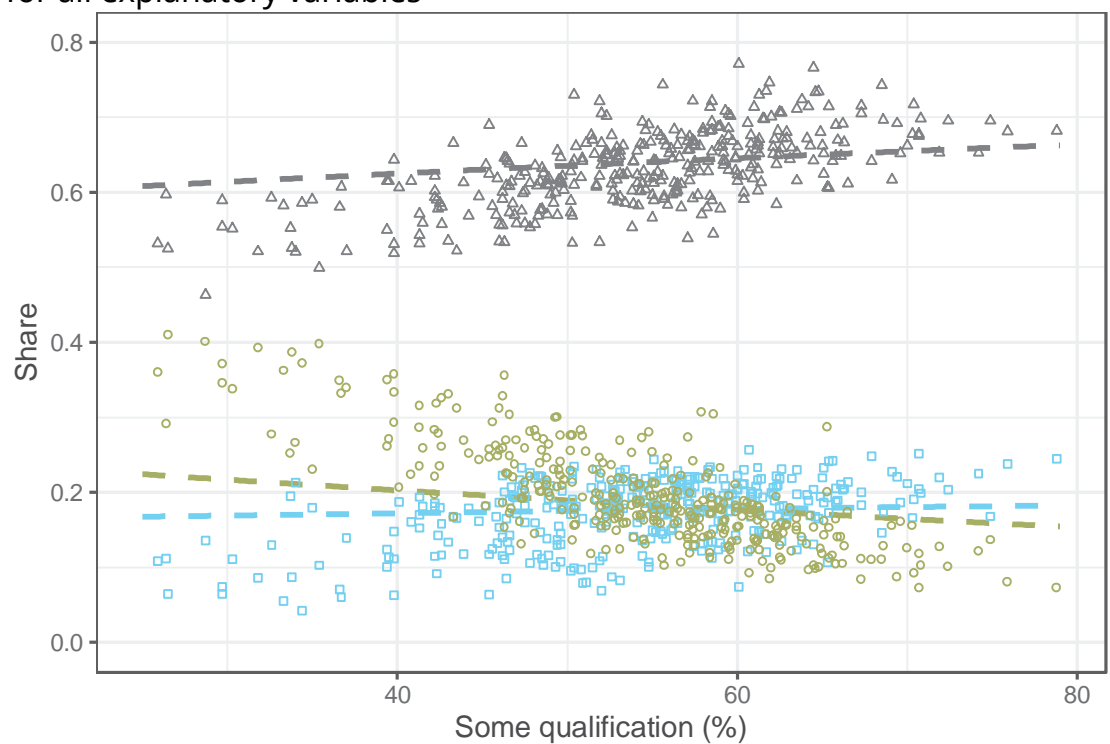

Figure 3S : Marginal plots for all explanatory variables

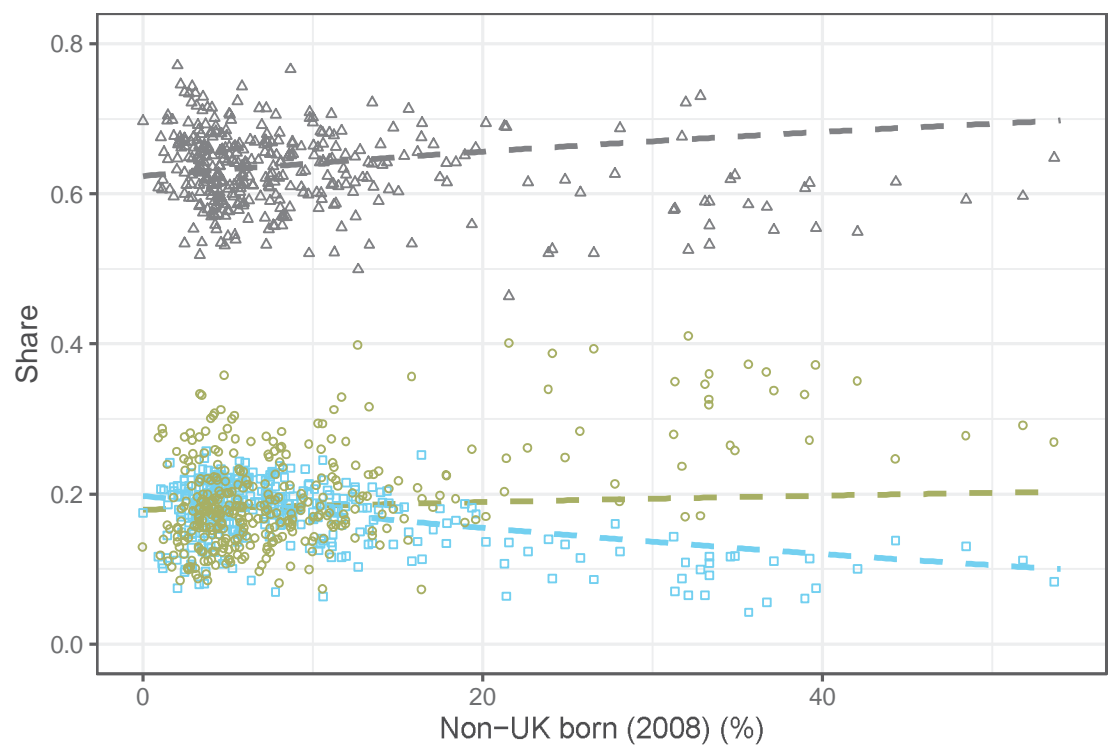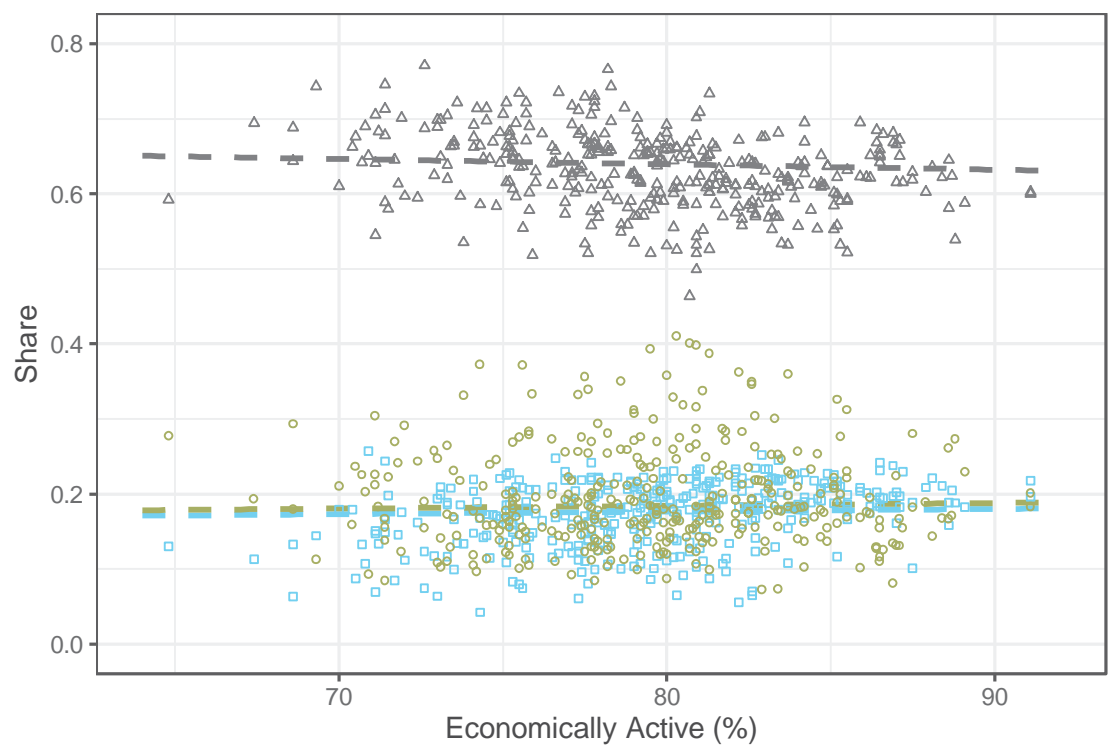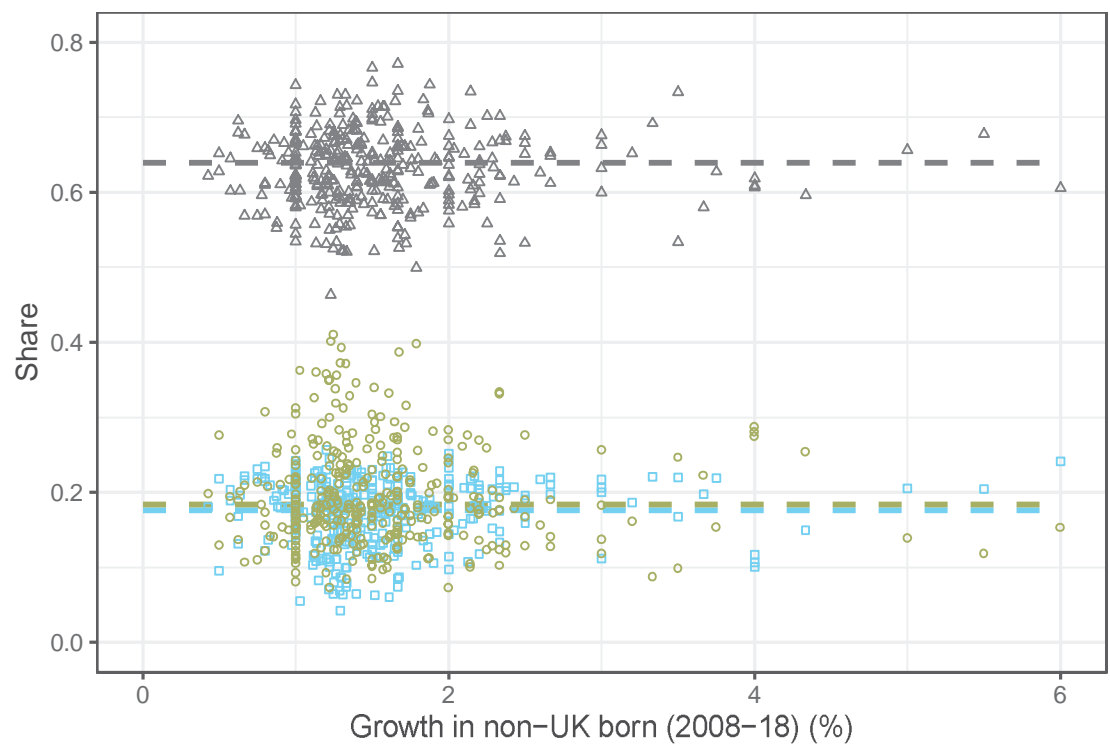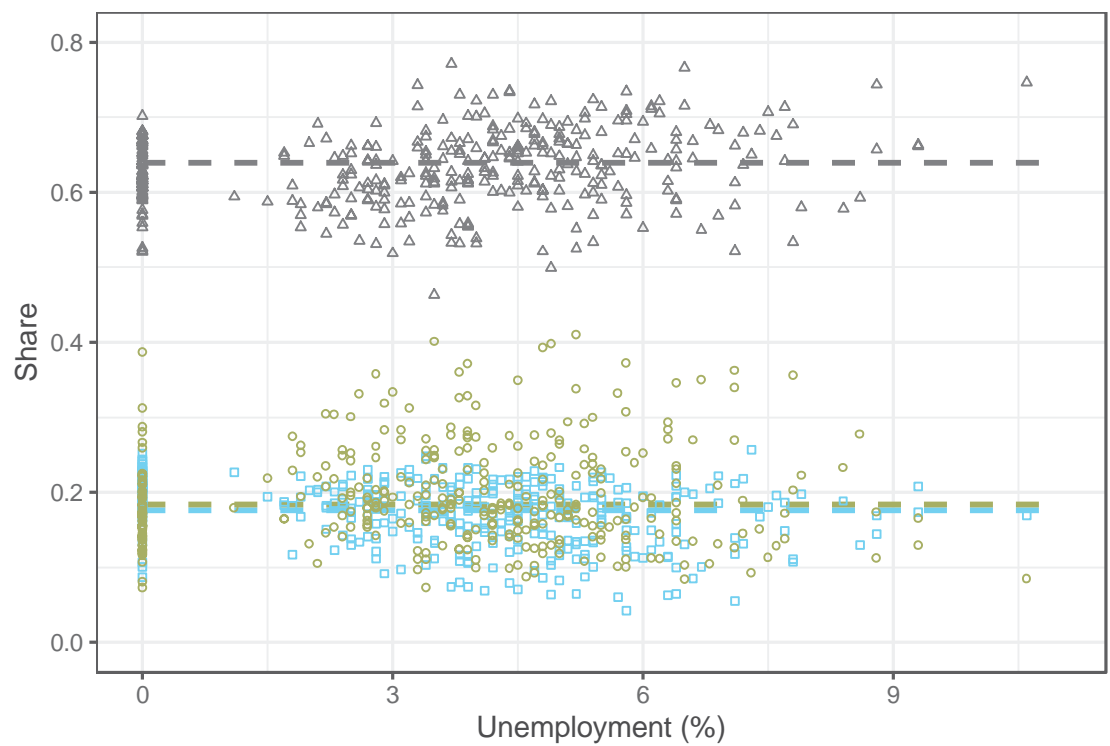

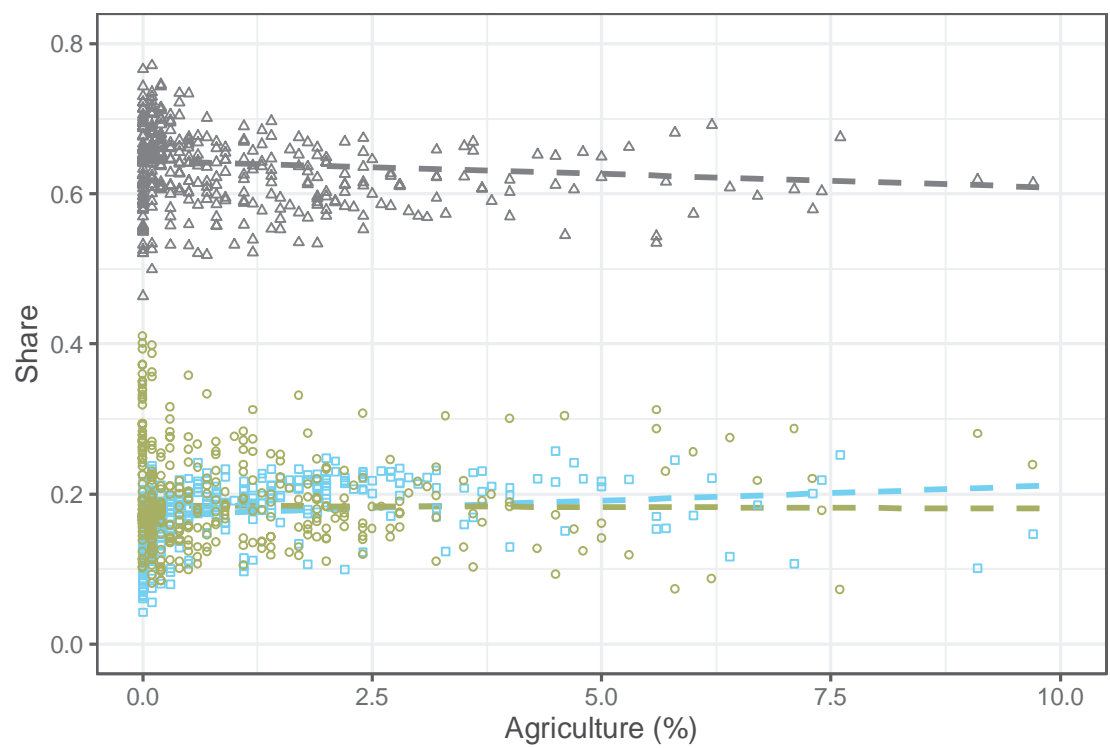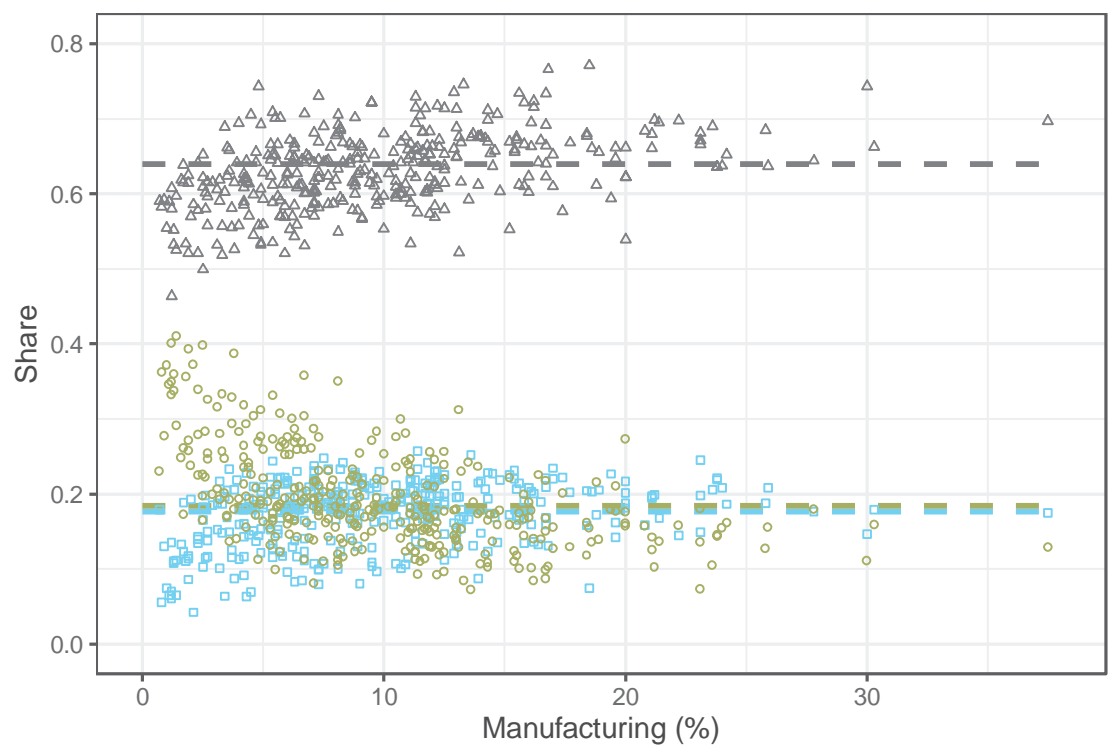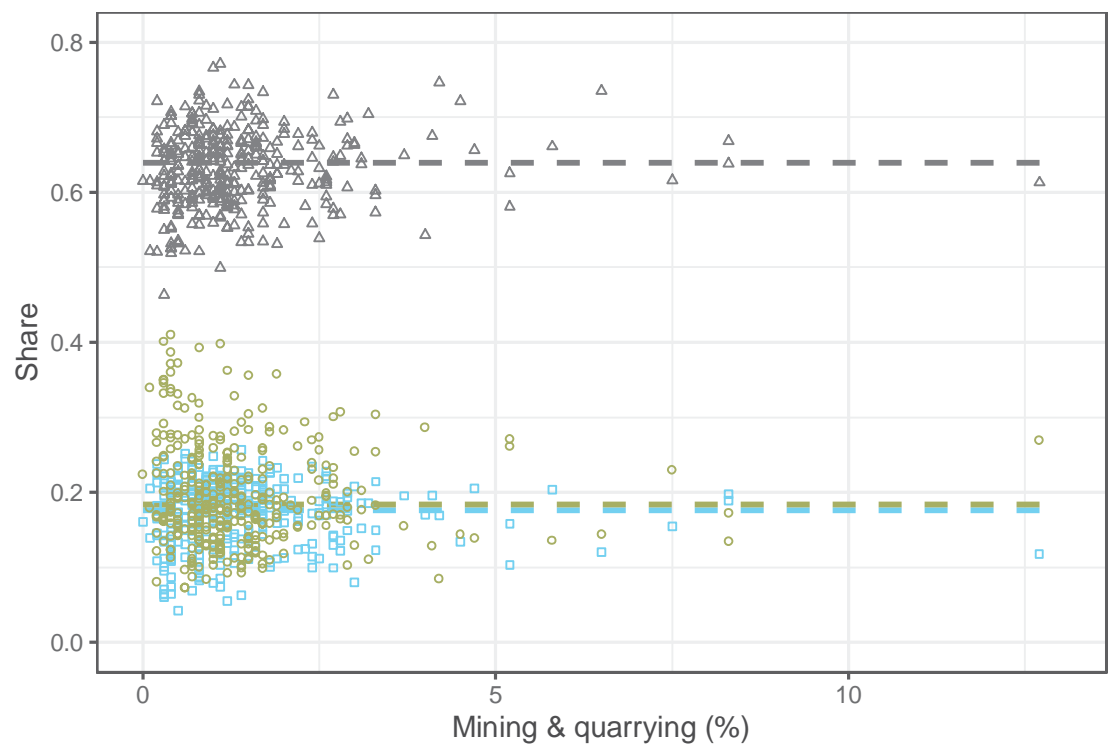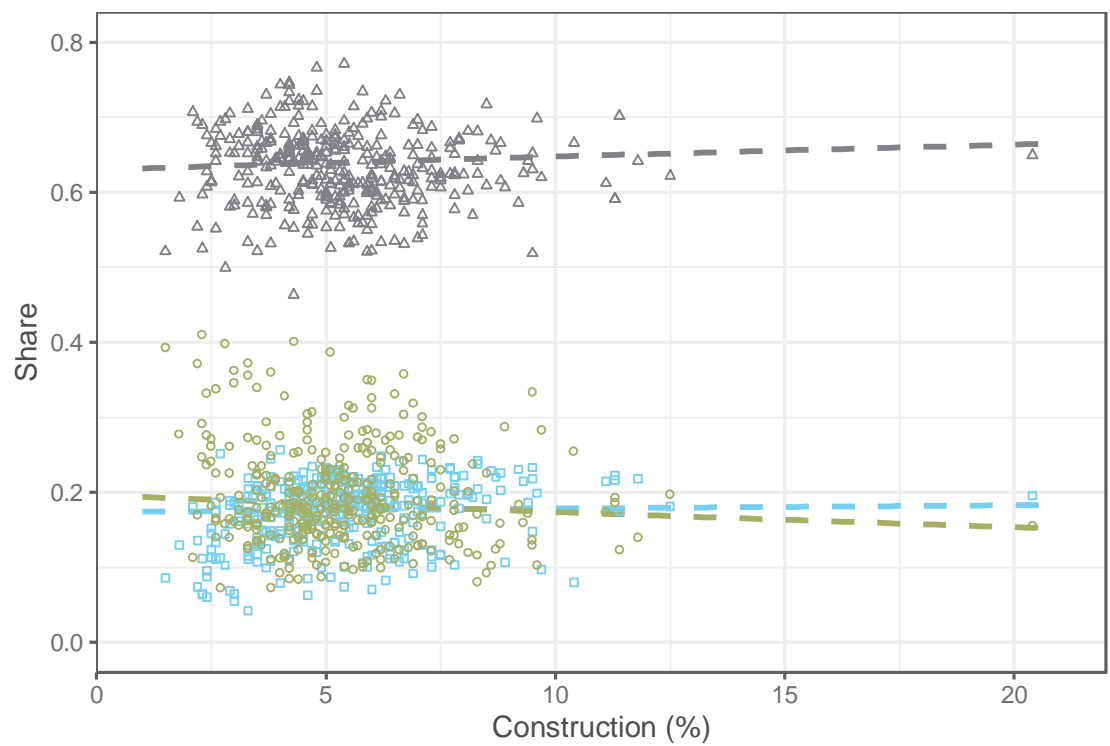

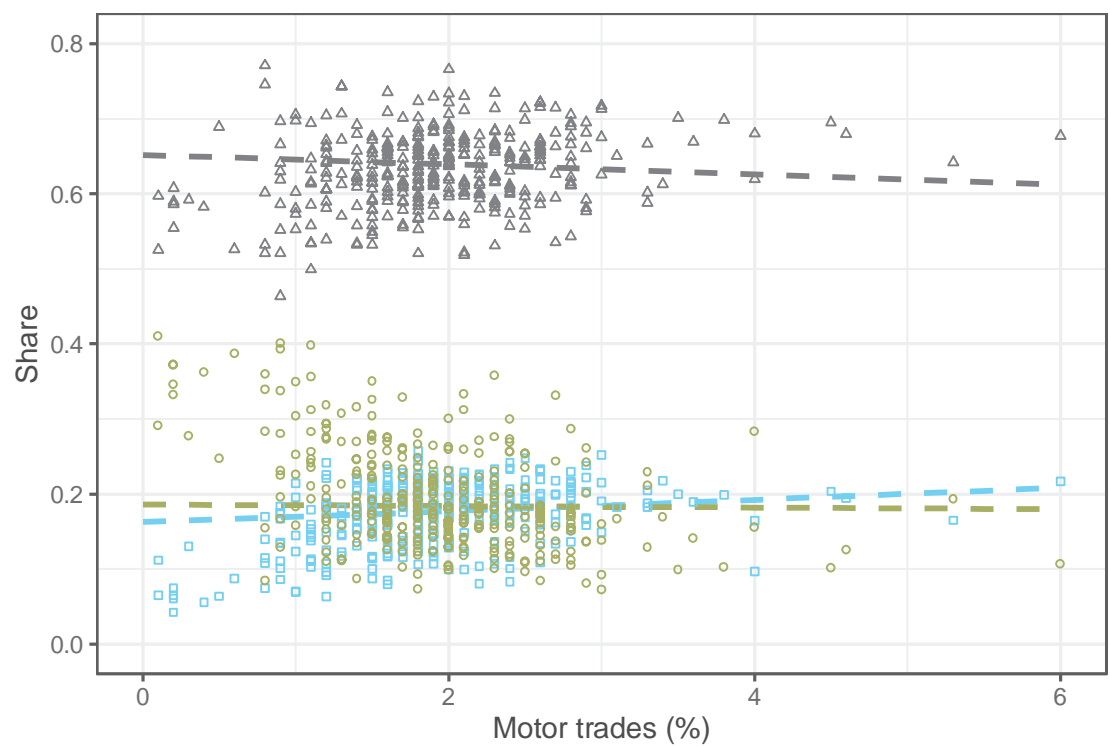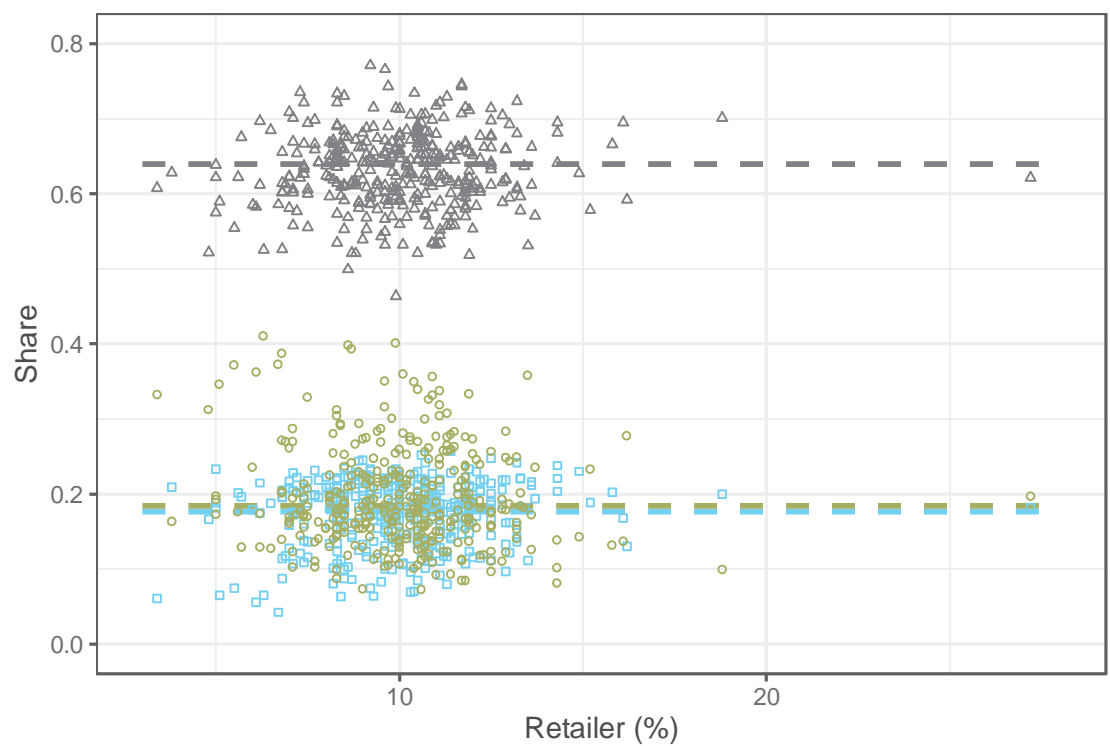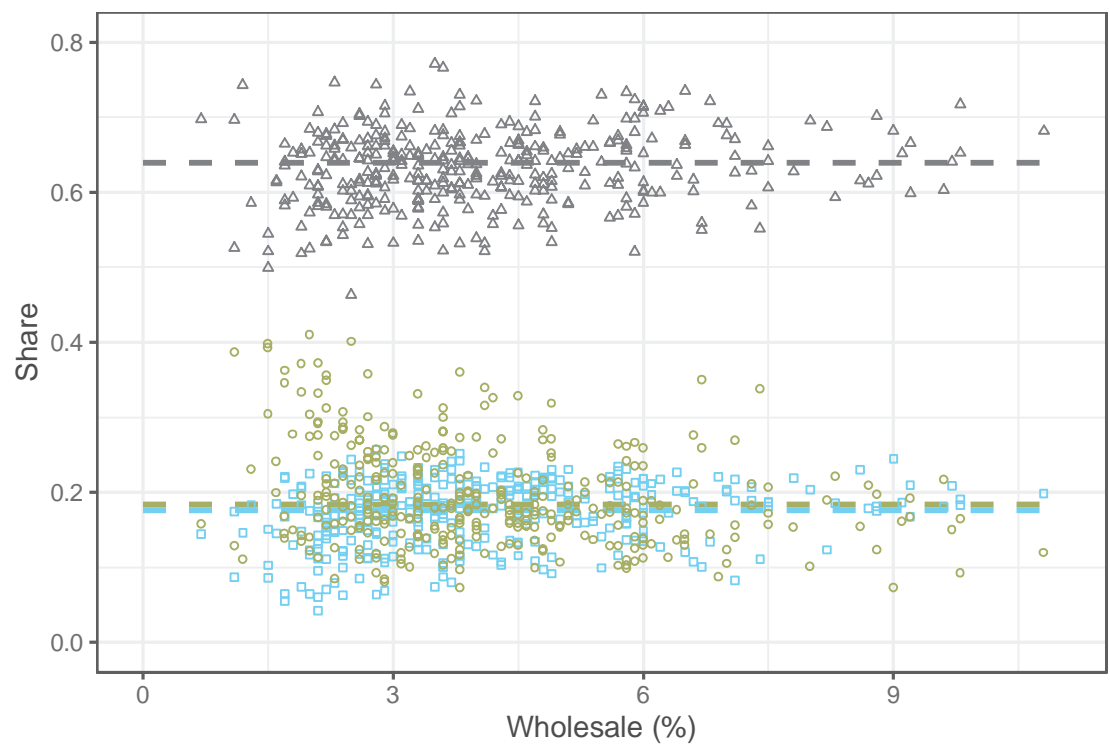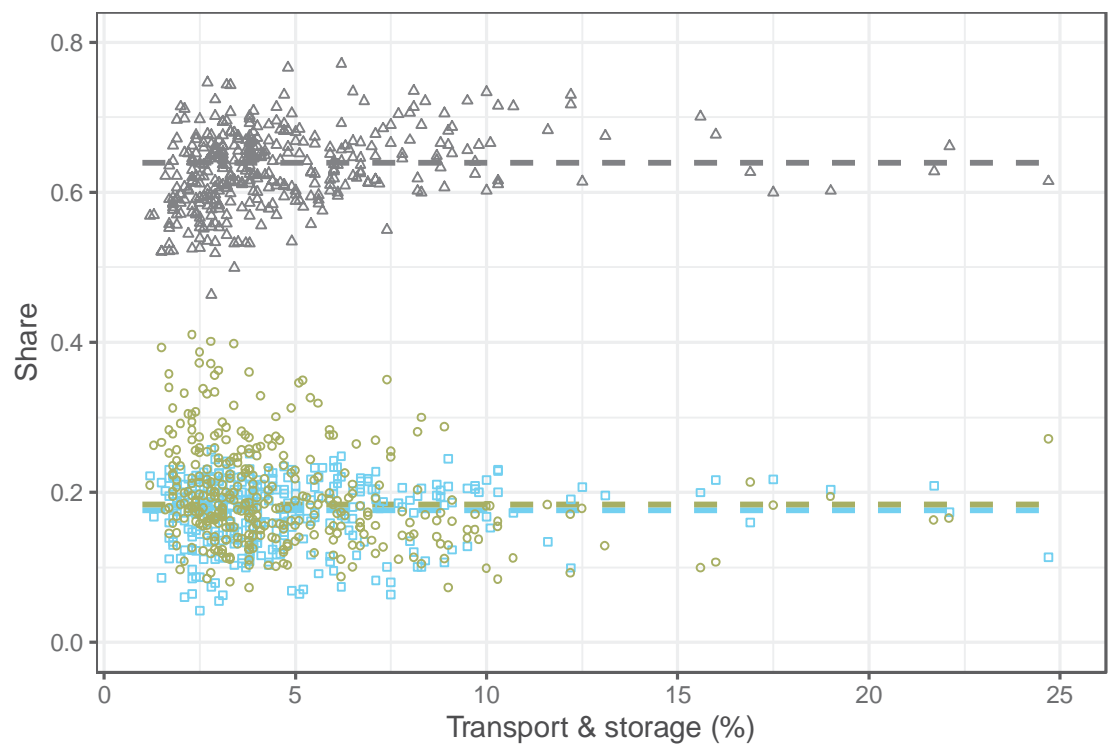

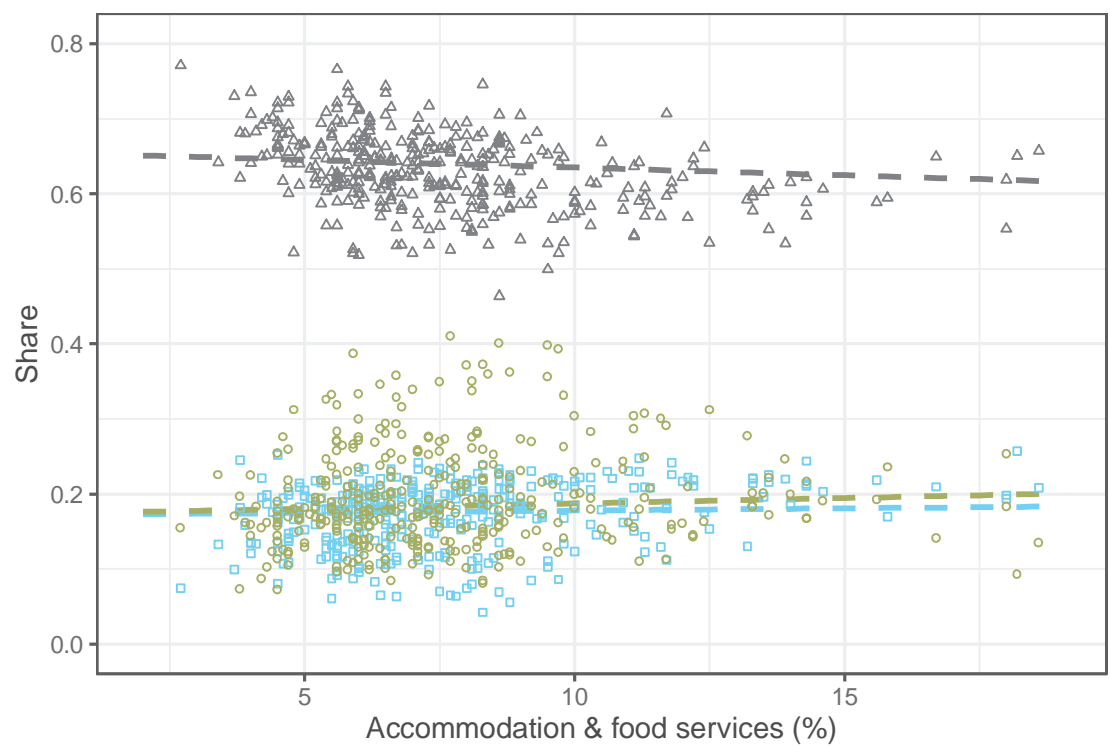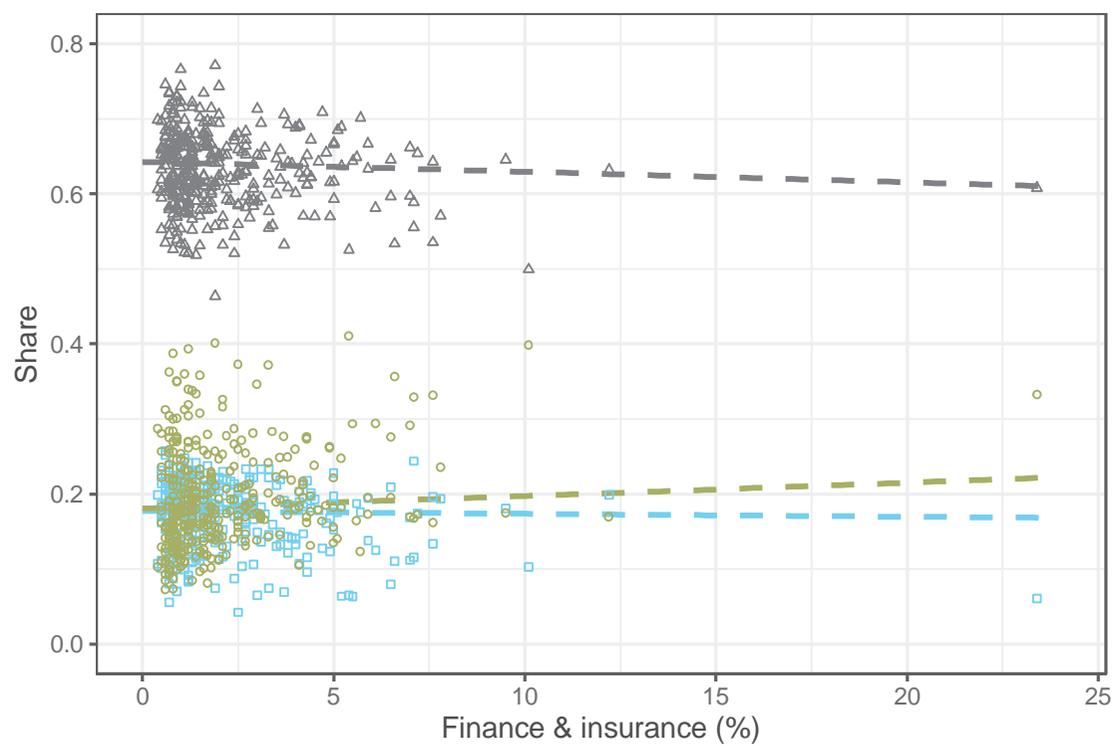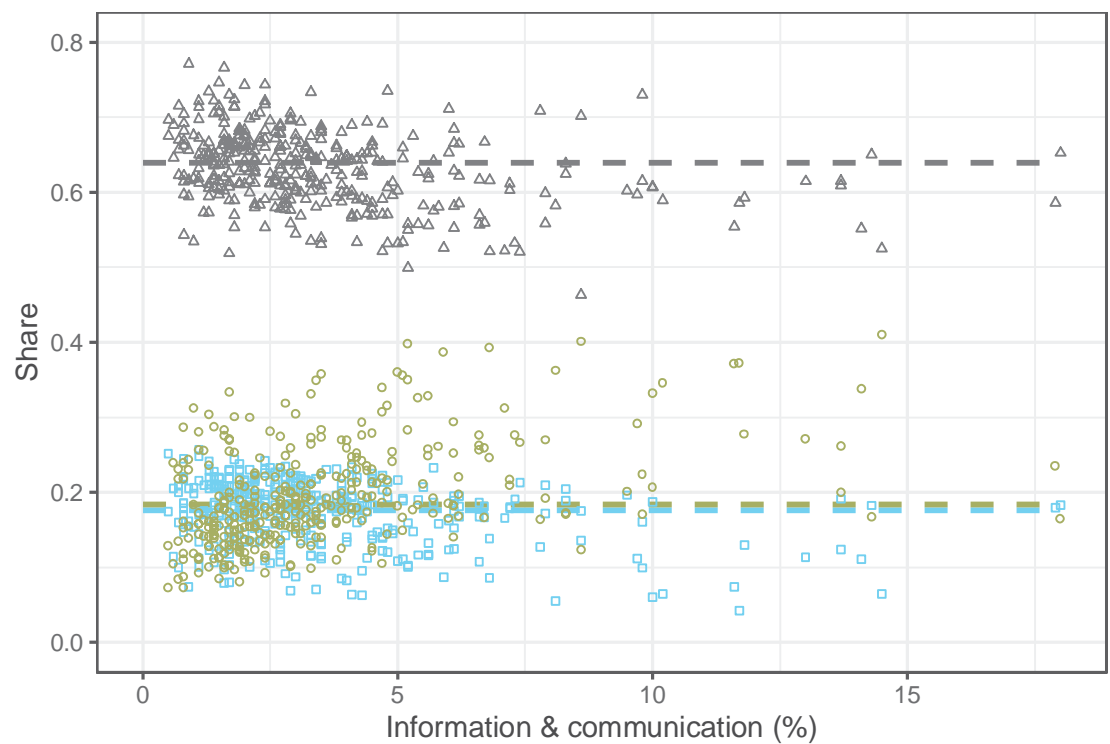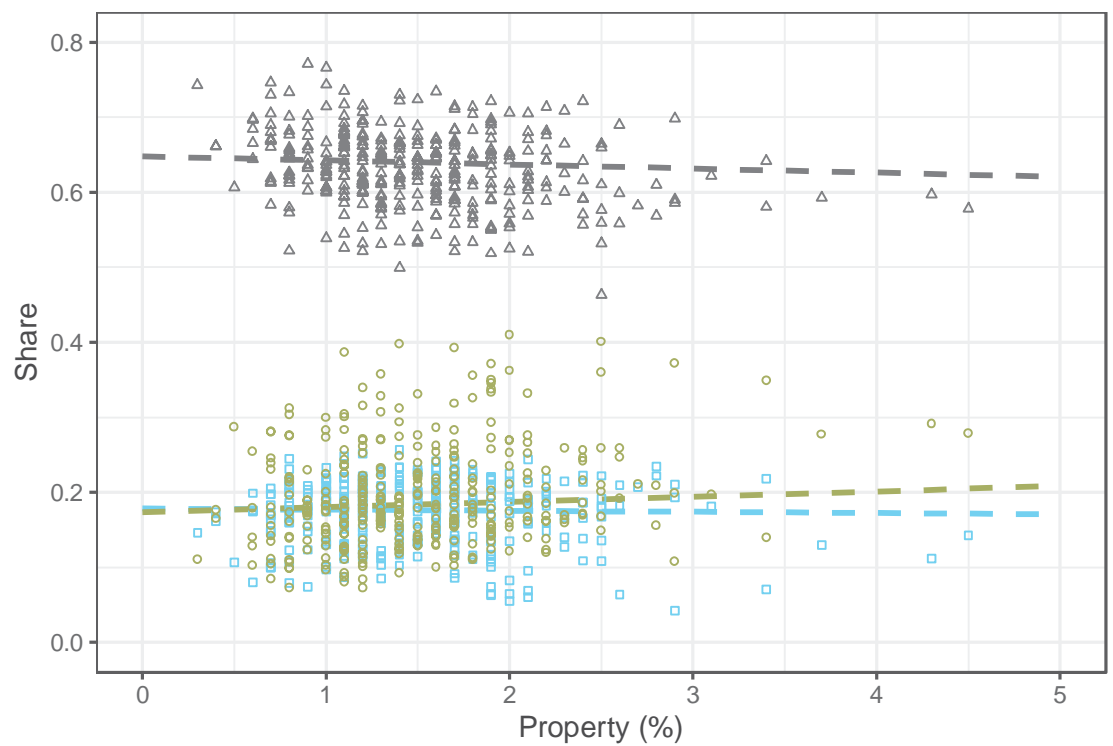

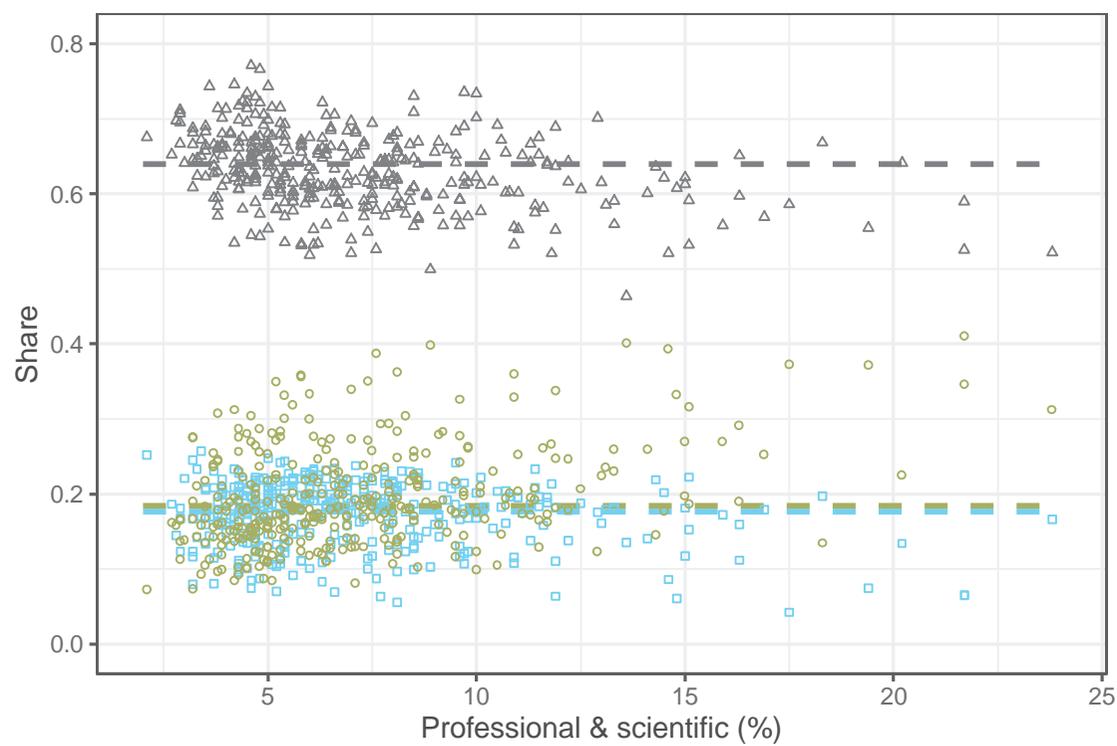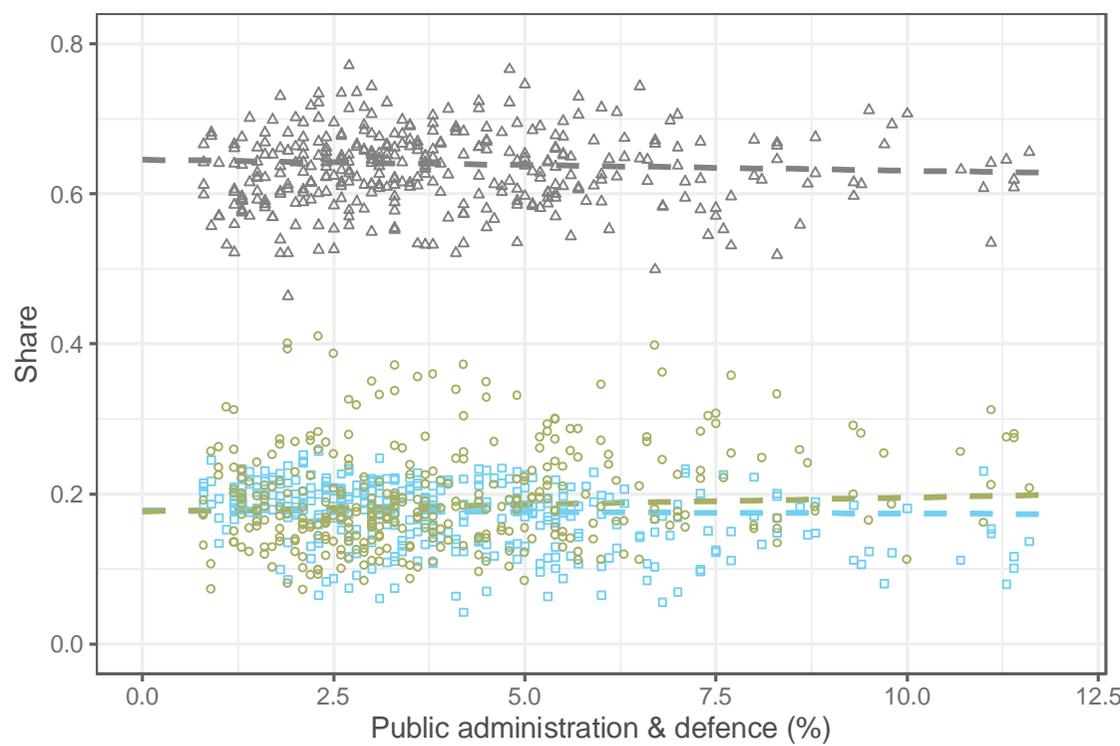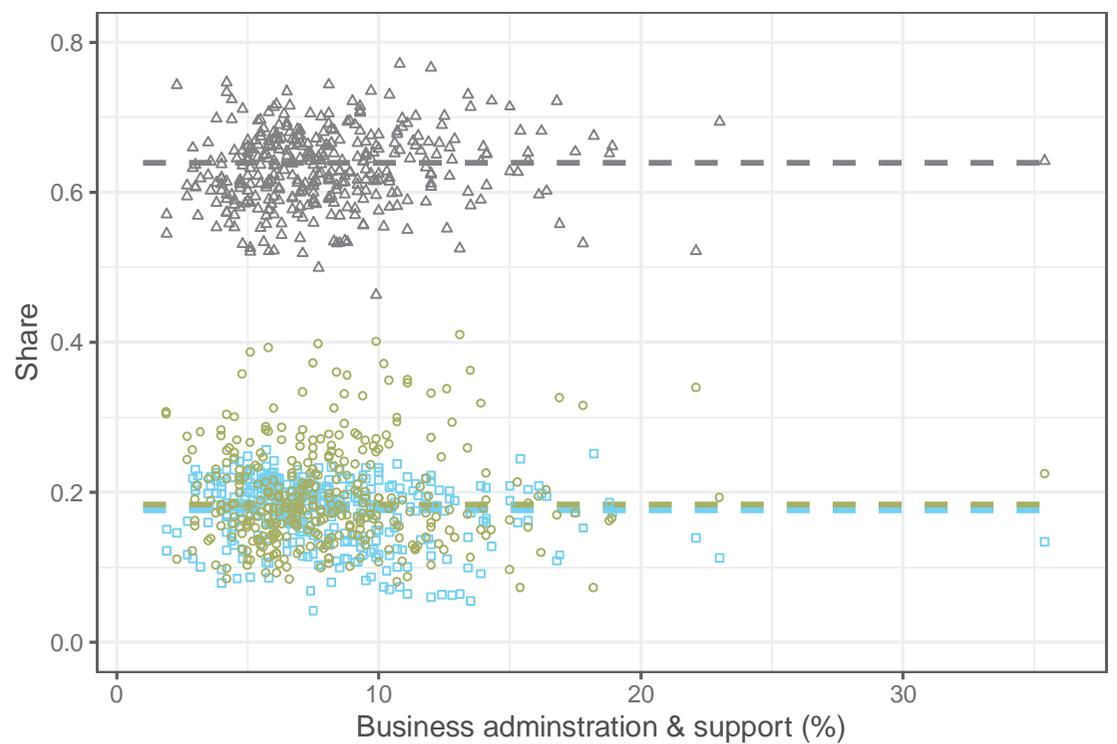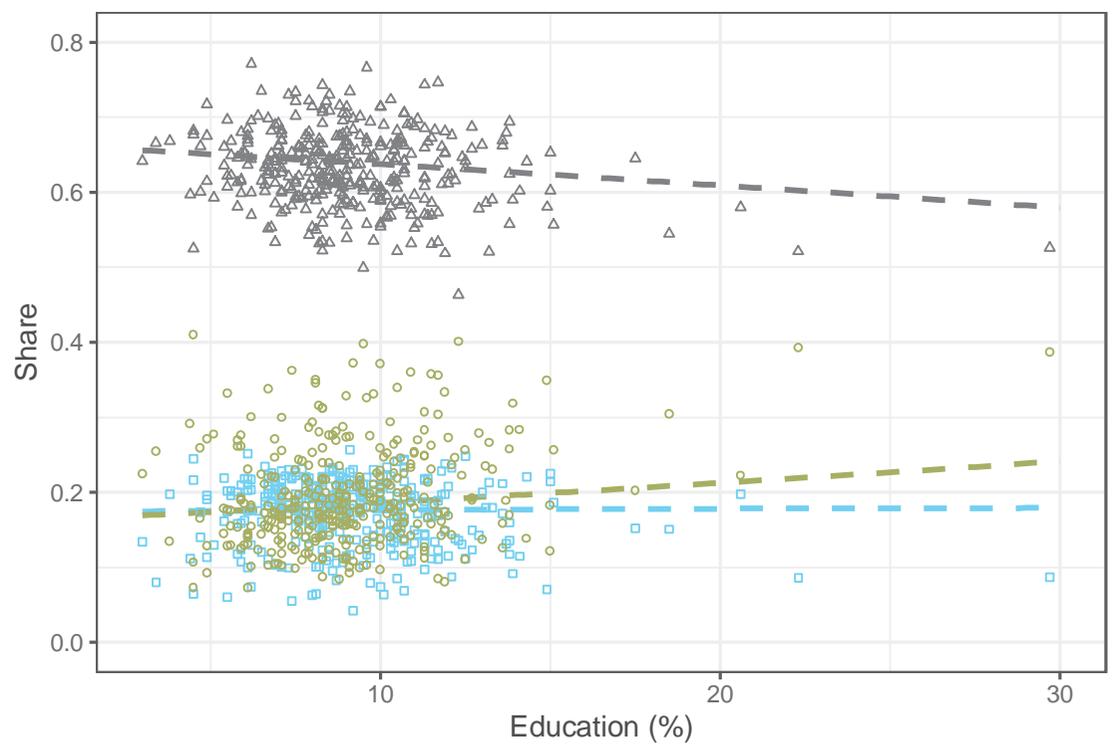

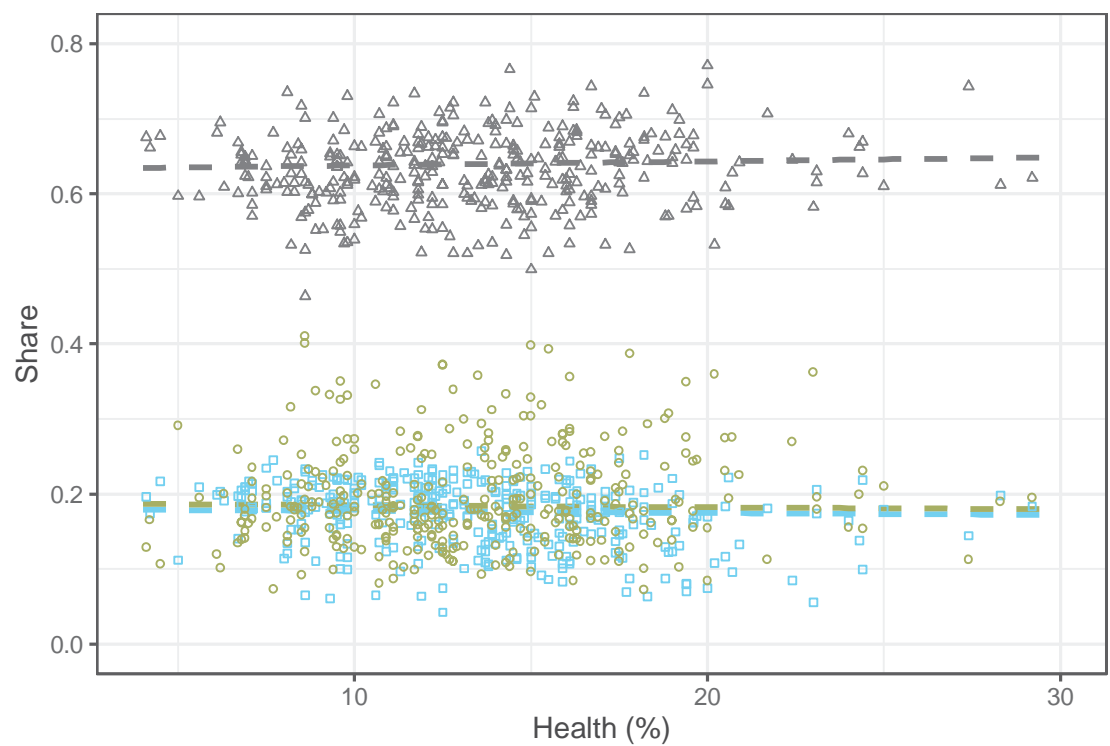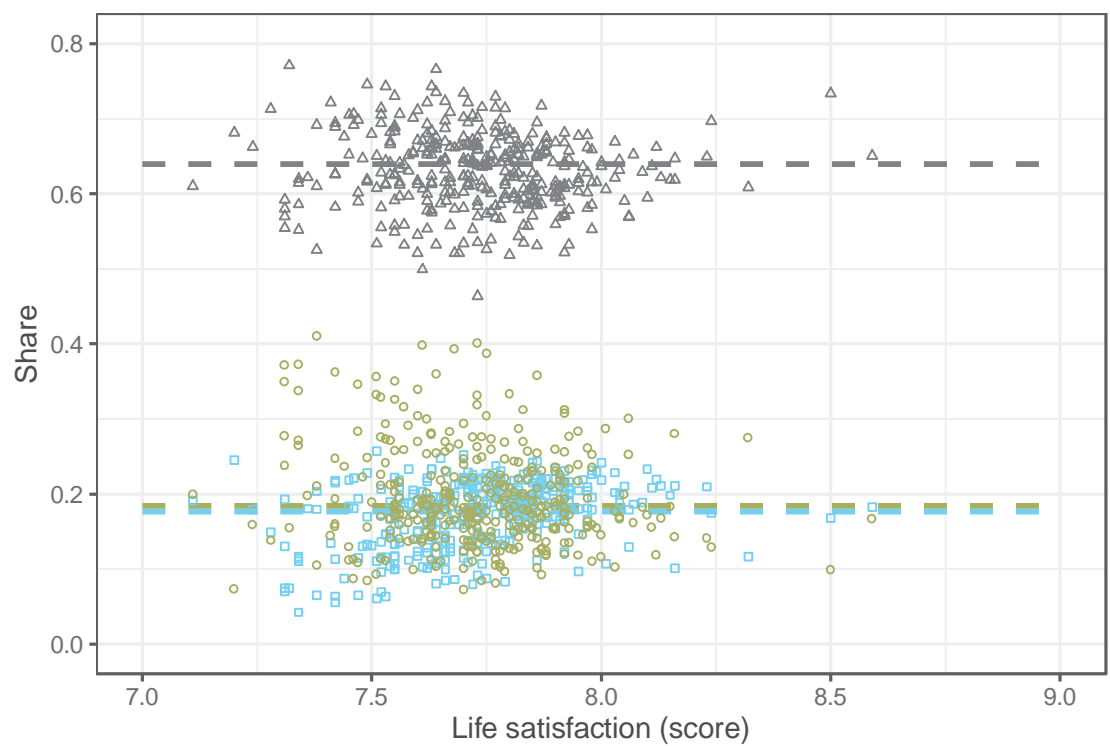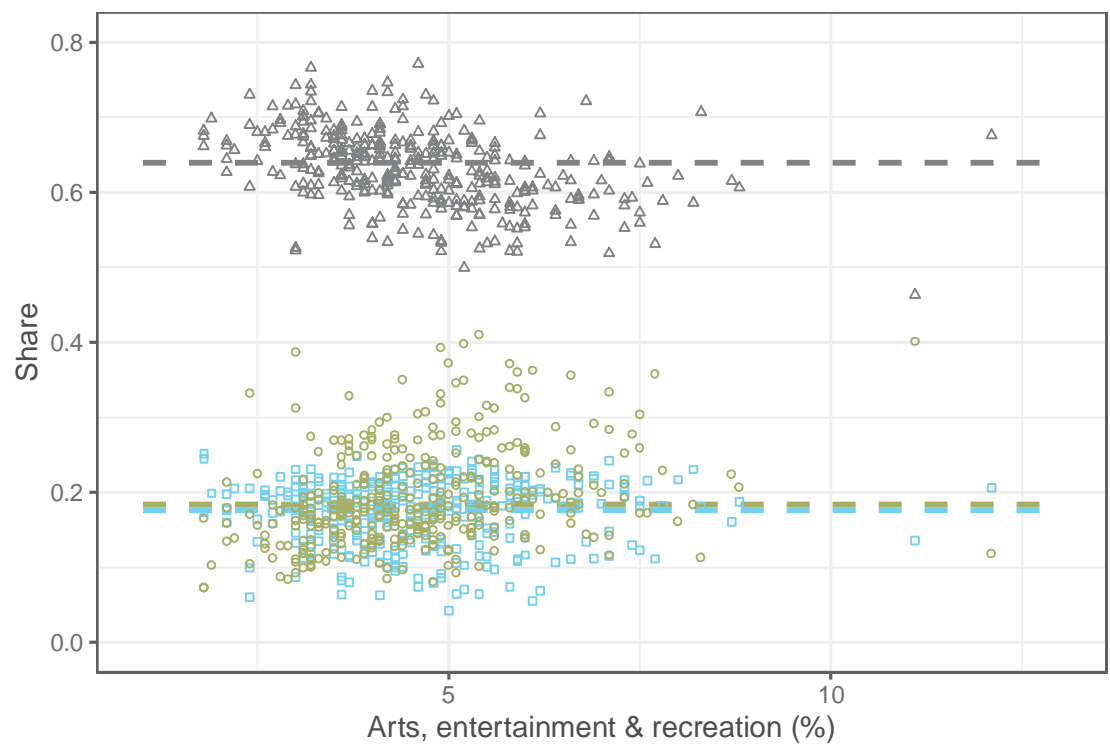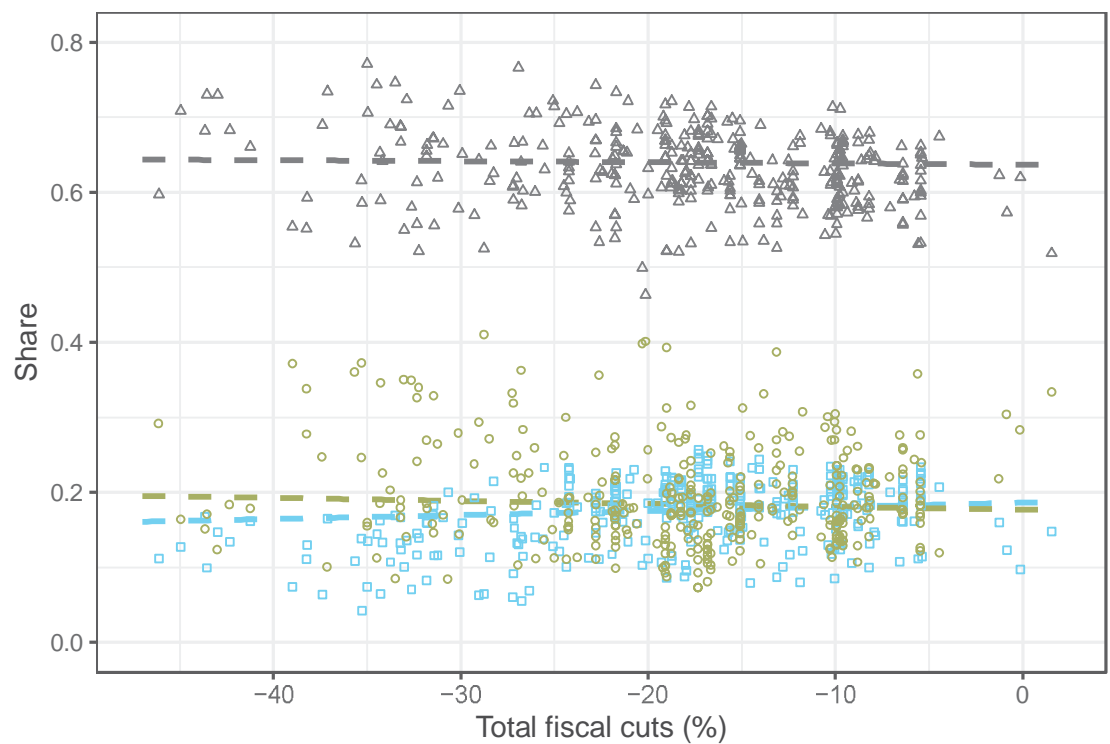

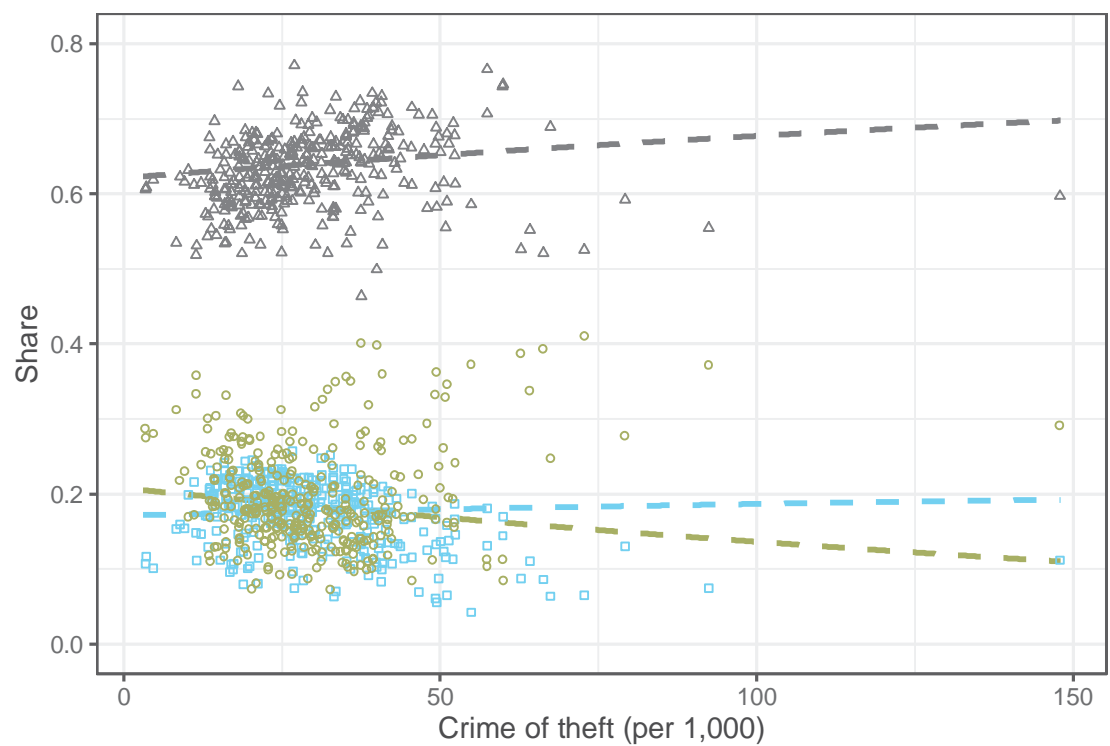
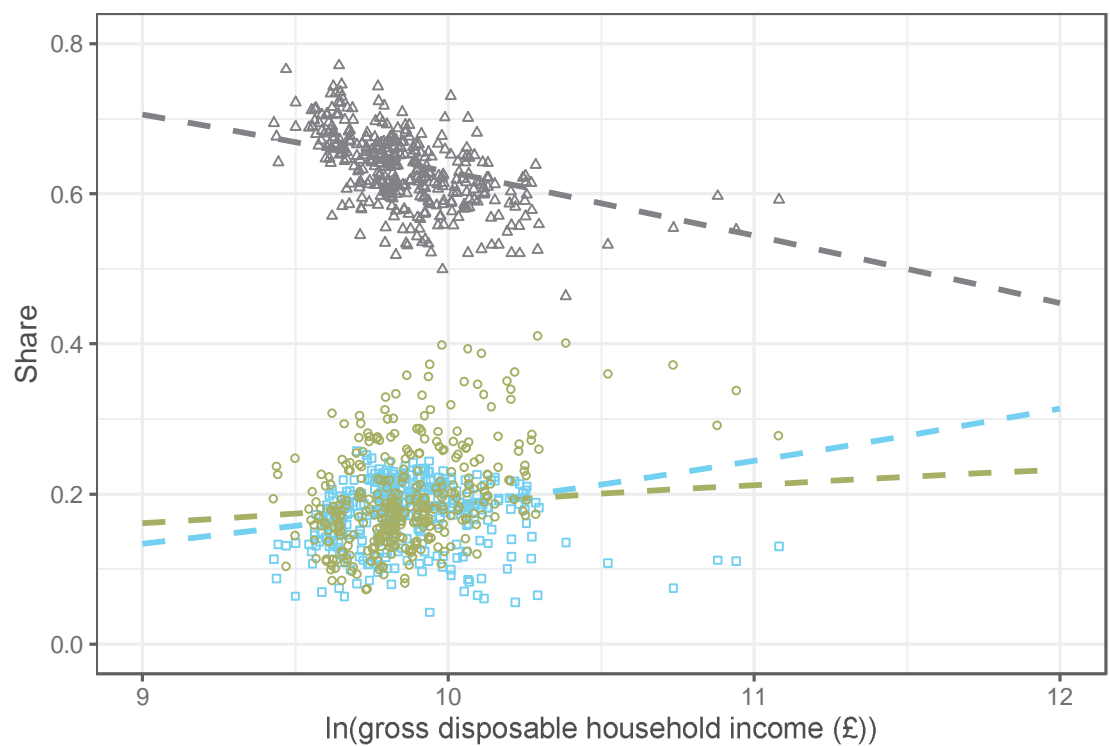
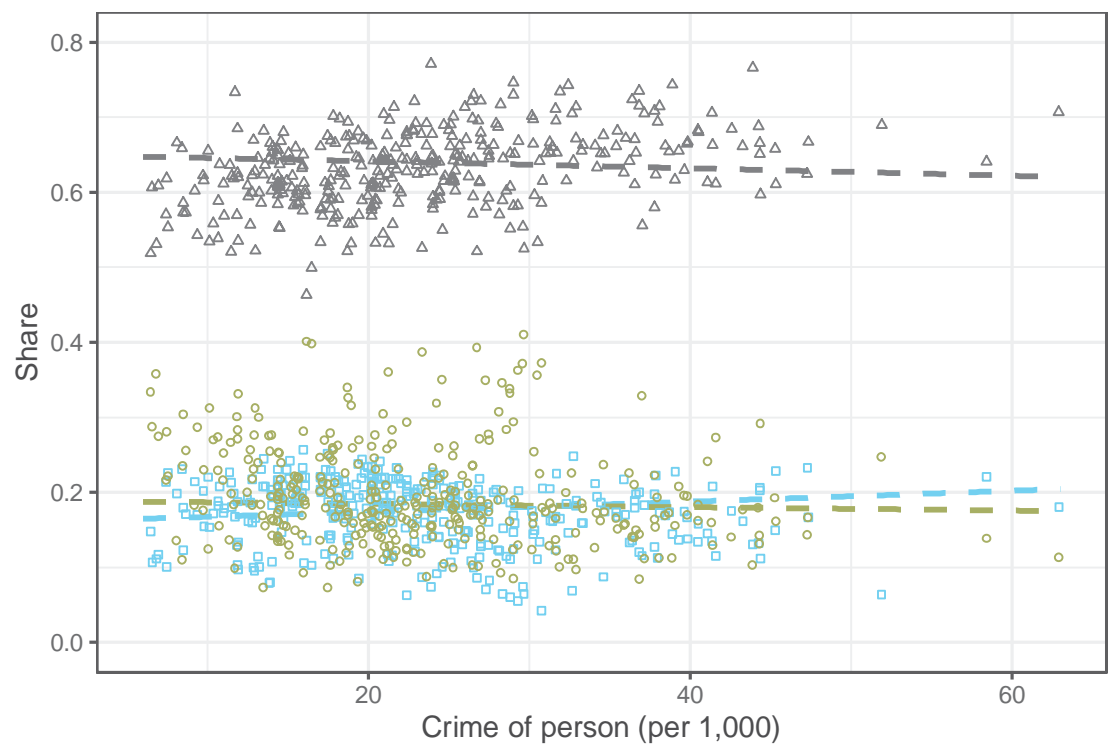